\documentclass[prd,twocolumn,superscriptaddress,floatfix,amsmath,amssymb,amsfonts,nofootinbib,longbibliography]{revtex4-1}
\usepackage{float} 
\usepackage[normalem]{ulem}
\usepackage[english]{babel}
\usepackage{graphicx}
\usepackage{dcolumn}
\usepackage{bm}
\usepackage{blindtext}
\usepackage{verbatim}
\usepackage{relsize}
\usepackage{mathrsfs}
\usepackage{amsmath}
\usepackage{amsfonts}
\usepackage{blindtext}
\usepackage{cancel}
\usepackage{physics}
\usepackage{epstopdf}
\usepackage{mathtools}
\usepackage{blindtext}
\usepackage{tensor}
\usepackage{color}
\usepackage[usenames,dvipsnames]{pstricks}
\usepackage{epsfig}
\usepackage{pst-grad} 
\usepackage{pst-plot} 
\usepackage{hyperref}
\usepackage{verbatim}
\usepackage{slashed}
\usepackage{dsfont}

\usepackage{booktabs}
\usepackage{multirow}
\usepackage{siunitx}   

\newcommand{\phihat}{\hat{\phi}}

\newcommand{\mf}{\mathsf}

\newcommand{\ii}{\mathrm{i}}

\newcommand{\s}{{\hat{\sigma}}}

\renewcommand{\ket}[1]{| {#1} \rangle}
\renewcommand{\bra}[1]{\langle {#1} |}
\renewcommand{\dd}{\mathrm{d}}
\renewcommand{\norm}[1]{|| {#1} ||}

\renewcommand{\braket}[2]{\langle {#1}|{#2}\rangle}

\begin{document}
\title{A review of applications of Quantum Energy Teleportation:\\ from experimental tests to thermodynamics and spacetime engineering}

\author{Boris Ragula}
\email{bragula@uwaterloo.ca}
\affiliation{Department of Applied Mathematics, University of Waterloo, Waterloo, Ontario, N2L 3G1, Canada}
\affiliation{Institute for Quantum Computing, University of Waterloo, Waterloo, Ontario, N2L 3G1, Canada}
\affiliation{Perimeter Institute for Theoretical Physics, Waterloo, Ontario, N2L 2Y5, Canada}

\author{Eduardo Mart\'{i}n-Mart\'{i}nez}
\email{emartinmartinez@uwaterloo.ca}
\affiliation{Department of Applied Mathematics, University of Waterloo, Waterloo, Ontario, N2L 3G1, Canada}
\affiliation{Institute for Quantum Computing, University of Waterloo, Waterloo, Ontario, N2L 3G1, Canada}
\affiliation{Perimeter Institute for Theoretical Physics, Waterloo, Ontario, N2L 2Y5, Canada}

\begin{abstract}
    Quantum energy teleportation (QET) exploits the existence of correlations to enable remote energy transfer without the need for physical energy carriers between emitter and receiver. This paper presents a review of the thermodynamic foundations of QET and reviews its first experimental demonstration (performed using Nuclear Magnetic Resonance), along with its implementation on publicly available superconducting quantum hardware. Additionally, we review an application of QET in the field of quantum thermodynamics  as an efficient algorithmic cooling technique to cool down individual parts of interacting systems. Finally, we will review how QET can be employed to optimally generate exotic quantum states characterized by negative average stress-energy densities, offering a new operational approach to engineering such states which are promising in the context of semiclassical gravity. 
\end{abstract}

\maketitle
\tableofcontents

\section{Introduction}

The laws of thermodynamics governing the flow of energy between interacting systems are, arguably, well understood in the classical regime. However, these classical laws are less straightforward when considering systems operating in regimes where quantum effects start to become be non-negligible. Protocols from quantum information and quantum thermodynamics have necessitated refined thermodynamic principles capable of more accurately describing energy transfer and extraction from quantum systems. Among these is the so-called Quantum Energy Teleportation (QET) protocol, introduced in a series of seminal works by Masahiro Hotta~\cite{Hotta2008,Hotta2009,HottaMinimal}. The QET protocol exploits correlations in many-body quantum systems to transfer energy between two subsystems without the need for any energy to physically propagate between them. As a consequence of this, the transfer can happen at rates that may exceed the speed at which energy propagates within the system, only limited by the speed at which two external parties can communicate. 

In this paper we review several results where QET is applied to different scenarios: from the first experimental verifications of QET, its application to algorithmic cooling and the creation of exotic matter in quantum field theoretical settings.

We will begin in Sec.~\ref{sec:Protocols} by providing a brief review of the notion of Strong Local Passivity~\cite{FreyFunoHottaSLP,iffSLP} (i.e., the inability to extract energy with local operations from systems with entangled ground states), and how QET breaks local passivity by allowing for local operations plus communication. We will then review two implementations of a minimal QET protocol, one using LOCC and another one that uses quantum communication (called the fully unitary QET protocol).

After this we will review in detail two experimental implementations  of QET. First, an implementation utilizing Nuclear Magnetic Resonance in Sec.~\ref{sec:IQCImplementation}, that constituted the first experimental proof that the protocol works in a laboratory setting~\cite{IQCQETExperiment}, transmitting information between two atoms in an organic molecule without energy propagating between them and in times much shorter than the time it would take for the energy to naturally propagate between the atoms. Second, in Sec.~\ref{sec:SCImplementation}, we will review an implementation of the protocol in IBM's quantum hardware that appeared shortly after the NMR experiment~\cite{Ikeda2023}.

After that, in Sec.~\ref{sec:AlgorithmicCooling} we will review how QET can be used in practice to outperform current algorithm cooling protocols when it comes to cooling down subsystems of a larger interacting system~\cite{QETAlgorithmicCooling}. \textit{Algorithmic cooling}~\cite{BoykinMorAC,NayeliACPRL,RaeisiMoscaAC,BaughLaflamme2005} is the generic name given to the family of protocols that---through the manipulation of quantum information---can cool down individual components of a larger quantum systems below the temperature of a thermal bath with which the system can be in contact. Most protocols to date were conceived for systems with little to no internal interactions~\cite{FernandezMorAC,PPA1,Schulman2007,Elias2006}. The challenge for many of these algorithmic cooling methods is that the interaction introduces a non-zero entropy in the subsystems even when the total system is cooled down to absolute zero due ground state entanglement. QET excels at using these correlations to cool down the individual subsystems. 

Finally, in Sec.~\ref{sec:Engineering} we will review how the QET protocol can be used to generate   custom-shaped violations of the weak energy condition when applied to a quantum field~\cite{nichoTeleport}. In particular, we will show that when we apply QET to break the local passivity of the vacuum state of a quantum field, the extraction of energy by Bob results in the generation of an area around Bob's measurement where the field's energy density becomes negative. We show that QET is optimal in creating negative energy densities and that it can saturate fundamental scaling limits for the violation of the weak energy condition. This is particularly interesting since it is expected that the gravitational properties of that kind of exotic matter may allow for exotic spacetime backreactions.

\section{A review of Strong Local Passivity and QET}\label{sec:Protocols}

Before we review the experimental implementations of QET protocols, its application in algorithmic cooling and the engineering of states of exotic matter, let us first present an overview of the notion of Strong Local Passivity and a brief summary of two different implementations of the QET protocol. First we will revisit the minimal QET protocol~\cite{Hotta2010,HottaMinimal}, where the communication between the parties is classical, and then the fully unitary QET protocol (as it was used for instance in~\cite{IQCQETExperiment,nichoTeleport,DerivativeQET,GuillaumeThesis}), where the communication between the parties is done through a quantum channel.

\subsection{Concepts of Strong Local Passivity}

QET can be thought of as a way to extract energy from states where apriori one would think energy cannot be extracted. In particular, we are talking about states that are passive under local operations. To fully understand why QET is a powerful protocol we need to have some surface understanding of the notion of Strong Local Passivity. This notion arises when trying to analyze energy flow within microscopic systems that are in, or near, their ground state, (where quantum effects start to become relevant) or near more general passive states. Consider a  multipartite system to which only local access to a single subsystem is granted. A state of the system is Strong Local Passive (SLP) if it is impossible to extract energy through only local operations (unitary or not) on the accessible subsystem. Specifically, one may be interested in understanding how, exactly, quantum effects and correlations within the system may influence the flow of energy into and out of the system. In very plain words: Is it possible to touch a very hot system and get frozen instead of burnt due to these effects? 

A pioneering analysis of SLP states was given in~\cite{FreyFunoHottaSLP}. The authors proved a set of sufficient conditions for when it is impossible to extract energy from multipartite systems where unrestricted local access to only a single subsystem is allowed. Concretely, they showed that a \textit{sufficient} condition for Strong Local Passivity in states that are diagonal in the full system Hamiltonian eigenbasis $\{\ket{E_n}\}$:
\begin{equation}
\hat \rho=\sum_{n}p_n\ket{E_n}\!\bra{E_n}
\end{equation}
is the existence of a non-degenerate entangled ground state with full Schmidt rank (we will call this the existence of a max-rank entangled ground state). If the system has a max-rank entangled ground state, as long as the  ground state population is above some threshold $p_0\ge p_0^*$, the state will be SLP. As a corollary of this they find that thermal states for systems with max-rank entangled ground states are SLP for temperatures below a threshold $T\le T^*$. 

In a later study, the authors of~\cite{iffSLP} found necessary and sufficient conditions for Strong Local Passivity,  strengthening the set of  sufficient conditions presented in~\cite{FreyFunoHottaSLP} by providing further physically motivated tightened bounds for the ground state populations and critical temperatures for which the system displays SLP. Here we briefly summarize the three main theorems related to the necessary and sufficient conditions for SLP states found in~\cite{iffSLP}.

\noindent\textbf{First SLP Theorem} \textit{The pair} \{$\hat \rho_\textsc{ab},\hat H_\textsc{ab}$\} \textit{is SLP with respect to subsystem A if and only if} Tr$_{\textsc{a}'}[d_\textsc{a}\ket{\Phi}\bra{\Phi}C_{\textsc{aa}'}]$ \textit{is Hermitian and}
\begin{equation}
        C_{\textsc{aa}'}-\mathrm{Tr}_{\textsc{a}'}[d_\textsc{a}\ket{\Phi}\bra{\Phi}C_{\textsc{aa}'}]\otimes\openone_{\textsc{a}'}\geq 0,
\end{equation}
\textit{where $\mathcal{H}_{\textsc{a}'}$ is a copy of the Hilbert space $\mathcal{H}_\textsc{a}$, $C_{\textsc{aa}'}\in\mathcal{H}_\textsc{a}\otimes\mathcal{H}_{\textsc{a}'}$ is a Hermitian operator defined as \mbox{$C_{\textsc{aa}'}\equiv\mathrm{Tr}[\hat\rho_{\textsc{ab}}^{\Gamma_\textsc{a}}\hat H_{\textsc{a}'\textsc{b}}]$}, with $\rho_{\textsc{ab}}^{\Gamma_\textsc{a}}$ the partial transpose with respect to A, and $d_\textsc{a}\ket{\Phi}\bra{\Phi}$ the maximally entangled Choi-Jamio\l kowski operator of the identity channel.}

The First SLP Theorem provides the necessary and sufficient conditions for the impossibility of energy extraction from a single subsystem of a multipartite system using only local operations. While powerful, this theorem is rather abstract and not easily connectable to the physical properties of a given system. Because of this, the authors of~\cite{iffSLP} also provide more physical, sufficient conditions for a state to be SLP that strengthen the ones first found in~\cite{FreyFunoHottaSLP}.

\noindent \textbf{Second SLP Theorem} \textit{Let the ground state} $\ket{E_0}$ \textit{of the Hamiltonian} $\hat H_\textsc{ab}$ \textit{be non-degenerate and with full Schmidt rank. All pairs} $\{\hat\rho_\textsc{ab},\hat H_\textsc{ab}\}$ \textit{with} $\hat\rho_\textsc{ab} = \sum_i p_i\ket{E_i}\!\bra{E_i}$ \textit{and} $p_0\geq p_*$ \textit{are SLP with respect to A, with the threshold ground state population bounded from above by}
\begin{equation}
    p_*\leq \left(1+\frac{E_1(q_{0,\text{min}}^\textsc{ab})^2}{\text{max}_{i\geq1}\left[E_i(q_{i,\text{max}}^\textsc{ab})^2\right]}\right)^{-1},
\end{equation}
\textit{where} $\{q_{i,\alpha}^\textsc{ab}\}_{\alpha=0}^{d_\textsc{a}-1}$ \textit{denotes the Schmidt coefficient of} $E_i$ \textit{and} $q_{0,\text{min}}^\textsc{ab}\equiv\text{min}_\alpha[q_{i,\alpha}^\textsc{ab}]$ \textit{and} $q_{0,\text{max}}^\textsc{ab}\equiv\text{max}_\alpha[q_{i,\alpha}^\textsc{ab}]$.

The Second SLP Theorem, gives a concrete value for the critical ground state population over which a system with a max-rank entangled energy-diagonal state is SLP. 

Finally, we present the Third SLP Theorem which tackles the particular case of thermal states, and provides bounds for the critical temperature under which a system is SLP. In order to introduce the Third SLP Theorem, it is helpful to first introduce the definition of clustering correlations.

\noindent \textbf{Clustering of Correlations} \textit{A state} $\hat \rho$ \textit{on a finite square lattice} $\mathbb{Z}^D$ \textit{has} $\epsilon(l)$-\textit{clustering of correlations if}
\begin{equation}
    \text{max}_{\hat M,\hat N}\bigg|\text{Tr}[\hat M\otimes \hat N \hat\rho]-\text{Tr}[\hat M\hat\rho]\text{Tr}[\hat N\hat\rho]\bigg|\leq\norm{\hat M}\,\norm{\hat N}\,\epsilon(l), 
\end{equation}
\textit{where the operator} $\hat M$ \textit{has support on region A and N on region B, and} $l\leq\text{dist}(A,B)$, \textit{with} $\text{dist}(A,B)$ \textit{the Euclidean distance on the lattice.}

\begin{figure}[h!]
    \centering
    \includegraphics[width=0.8\linewidth]{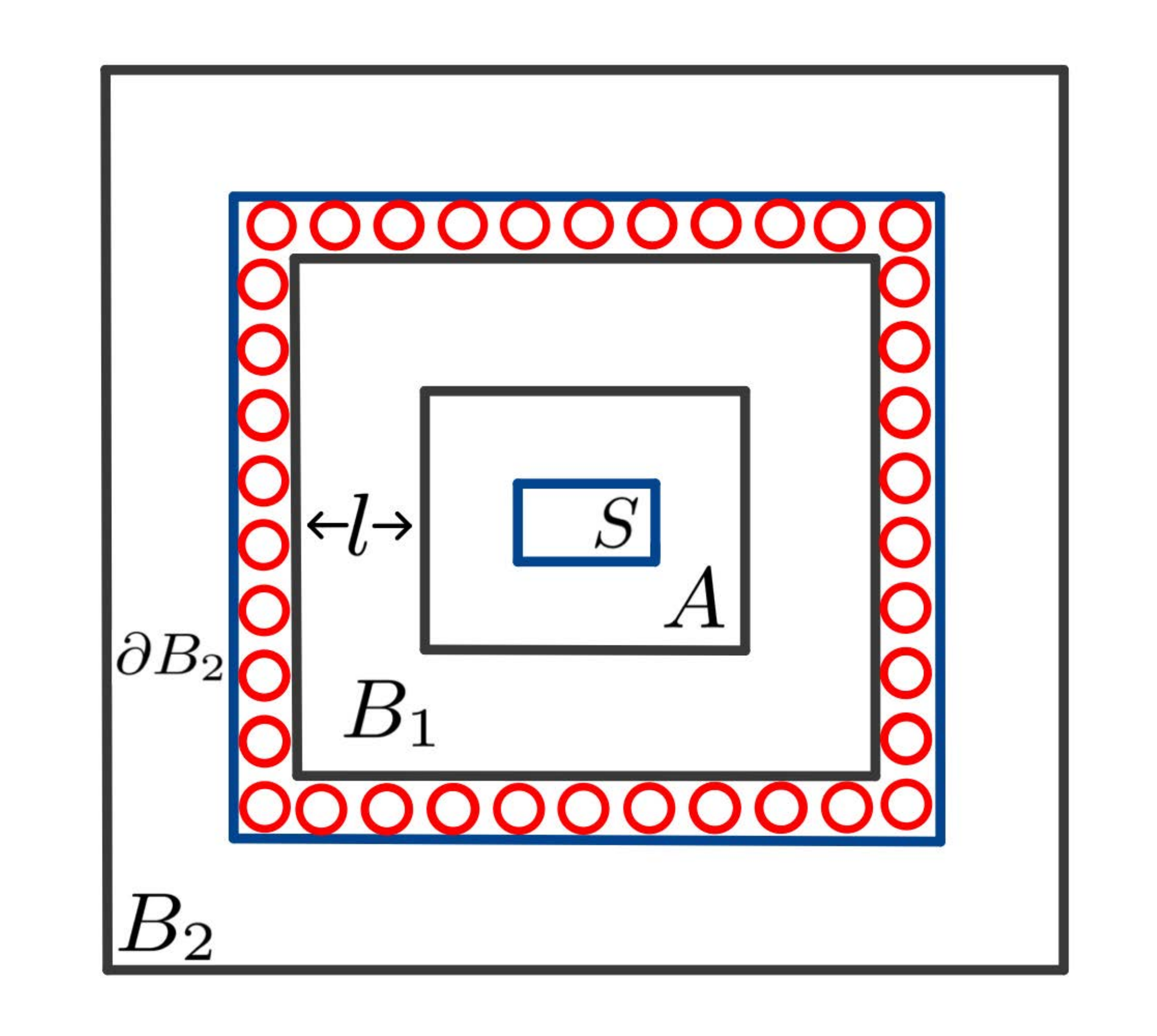}
    \caption{Pictographic representation of the lattice for clustering of correlations and the Third SLP Theorem. Here, $S\subset A$ is the region on which the local operations act. $A$ is separated from B$_2$ by B$_1$ by a distance of $l=\text{dist}(A,B_1)$, and the size of the boundary of B$_2$, $|\partial$B$_2|$ is given by the number of sites on the boundary between B$_1$ and B$_2$. This is a modified version of Fig. 1 in~\cite{iffSLP}}
    \label{fig:Clustering}
\end{figure}

\noindent We further define a Hamiltonian on a $d-$dimensional finite square lattice given by
\begin{equation}\label{eq:SLPLatticeHam}
    \hat H_\textsc{ab} = \hat H_{\textsc{ab}_1} + \hat V_{\textsc{b}_1\textsc{b}_2} + \hat H_{\textsc{b}_2},
\end{equation}
where 
\begin{equation}
    \hat H_{\textsc{ab}_1} = \hat H_\textsc{a} + \hat V_{\textsc{ab}_1} +\hat H_{\textsc{b}_1},
\end{equation}
B$_1$ and B$_2$ are any splitting of B with $l\equiv\text{dist}(\text{A},\text{B}_2)$, and B$_1$ and B$_2$ have a boundary between them of size $|\partial\text{B}_2|$. In Fig.~\ref{fig:Clustering} we give a pictographic representation of each of the regions for illustrative purposes.
Using this definition we can state the following theorem: 

\noindent \textbf{Third SLP Theorem} \textit{Consider a Hamiltonian} $\hat H_\textsc{ab}$ \textit{of the form~\eqref{eq:SLPLatticeHam} and let} $\hat\tau^\beta_{\textsc{ab}} = e^{-\beta\hat H_\textsc{ab}}/Z_\textsc{ab}$ \textit{be its thermal state with} $\epsilon(l)$-\textit{clustering of correlations. There exists a finite temperature} $\beta_*$ \textit{such that all pairs} $\{\hat\tau^\beta_\textsc{ab},\hat H_\textsc{ab}\}$ \textit{with} $\beta\geq\beta_*$ \textit{are SLP with respect to local operations on S if the regions} B$_1$, B$_2$ \textit{can be chosen such that}
\begin{equation}
    E_1^{\textsc{ab}_1}(q_{0,\text{min}}^{\textsc{ab}_1})^2 > \lambda(l),
\end{equation}
\textit{where}
\begin{equation}
    \lambda(l) = Kd^2_\textsc{a}\norm{\hat H_\textsc{a}}|\partial B_2|(\epsilon(l/2) + c_1e^{-c_2l}).
\end{equation}
\textit{Moreover,} $\beta_*$ \textit{is such that}
\begin{align}
    \text{Tr}[e^{-\beta_*\hat H_{\textsc{ab}_1}}]^{-1}&\leq\left(1+\frac{\lambda(l)}{\text{max}_{i\geq1}[E_i^{\textsc{ab}_1}(q_{i,\text{max}}^{\textsc{ab}_1})^2]}\right)\nonumber\\
    &\times \left(1+\frac{E_1^{\textsc{ab}_1}(q_{0,\text{min}}^{\textsc{ab}_1})^2}{\text{max}_{i\geq1}[E_i^{\textsc{ab}_1}(q_{i,\text{max}}^{\textsc{ab}_1})^2]}\right)^{-1},
\end{align}
\textit{where} $K,c_1,c_2>0$ \textit{are constants.}

The proofs of these theorems are quite involved and require the use of semi-definite programming techniques~\cite{Semidefinite1,Semidefinite2}. The full detailed proofs of these theorems can be found in~\cite{iffSLP}.

The three SLP theorems provide not only the necessary and sufficient conditions for a system to exhibit SLP, but also provide physical parameters that allow one to determine if the system is SLP. However, in~\cite{iffSLP}, the authors provide these theorems under the assumption that only local operations on a subsystem are permitted. This leads to the idea that Strong Local Passivity could be broken through other means. One question that comes to mind is how much does one have to violate the assumptions to break Strong Local Passivity. For example, is it enough to allow for LOCC? (that is, add the ability to perform classical communication with the ability to carry out local operations). This is precisely what the QET protocol achieves: two actors that not only have the ability to perform local operations on subsystems of an SLP state, but also have access to classical communication can unlock energy extraction and overcome the limitations of Strong Local Passivity. We will see how in the summary of the following Subsections, where we will review the QET protocol.

\subsection{Minimal QET}\label{sec:minimalQET}

Consider that Alice and Bob have access each to a qubit of a bipartite interacting system. The qubits interact through a Hamiltonian
\begin{equation}\label{eq:HamiltonianQET}
    \hat{H} = \hat{H}_{\textsc{a}}\otimes\openone_\textsc{b} + \openone_\textsc{a}\otimes\hat{H}_\textsc{b}+\hat{V}_\textsc{ab},
\end{equation}
where the interaction term $\hat V_\textsc{ab}$ does not commute with the local Hamiltonians $\hat{H}_{\textsc{a}}\otimes\openone_\textsc{b} + \openone_\textsc{a}\otimes\hat{H}_\textsc{b}$, so that the ground state of such a Hamiltonian will be entangled.

Let us assume that we begin in the ground state. Then, the minimal QET protocol consists of three steps:  Alice will first measure qubit A through a POVM that does not commute with the interaction term of the Hamiltonian. Next, through classical communication, Alice sends the results of her measurement to Bob, typically in a timescale less than the time it takes for energy to propagate through the system naturally\footnote{This is to ensure that no energy injected through Alice's measurement is capable of propagating to Bob before he receives the result of Alice's measurement.}. The final step involves Bob implementing a unitary on his subsystem that depends on the result of Alice's measurement. If the QET protocol is successful Bob will be able to extract energy from the application of the local unitary.

Notice a crucial aspect of the protocol: if the protocol is executed ``fast enough", the action of the measurement of Alice does not have time to propagate to Bob. In other words, if Bob does not receive the result of the measurement, no local operations can tell that Alice has done anything, and the state's local statistics look exactly like the ground state for any local measurements he can perform. Seen from his perspective it would look like knowing the information that Alice sends allows him to unlock energy extraction from (what to him looks like) the ground state. This is why this protocol is sometimes talked about in terms of \textit{zero-point energy activation}~\cite{IQCQETExperiment} even though this is not at all an extraction of zero-point energy in the literal (second law violating) sense.

Let us now go through the steps for a particular example to show that indeed Bob can extract energy from the system before any energy deposited in the system by Alice's measurement has reached him. Let us particularize the Hamiltonians as follows
\begin{equation}\label{eq:minimalHamiltonian}
    \hat{H} = \hat{H}_{\textsc{a}} + \hat{H}_\textsc{b}+\hat{V}_\textsc{ab},
\end{equation}
where we dropped the identities for notational simplicity. Let us choose
\begin{gather}
    \hat{H}_i = h\s^i_z + f(h,k)\openone,\\
    \hat{V}_\textsc{ab} = 2\left(k\s_{x}^\textsc{a}\s_x^\textsc{b} +\frac{k^2}{h^2}f(h,k)\openone\right),
\end{gather}
and $i\in\{\textsc{A},\textsc{B}\}$. To simplify the analysis, let us fix the irrelevant multiple of identity $f(h,k)$ so that the ground state, $\ket{g}$, of the full Hamiltonian satisfies 
\begin{equation}\label{eq:GroundStateExp}
    \bra{g}\hat{H}_\textsc{a}\ket{g} = \bra{g}\hat{H}_\textsc{b}\ket{g} =\bra{g}\hat{V}_\textsc{ab}\ket{g} =\bra{g}\hat{H}\ket{g} = 0.
\end{equation}
which sets 
\begin{equation}\label{eq:fhk}
    f(h,k) = \frac{h^2}{\sqrt{h^2+k^2}}.
\end{equation}
 Working in the eigenbasis\footnote{We use the convention that $\s_z\ket{0}=\ket{0}$ and $\s_z\ket{1}=-\ket{1}$.} of $\s_z^\textsc{a}$ for qubit A and $\s_z^\textsc{b}$ for qubit B, we can write the ground state as
\begin{equation}\label{eq:MinimalEntangledGround}
    \ket{g} = \frac{1}{\sqrt{2}}\left(C^-\ket{1_\textsc{a}1_\textsc{b}} - C^{+}\ket{0_\textsc{a}0_\textsc{b}}\right),
\end{equation}
where 
\begin{equation}
    C^\pm = \sqrt{1\pm \frac{f(h,k)}{h}}.
\end{equation}

With the total system in its ground state, Alice will perform a POVM on her qubit, whose measurement operators $\hat M_\textsc{a}(\alpha)$ commute with the interaction term of the Hamiltonian in Eq.~\eqref{eq:minimalHamiltonian}, with measurement outcomes $\alpha = \pm 1$. The commutativity between the measurement operators and the interaction term ensures that the measurement does not raise the energy of subsystem B~\cite{HottaMinimal,Hotta2010}. For this minimal example let us simplify to the case that the measurements are projector-valued measures (PVMs). In particular, the projectors implementing the state update for the measurement given an outcome $\alpha$ are
\begin{equation}\label{eq:Palpha}
    \hat P_\textsc{a}(\alpha) = \frac{1}{2}\left(\openone + \alpha\s_x^\textsc{a}\right).
\end{equation}
Consider that we perform this measurement on an ensemble of ground states. Then, the resulting, post-measurement state is given by the density operator
\begin{align}\label{projsum}
    \hat{\rho}_1 &= \sum_{\alpha = \pm 1} \hat P_\textsc{a}(\alpha)\ket{g}\!\bra{g} \hat P_\textsc{a}(\alpha).
\end{align}
In expanding the summand of Eq. \eqref{projsum}, we find that
\begin{align}
    P_\textsc{a}(\alpha)\ket{g}\!\bra{g} \hat P_\textsc{a}(\alpha) &= \frac{1}{4}\bigg( \alpha(\ket{g}\!\bra{g}\s_x^\textsc{a}+\s_x^\textsc{a}\ket{g}\!\bra{g}) \nonumber\\
    &+\ket{g}\!\bra{g} + \alpha^2\s_x^\textsc{a}\ket{g}\!\bra{g}\s_x^\textsc{a}\bigg)
\end{align}
so that Eq.~\eqref{projsum} becomes 
\begin{equation}\label{eq:MinimalPostMeasurement}
    \hat \rho_1 = \frac{1}{2}\ket{g}\!\bra{g} + \frac{1}{2}\s_x^\textsc{a}\ket{g}\!\bra{g}\s_x^\textsc{a}
\end{equation}
Since we started from the ground state that had zero energy, we can easily calculate the expectation of the energy injected into the field by performing this measurement:
\begin{align}\label{eq:rho1H}
    E_{P_\textsc{a}} &= \text{Tr}(\hat{\rho}_1\hat{H})\nonumber\\
    &= \sum_{\alpha=\pm 1}\bra{g}\hat{P}_{\textsc{a}}(\alpha)\hat{H}_\textsc{a}\hat{P}_{\textsc{a}}(\alpha)\ket{g} \nonumber\\
    &+ \sum_{\alpha=\pm 1}\bra{g}\hat{P}_{\textsc{a}}(\alpha)\hat{H}_\textsc{b}\hat{P}_{\textsc{a}}(\alpha)\ket{g} \nonumber\\
    &+ \sum_{\alpha=\pm 1}\bra{g}\hat{P}_{\textsc{a}}(\alpha)\hat{V}_\textsc{ab}(\alpha)\hat P_\textsc{a}(\alpha)\ket{g}.
\end{align}
 We can simplify this using that $\big[\hat{P}_\textsc{a}(\alpha),\hat{H}_\textsc{b}\big] = 0$, $\big[\hat{P}_\textsc{a}(\alpha),\hat{V}_\textsc{ab}\big]= 0$ and the fact that $\hat{P}_\textsc{a}(\alpha)$ is a projector so it squares to itself. With this we can show that
\begin{align}
    &\sum_{\alpha = \pm 1}\bra{g}\hat{P}_{\textsc{a}}(\alpha)\hat{H}_\textsc{b}\hat{P}_{\textsc{a}}(\alpha)\ket{g} = \nonumber\\
    &\frac{1}{2}\sum_{\alpha = \pm 1}\alpha h \bra{g}\s_z^\textsc{b}\s_x^\textsc{a}\ket{g} + \alpha f(h,k)\bra{g}\s_x^\textsc{a}\ket{g} = 0,
\end{align}
where we also used the fact that $\bra{g}\s_x^{\textsc{a}}\ket{g} = 0$ as per Eq. \eqref{eq:GroundStateExp}. Similarly, we can show
\begin{align}
    &\sum_{\alpha=\pm 1}\bra{g}\hat{P}_{\textsc{a}}(\alpha)\hat{V}_\textsc{ab}\hat P_\textsc{a}(\alpha)\ket{g} = \nonumber\\
    &\sum_{\alpha = \pm 1}\alpha k \bra{g}\s_x^\textsc{b}\ket{g} + \alpha\frac{k^2}{h^2}f(h,k)\bra{g}\s_x^\textsc{a}\ket{g}= 0.
\end{align}
For the only non-zero term, it is helpful to note that 
\begin{align}
    \hat{P}_\textsc{a}(\alpha)\hat{H}_\textsc{a}\hat{P}_\textsc{a}(\alpha) = \frac{f(h,k)}{4}\bigg((\alpha+2)\openone + \alpha\s_x^\textsc{a}\bigg).
\end{align}
Thus, we have 
\begin{align}\label{eq:MinAliceDeposit}
    &\sum_{\alpha = \pm 1} \bra{g}\hat{P}_\textsc{a}(\alpha)\hat{H}_\textsc{a}\hat{P}_\textsc{a}(\alpha)\ket{g} = \nonumber\\
    &\frac{1}{2}\sum_{\alpha =\pm 1} f(h,k)\bra{g}\openone\ket{g}
    +\frac{1}{4} \sum_{\alpha=\pm 1} \alpha f(h,k)\bra{g}\openone\ket{g}\nonumber\\
    &+\frac{1}{4}\sum_{\alpha=\pm1} \alpha f(h,k) \bra{g} \s_x^\textsc{a}\ket{g}= f(h,k) > 0.
\end{align}

As expected, the measurement Alice makes on her qubit injects energy into the system. Specifically, if Alice makes her measurement over an ensemble of identical setups, the average amount of energy that she adds to the field is $f(h,k)$, which, from Eq. \eqref{eq:fhk} is always going to be positive. It is important to note that this measurement is local on the system A: The apparatus used to make the measurement only couples to A and leaves B untouched.

We now proceed to the next steps of utilizing classical communication and a local unitary on B to extract energy from the system. Given the result, $\alpha$, of Alice's measurement, Alice then sends this result to Bob who performs the local unitary given by
\begin{equation}\label{eq:ClassCommUni}
    \hat U_\textsc{b}(\alpha) = \cos(\theta)\openone -\ii\alpha\sin(\theta)\s_y^\textsc{b},
\end{equation}
where $\theta$ is chosen such that 
\begin{align}
    \cos(2\theta) &= \frac{h^2 + 2k^2}{\sqrt{(h^2+2k^2)^2 + h^2k^2}},\nonumber\\
    \sin(2\theta) &= \frac{hk}{\sqrt{(h^2+2k^2)^2 + h^2k^2}}.
\end{align}

If we consider that Alice and Bob repeat the protocol many times~\cite{HottaMinimal,Hotta2010} and Bob performs this unitary on the ensemble of states on which Alice has measured and communicated the result of the measurement, the final state of the ensemble is
\begin{equation}\label{eq:MinimalRhoB}
    \hat{\rho}_2 = \sum_{\alpha = \pm 1} \hat U_\textsc{b}(\alpha)\hat P_\textsc{a}(\alpha)\ket{g}\!\bra{g}\hat P_\textsc{a}(\alpha)\hat U^\dagger_\textsc{b}(\alpha).
\end{equation}
In order to find the average energy cost of Bob's unitary, we compute the difference between the expectation of the full Hamiltonian before and after the unitary. That is, 
\begin{equation}\label{eq:UnitaryCost}
    E_{U_\textsc{b}} \coloneqq \text{Tr}(\hat{\rho}_2\hat{H}) - \text{Tr}(\hat{\rho}_1 \hat{H}).
\end{equation}
We can then expand the first term of Eq.~\eqref{eq:UnitaryCost} to find
\begin{align}\label{eq:EnergyCost}
    &\text{Tr}(\hat{\rho}_2\hat{H}) = \sum_{\alpha=\pm1}\bra{g}\hat P_\textsc{a}(\alpha)\hat U^\dagger_\textsc{b}(\alpha)\hat H_\textsc{a}\hat U_\textsc{b}(\alpha)\hat P_\textsc{a}(\alpha)\ket{g} \nonumber\\
    &+ \sum_{\alpha=\pm1}\bra{g}\hat P_\textsc{a}(\alpha)\hat U^\dagger_\textsc{b}(\alpha)\hat H_\textsc{b}\hat U_\textsc{b}(\alpha)\hat P_\textsc{a}(\alpha)\ket{g}\nonumber\\
    &+ \sum_{\alpha=\pm1}\bra{g}\hat P_\textsc{a}(\alpha)\hat U^\dagger_\textsc{b}(\alpha)\hat V_\textsc{ab}\hat U_\textsc{b}(\alpha)\hat P_\textsc{a}(\alpha)\ket{g}\nonumber\\
    &= E_{P_\textsc{a}} 
    + \sum_{\alpha = \pm 1}\bra{g}\hat P_\textsc{a}(\alpha)\hat U^\dagger_{\textsc{b}}(\alpha)\left(\hat{H}_\textsc{b} + \hat{V}_\textsc{ab}\right)\hat U_{\textsc{b}}(\alpha)\hat P_\textsc{a}(\alpha)\ket{g}
\end{align}
The first term can be seen from the fact that $\big[\hat H_\textsc{a}(\alpha),\hat{U}_\textsc{b}\big] = 0$ and recalling that the first summand is the only non-zero term of $\text{Tr}(\hat \rho_1\hat H)$  in Eq.~\eqref{eq:rho1H}. Then, using the fact that $\bra{g}\s_x^\nu\ket{g} = 0$ for $\nu\in\{\text{A},\text{B}\}$, it is moderately lengthy but straightforward to show that 
\begin{equation}\label{eq:energycostunitary}
    E_{U_\textsc{b}} = \frac{-1}{\sqrt{h^2 + k^2}}\left(hk\sin(2\theta) - (h^2 +2k^2)(1-\cos(2\theta)\right).
\end{equation}
Choosing a small enough value of $\theta$ (i.e $0<\theta\ll1$) we find that
\begin{equation}
    E_{U_\textsc{b}} \approx \frac{-2hk\theta}{\sqrt{h^2+k^2}}<0.
\end{equation}
With this, we can see that the local operation $\hat U_\textsc{b}(\alpha)$ has a negative energy cost. Importantly, the energy given away is always less than or equal to the energy that was injected by the measurement performed by Alice. 

To interpret these results it is helpful to consider two further points. Let us first discuss whether the energy that Bob sees is coming from the propagation of the energy injected by Alice. To do this, let us estimate how fast the energy that is injected in subsystem A propagates through the interaction term to subsystem B.  

Concretely, making a measurement on the local system A will inject energy into the system. This will take the system out of the ground state and this energy will flow towards B. We can compute the average energy in system B after the measurement on A is performed by time evolving the resulting state after Alice's measurement. Explicitly, we have 
\begin{align}
    \langle\hat{H}_\textsc{b}(t)\rangle &= \sum_{\alpha = \pm 1}\bra{g}\hat P_\textsc{a}(\alpha)e^{\ii\hat{H}t}\hat{H}_\textsc{b}e^{-\ii\hat{H}t}\hat P_\textsc{a}(\alpha)\ket{g}\nonumber\\
    &= \frac{1}{2}f(h,k)(1-\cos(4kt)).
\end{align}
Notice that analyzing this local term is enough because during the whole evolution, the energy expectation of the interaction remains zero, $\langle\hat{V}_\textsc{ab}(t)\rangle = 0$ for all times. Thus, we can interpret the energy flowing from subsystem A to subsystem B in a characteristic time $t_c = k^{-1}$. If classical communication and the subsequent measurement by Bob are faster than this timescale it is clear that the energy from Alice's measurement does not have time to flow to Bob before he performs his measurement.

A second important point of interpretation is to note that the classical communication phase is crucial for the energy extraction. We can prove (see \cite{HottaMinimal}) that Bob cannot, on average, extract any energy from his subsystem if Alice does not communicate the outcome to him. If one considers an identical protocol where Alice measures first and then Bob applies an arbitrary unitary operation without Alice revealing the result of her measurement to him, Bob would not be able to extract any energy. This is not too difficult to see: the only calculation that changes in this scenario is that in Eq.~\eqref{eq:EnergyCost}. Let Bob apply a local unitary $\hat W_\textsc{b}$ that is $\alpha$-independent, the resulting state is given by 
\begin{equation}
    \hat{\rho}^W_2 = \hat W_\textsc{b}\hat{\rho}_1\hat W_\textsc{b}^\dagger.
\end{equation}
If we now compute the energy cost of this local unitary as we did in Eq.~\eqref{eq:UnitaryCost} we find that
\begin{equation}\label{eq:UnitaryCost2}
    E_{W} = \text{Tr}(\hat{\rho}^W_2\hat{H}) - \text{Tr}(\hat{\rho}_1 \hat{H}).
\end{equation}
From a calculation similar to Eq.~\eqref{eq:EnergyCost} we get that
\begin{equation}\label{eq:EnergyCost2}
    \text{Tr}(\hat{\rho}^W_2\hat H) \!=\! E_{P_\textsc{a}}\!+\!\!\!\sum_{\alpha = \pm 1} \!\!\bra{g}\hat P_\textsc{a}(\alpha)\hat W^\dagger_\textsc{b}\big(\hat H_\textsc{b} \!+\! \hat V_\textsc{ab}\big)\hat W_\textsc{b}\hat P_\textsc{a}(\alpha)\ket{g}.
\end{equation}
Since $\hat P_\textsc{a}(\alpha)$ is a projector we know $\sum_\alpha \hat P_\textsc{a}(\alpha) = \openone$. Moreover, since the operation is local to B, $\big[\hat W_\textsc{b},\hat H_\textsc{a}\big]=\big[\hat W_\textsc{b},\hat P_\textsc{a}(\alpha)\big] = 0$, and  we can write Eq.~\eqref{eq:UnitaryCost2} as 
\begin{align}\label{eq:EW}
    E_W&= \bra{g}\hat W^\dagger_\textsc{b}\left[ \hat H_\textsc{b} + \hat V_\textsc{ab}\right]\hat W_\textsc{b}\ket{g}.
\end{align}
However, from the commutation of $\hat W_\textsc{b}$ and $\hat H_\textsc{a}$ together with Eq.~\eqref{eq:GroundStateExp}
\begin{equation}
\bra{g}\hat W^\dagger_\textsc{b}\hat H_\textsc{a}\hat W_\textsc{b}\ket{g}=\bra{g}\hat H_\textsc{a}\ket{g}=0
\end{equation}
and that means that we can add a null term to \eqref{eq:EW} to write it as
\begin{align}
    E_W&= \bra{g}\hat W^\dagger_\textsc{b}\left[ \hat H_\textsc{a}+\hat H_\textsc{b} + \hat V_\textsc{ab}\right]\hat W_\textsc{b}\ket{g}=\bra{g} W^\dagger_\textsc{b}\hat HW_\textsc{b}\ket{g}.
\end{align}

Since we know that $\hat H$ is a non-negative operator, we have that $\bra{g}\hat W^\dagger_\textsc{b}\hat H \hat W_\textsc{b}\ket{g} \geq 0$. That is, Bob's local unitary will, on average, cost energy if he has no information about the outcome of Alice's measurement.

\subsection{Fully Unitary QET}\label{sec:FullyUnitaryQET}

Given the minimal QET model presented in Sec.~\ref{sec:minimalQET}, we will demonstrate how the same results can be achieved through local operations and quantum communication (LOQC). This process of using QET with LOQC is what is commonly referred to as ``fully unitary QET". The main idea behind fully unitary QET is that we want to be able to replicate each of the minimal QET steps by a set of unitary operations that can be implemented experimentally. We will make some adaptations to the minimal QET model from Sec.~\ref{sec:minimalQET} that will be important when discussing the experimental verification of QET. 

Similar to the minimal QET model, we have the full Hamiltonian for our system given by 
\begin{equation}\label{eq:ExperimentHam}
    \hat H = \hat H_\textsc{a} + \hat H_\textsc{b} + \hat V_\textsc{ab},
\end{equation}
where $\hat H_\nu = -h_\nu\s_z^\nu +h_\nu f(h_\textsc{a},h_\textsc{b},k)\openone$ for $\nu\in\{\text{A},\text{B}\}$, and 
\begin{equation}
    \hat V_\textsc{ab} = 2k\s_x^\textsc{a}\s_x^\textsc{b} +\frac{4k^2}{h_\textsc{a}+h_\textsc{b}}f(h_\textsc{a},h_\textsc{b},k)\openone,
\end{equation}
with a slightly different multiple of the identity in the Hamiltonian that is more convenient for this scenario: 
\begin{equation}
f(h_\textsc{a},h_\textsc{b},k) = \left(\frac{4k^2}{(h_\textsc{a}+h_\textsc{b})^2}+1\right)^{-1/2}.
\end{equation}
Same as in the previous Subsection, this Hamiltonian admits a non-degenerate max-rank entangled ground state given by
\begin{equation}\label{eq:UnitaryGround}
    \ket{g} = \frac{1}{\sqrt{2}}\left(F_+\ket{0_\textsc{a}0_\textsc{b}} - F_-\ket{1_\textsc{a}1_\textsc{b}}\right),
\end{equation}
where $F_\pm = \sqrt{1\pm f(h_\textsc{a},h_\textsc{b},k)}.$

Given this ground state, we recall that the first part of the procedure is for Alice to make a measurement on her subsystem and communicate this information to Bob. When using quantum communication between Alice and Bob, this transfer of information is implemented through the local interactions with an ancillary system ($\text{An}$). In particular, we assume that the ancilla starts in the positive eigenvalue eigenstate of $\hat\sigma_{z}^{\text{An}}$, notated as $\ket{0_{\textsc{a}_n}}$. With this, our joint state between A, B, and $\text{An}$ is given by
\begin{equation}\label{eq:UnitaryQETSLPstate} \ket{g}\otimes\ket{0_{\textsc{a}_n}} = \frac{1}{\sqrt{2}}\left(F_+\ket{0_\textsc{a}0_\textsc{b}} - F_-\ket{1_\textsc{a}1_\textsc{b}}\right)\otimes\ket{0_{\textsc{a}_n}}.
\end{equation}
In this case, the measurement is replaced  by an interaction of the ancilla with subsystem A. Through this interaction the ancilla gains information about A.

The interaction is implemented through a joint unitary between qubit A and the ancilla, $\hat U_{\textsc{a}_n\textsc{a}}$. In order to maximize the amount of information that the ancilla extracts from subsystem A, we perform the unitary that maximizes the mutual information between the two. Additionally, the unitary must satisfy $\big[\hat U_{\textsc{a}_n\textsc{a}},\hat V_\textsc{ab}\big] = 0$, which, together with the fact that it is local to the subsystem A-An (which implies that $\big[\hat U_{\textsc{a}_n\textsc{a}},\hat H_\textsc{b}\big] = 0$), is the condition that the measurement does not raise the energy of qubit B. The unitary that maximizes the mutual information between A and An  under the constraint that it commutes with $\hat{V}_\textsc{ab}$ is given by
\begin{equation}\label{eq:anicllaAliceunitary}
    \hat U_{\textsc{a}_n\textsc{a}} = \frac{1}{\sqrt{2}}
    \begin{pmatrix}
        1 & 0 & 0 & 1 \\
        0 & 1 & 1 & 0 \\
        0 & -1 & 1 & 0 \\
        -1 & 0 & 0 & 1
    \end{pmatrix},
\end{equation}
To show that this unitary indeed maximizes the mutual information between A and the ancilla, we show that the computational basis $\{\ket{0_\textsc{a}0_{\textsc{a}_n}},\ket{1_\textsc{a}0_{\textsc{a}_n}},\ket{0_\textsc{a}1_{\textsc{a}_n}},\ket{1_\textsc{a}1_{\textsc{a}_n}}\}$ is mapped to the Bell states under the action of Eq.~\eqref{eq:anicllaAliceunitary}. Specifically, we have
\begin{align}
    \hat U_{\textsc{a}_n\textsc{a}}\ket{0_{\textsc{a}_n}0_\textsc{a}} &= \frac{1}{\sqrt{2}}\left(\ket{0_{\textsc{a}_n}0_\textsc{a}} - \ket{1_{\textsc{a}_n}1_\textsc{a}}\right) = \ket{\Phi^-_{\textsc{a}_n\textsc{a}}}\\
    \hat U_{\textsc{a}_n\textsc{a}}\ket{1_{\textsc{a}_n}1_\textsc{a}} &= \frac{1}{\sqrt{2}}\left(\ket{0_{\textsc{a}_n}0_\textsc{a}} + \ket{1_{\textsc{a}_n}1_\textsc{a}}\right) = \ket{\Phi^+_{\textsc{a}_n\textsc{a}}}\\
    \hat U_{\textsc{a}_n\textsc{a}}\ket{0_{\textsc{a}_n}1_\textsc{a}} &= \frac{1}{\sqrt{2}}\left(\ket{0_{\textsc{a}_n}1_\textsc{a}} - \ket{1_{\textsc{a}_n}0_\textsc{a}}\right) = \ket{\Psi^-_{\textsc{a}_n\textsc{a}}}\\
    \hat U_{\textsc{a}_n\textsc{a}}\ket{1_{\textsc{a}_n}0_\textsc{a}} &= \frac{1}{\sqrt{2}}\left(\ket{0_{\textsc{a}_n}1_\textsc{a}} + \ket{1_{\textsc{a}_n}0_\textsc{a}}\right) = \ket{\Psi^+_{\textsc{a}_n\textsc{a}}}.
\end{align}

The resulting ``post-measurement" state of the entire system is then 
\begin{equation}
    \ket{\Psi_{\textsc{b}\textsc{a}_n\textsc{a}}} = \frac{1}{\sqrt{2}}\left(F_+\ket{0_\textsc{b}}\ket{\Phi^-_{\textsc{a}_n\textsc{a}}}-F_-\ket{1_\textsc{b}}\ket{\Psi^-_{\textsc{a}_n\textsc{a}}}\right).
\end{equation}
Importantly, it is possible to design this unitary operation through a series of gates as follows: apply a CNOT gate with the ancilla as the control, then apply a Hadamard gate on the ancilla, then apply another CNOT gate with the ancilla as the control, and finally perform a Z gate on the ancilla. We will denote the CNOT gate with $C_\textsc{not}$ and, when applied to the states $\ket{00}_{\textsc{a}\textsc{a}_n}$ and $\ket{10}_{\textsc{a}\textsc{a}_n}$, we have 
\begin{align}\label{eq:MeasurementDecomp}
    (Z\otimes \openone) &C_\textsc{not} (H\otimes\openone)C_\textsc{not}\ket{0_{\textsc{a}_n}0_\textsc{a}}\nonumber\\
    &= (Z\otimes\openone) C_\textsc{not} (H\otimes\openone)\ket{0_{\textsc{a}_n}0_\textsc{a}}\nonumber\\
    &= \frac{1}{\sqrt{2}}(Z\otimes\openone) C_\textsc{not}\left(\ket{0_{\textsc{a}_n}0_\textsc{a}} +\ket{1_{\textsc{a}_n}0_\textsc{a}}\right)\nonumber\\
    &= \frac{1}{\sqrt{2}}\left(\ket{0_{\textsc{a}_n}0_\textsc{a}} - \ket{1_{\textsc{a}_n}1_\textsc{a}}\right) = \ket{\Phi^-_{\textsc{a}_n\textsc{a}}},
\end{align}
and
\begin{align}
    (Z\otimes\openone)&C_\textsc{not} (H\otimes\openone) C_\textsc{not}\ket{0_{\textsc{a}_n}1_\textsc{a}} \nonumber\\
    &=(Z\otimes\openone)C_\textsc{not} (H\otimes\openone)\ket{0_{\textsc{a}_n}1_\textsc{a}}\nonumber\\
    &= \frac{1}{\sqrt{2}}(Z\otimes\openone)C_\textsc{not}\left(\ket{0_{\textsc{a}_n}1_\textsc{a}}+\ket{1_{\textsc{a}_n}1_\textsc{a}}\right)\nonumber\\
     &= \frac{1}{\sqrt{2}}\left(\ket{0_{\textsc{a}_n}1_\textsc{a}} - \ket{1_{\textsc{a}_n}0_\textsc{a}}\right) = \ket{\Psi^-_{\textsc{a}_n\textsc{a}}}.
\end{align}
and analogously $(Z\otimes\openone)C_\textsc{not} (H\otimes\openone) C_\textsc{not}\ket{11}=\ket{\Phi^+_{\textsc{a}_n\textsc{a}}}$ and $(Z\otimes\openone)C_\textsc{not} (H\otimes\openone) C_\textsc{not}\ket{10}=\ket{\Psi^+_{\textsc{a}_n\textsc{a}}}$.

After this measurement, the ancilla is sent to qubit B in a time shorter than the coupling time scale to avoid energy propagating from the measurement of qubit A to qubit B. For now, we can assume that this corresponds to gate speeds much faster than the characteristic coupling timescale (inversely proportional to the coupling strength as discussed in the previous subsection) between systems A and B. This assumption is important for the experimental verification and will be paid special attention when we analyze the actual experiments in Sec.~\ref{sec:IQCImplementation}. 

Now for Bob to extract energy from the system given the results of the measurement from Alice, we need to transfer the information of the quantum measurement to Bob. We do this by coupling B to the ancilla through a unitary, $\hat U_{\textsc{b}\textsc{a}_n}$. In particular, we perform a unitary on B, $\hat U_\textsc{b}(\alpha)$, conditioned to the state of the ancilla, that is, 
\begin{equation}
    \hat U_{\textsc{b}\textsc{a}_n} = \hat U_\textsc{b}(0) \otimes\ket{0_{\textsc{a}_n}}\!\bra{0_{\textsc{a}_n}} + \hat U_\textsc{b}(1) \otimes \ket{1_{\textsc{a}_n}}\!\bra{1_{\textsc{a}_n}}.
\end{equation}
Given this joint unitary, we can now compute the state of subsystem B after applying the joint unitary. In this case, we obtain
\begin{align}\label{eq:UnitaryRhoB}
    \hat{\rho}_\textsc{b} &= \text{Tr}_{\textsc{a}\textsc{a}_n}\left(\hat U_{\textsc{b}\textsc{a}_n}\ket{\Psi_{\textsc{b}\textsc{a}_n\textsc{a}}}\!\bra{\Psi_{\textsc{b}\textsc{a}_n\textsc{a}}}\hat U_{\textsc{b}\textsc{a}_n}^\dagger\right)\nonumber\\
    &= \text{Tr}_\textsc{a}\left[\sum_{\alpha}\hat U_\textsc{b}(\alpha)\braket{\alpha_{\textsc{a}_n}}{\Psi_{\textsc{b}\textsc{a}_n\textsc{a}}}\!\braket{\Psi_{\textsc{b}\textsc{a}_n\textsc{a}}}{\alpha_{\textsc{a}_n}}\hat U_\textsc{b}^\dagger(\alpha)\right],
\end{align}
We can further simplify this expression by evaluating $\braket{\alpha_{\textsc{a}_n}}{\Psi_{\textsc{b}\textsc{a}_n\textsc{a}}}$. In doing so we obtain, for $\alpha = 0$
\begin{align}\label{eq:mu0}
    \braket{0_{\textsc{a}_n}}{\Psi_{\textsc{b}\textsc{a}_n\textsc{a}}}\! &=\! \frac{1}{\sqrt{2}}\left(F_+\ket{0_\textsc{b}}\!\langle{0_{\textsc{a}_n}}|\Phi^-_{\textsc{a}_n\textsc{a}}\rangle - F_-\ket{1_\textsc{b}}\!\langle{0_{\textsc{a}_n}}|\Psi^-_{\textsc{a}_n\textsc{a}}\rangle\right)\nonumber\\
    &= \frac{1}{\sqrt{2}}\left[\frac{1}{\sqrt{2}}\left(F_+\ket{0_\textsc{a}0_\textsc{b}} - F_-\ket{1_\textsc{a}1_\textsc{b}}\right)\right]
    \!=\! \frac{1}{\sqrt{2}}\ket{g},
\end{align}
where we recall $\ket{g}$ is the ground state given in Eq.~\eqref{eq:UnitaryGround}. Analogously, for $\alpha = 1$ we have
\begin{align}\label{eq:mu1}
    \braket{1_{\textsc{a}_n}}{\Psi_{\textsc{b}\textsc{a}_n\textsc{a}}} \!&=\! \frac{1}{\sqrt{2}}\!\left(F_+\ket{0_\textsc{b}}\braket{1_{\textsc{a}_n}}{\Phi^-_{\textsc{a}_n\textsc{a}}} \!-\! F_-\ket{1_\textsc{b}}\braket{1_{\textsc{a}_n}}{\Psi^-_{\textsc{a}_n\textsc{a}}}\right)\nonumber\\
    &= -\frac{1}{\sqrt{2}}\left[\frac{1}{\sqrt{2}}\left(F_+\ket{1_\textsc{a}0_\textsc{b}} - F_-\ket{0_\textsc{a}1_\textsc{b}}\right)\right]\nonumber\\
    &= -\frac{1}{\sqrt{2}}\left[\frac{1}{\sqrt{2}}\left(F_+\s_x^\textsc{a}\ket{0_\textsc{a}0_\textsc{b}} - F_-\s_x^\textsc{a}\ket{1_\textsc{a}1_\textsc{b}}\right)\right]\nonumber\\
    &= -\frac{1}{\sqrt{2}}\s_x^\textsc{a}\ket{g}.
\end{align}

Taking into account Eqs.~\eqref{eq:mu0} and \eqref{eq:mu1}, the trace over the ancilla on the state $\ket{\Psi_{\textsc{b}\textsc{a}_n\textsc{a}}}$ is 
\begin{align}\label{eq:projectorrelation}  
\Tr_{\textsc{a}_n} \ket{\Psi_{\textsc{b}\textsc{a}_n\textsc{a}}} \!\bra{\Psi_{\textsc{b}\textsc{a}_n\textsc{a}}} &=\sum_{\alpha\in\{0,1\}} \braket{\alpha_{\textsc{a}_n}}{\Psi_{\textsc{b}\textsc{a}_n\textsc{a}}}\!\braket{\Psi_{\textsc{b}\textsc{a}_n\textsc{a}}}{\alpha_{\textsc{a}_n}} \nonumber\\ 
&= \frac{1}{2}\ket{g}\!\bra{g} + \frac{1}{2}\s_x^\textsc{a}\ket{g}\!\bra{g}\s_x^\textsc{a},
\end{align}
which is the same state as that obtained in the minimal QET model after applying the projective measurement as shown in Eq.~\eqref{eq:MinimalPostMeasurement}. Therefore, $\hat \rho_\textsc{b}$ in Eq.~\eqref{eq:UnitaryRhoB} is exactly the same state as that obtained after the  measurement step and application of the local unitary $\hat U_\textsc{b}$ in the projective version of the minimal QET scenario in Eq.~\eqref{eq:MinimalRhoB}. Thus, illustrating the equivalence between the minimal QET model and the fully unitary implementation of QET in the projective measurement case. Notice that by choosing different $\hat U_{\textsc{a}_n\textsc{a}}$ one would be able to reproduce the different choices of POVMs in the minimal QET model.~\cite{nichoTeleport,IQCQETExperiment,DerivativeQET,GuillaumeThesis}. 

If we now follow the same procedure outlined in Sec.~\ref{sec:minimalQET}, we will again find that the expected energy cost of this unitary is
\begin{equation}
    E_{U_\textsc{b}} = \text{Tr}[(\hat H_\textsc{b} + \hat V_\textsc{ab})\hat{\rho}_\textsc{b}]\leq 0.
\end{equation}
In this case, we can see that the amount of extracted energy is then bounded as follows~\cite{IQCQETExperiment,Hotta2010,NayeliThesis}
\begin{equation}
   E^{\textsc{ab}}_0 \leq E_{U_\textsc{b}} 
   \leq 0,
\end{equation}
where $E^{\textsc{ab}}_0$ is the most negative eigenvalue of $\hat H_\textsc{b} + \hat V_\textsc{ab}$. The upper bound is tight when the measurement operators of the POVM is proportional to a projector, as it is the case in this scenario. Hence, the maximum extractable energy from applying a controlled unitary $\hat U_\textsc{b}(\alpha)$ turns out to be bounded as follows 
\begin{equation}\label{eq:UnitaryMaxEnergy}
    E_{U_\textsc{b}} \leq \sqrt{h_\textsc{b}^2 + 4k^2} - \frac{h_\textsc{b}(h_\textsc{a}+h_\textsc{b}) + 4k^2}{\sqrt{(h_\textsc{a}+h_\textsc{b})^2 + 4k^2}}.
\end{equation}
One can check that this average extracted energy is always going to be less than the amount of average energy that we inject into subsystem A with the measurement. In our particular case, this is $E_\textsc{a} = h_\textsc{a}/\sqrt{1+\frac{4k^2}{(h_\textsc{a}+h_\textsc{b})^2}}$.

The fully unitary QET model lends itself to being implemented experimentally as it  can be formulated in terms of a set of unitaries comprised of CNOT, Pauli-Z and Hadamard gates. 

After having reviewed the theory, we will now review two different experimental implementations of QET, one of which uses LOQC and the other which uses LOCC. The first implementation that will be reviewed was also the first experimental test of QET and it was implemented through Nuclear Magnetic Resonance (NMR) to teleport energy between two loci in a transcrotonic acid molecule~\cite{IQCQETExperiment}. The second implementation runs the QET protocol on several of IBM's publicly available quantum hardware~\cite{Ikeda2023}. 

\section{Experimental implementation in Nuclear Magnetic Resonance}\label{sec:IQCImplementation}

In this section, we discuss how the fully unitary QET protocol was experimentally verified~\cite{IQCQETExperiment}. Namely, we will introduce the experimental testbed used for this implementation, discuss the devices that were used to implement each of the necessary steps of the protocol, and review the analysis of the parameters and duration of the different stages of the experiment that guarantees that the results constitute valid experimental evidence of QET. Concretely, we will discuss that the experiment is done in a ``fast enough" fashion so that no energy from A is capable of propagating to B before the protocol is complete. 

We begin by discussing the experimental setup of the NMR implementation of the QET protocol. In this case the experiment was performed using nuclear magnetic resonance  on a Bruker Avance III 700 MHz NMR spectrometer~\cite{BrukerNMR}. This machine was used to perform all of the necessary gate operations to construct the required unitaries described in Sec.~\ref{sec:FullyUnitaryQET}. 

The next requirement is selecting a system that can model the Hamiltonian in Eq.~\eqref{eq:ExperimentHam} while also maintaining the requirement that energy propagation between the qubits is kept much slower than the gate operations. The particular choice of molecule that satisfies all of these requirements is ${}^{13}\text{C}$-labeled transcrotonic acid dissolved in acetone-d6. 
\begin{figure}[h!]
    \centering
    \textbf{Transcrotonic Acid Molecule}\par\medskip
    \includegraphics[width=6.6cm]{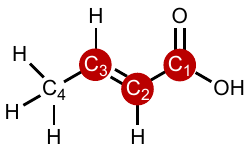}
    \medskip\medskip

    \begin{tabular}{|c||c|c|c|c|  }
        \hline
        \multicolumn{5}{|c|}{Chemical Shifts and Coupling Strengths (Hz)} \\
        \hline
        {} & $C_1$ & $C_2$ & $C_3$ & $C_4$\\
        \hline
        $C_1$   &  -29343.19   & {}         &   {} & {}\\
        $C_2$   &    72.27     & -21591.54  &  {} &{}\\
        $C_3$   &    1.16      & 69.68      &  -25463.29&{}\\
        $C_4$   &    7.04      & 1.44       &  41.65&-2991.62\\
        \hline
    \end{tabular}
    
    \caption{The transcrotonic acid molecule and the labeled carbons that will be used as the qubits for subsystems A, B, and $\text{A}_n$. The table gives the experimental parameters associated with each of the subsystems found by the authors of~\cite{IQCQETExperiment}. The diagonal elements correspond to the chemical shifts, while the off-diagonal elements are the J-coupling values.}
    \label{fig:Molecule}
\end{figure}
Fig.~\ref{fig:Molecule} shows the transcrotonic acid molecule, where we have labeled three of the carbons as $\text{C}_1$, $\text{C}_2$, and $\text{C}_3$. These three carbons correspond to the subsystems B, $\text{A}_{n}$ and A, respectively. Additionally, there is a fourth labeled carbon which is also coupled to the qubits that participate in the experiment. Carbon C$_4$ does not play a role in the protocol but we need to take it into account when analyzing the nuclear spin dynamics of the molecule. 

From the table in Fig.~\ref{fig:Molecule}, we can see that the coupling strength between subsystem A and subsystem B is 1.16Hz. This means that the time between the measurement made by Alice on subsystem A to the extraction of energy by Bob on subsystem B should be less than the timescale associated with energy propagation through the molecule, that is 
\begin{equation}\label{eq:MaxTime}
    t_\textsc{qet} \ll t_c=\frac{1}{1.16}\text{s} \approx 862\text{ms}.
\end{equation}
As we will see in a later discussion, the time it took to perform all of the gate operations necessary to fulfill the fully unitary QET protocol is much smaller than this bound.

\subsection{Pseudopure and ground state preparation}

In order to implement the fully unitary QET model, we must first prepare the necessary ground state of the Hamiltonian in Eq.~\eqref{eq:ExperimentHam}. This is done by having the full system start in the pseudopure state\footnote{A pseudopure state is an effective approximation of a pure quantum state that is constructed from a highly mixed thermal state. NMR quantum computers inherently work with ensembles of molecules, typically at room temperature, resulting in equilibrium states that are highly mixed. To run quantum algorithms, it is crucial to start from a state that behaves approximately like a pure state, at least for computational purposes. Physically, a pseudopure state still represents a highly mixed ensemble. However, the dynamics and outcomes of experiments conducted starting from pseudopure states are effectively identical to those that would be obtained from a pure state, provided that measurement observables and sequences are chosen correctly.}~\cite{PsuedopureStates} $\ket{0_\textsc{a}0_{\textsc{a}_n}0_\textsc{b}}$, and acting with a preparation unitary. Explicitly, this preparation unitary is a rotation  on subsystem B followed by a CNOT gate between A and B, where the rotation is given by 
\begin{equation}
    \hat Y(\theta) = \cos(\theta)\openone -\ii\sin(\theta)\s_y^\textsc{b},
\end{equation}
where
\begin{equation}
    \cos(\theta)=\frac{F_+}{\sqrt{2}},\qquad\sin(\theta)=\frac{F_-}{\sqrt{2}},
\end{equation}
and 
\begin{equation}
    F_\pm = \sqrt{1 \pm \frac{h_\textsc{a}+h_\textsc{b}}{\sqrt{4k^2 + (h_\textsc{a}+h_\textsc{b})^2}}}.
\end{equation}
The result of this preparation unitary is precisely the fully entangled state given in Eq.~\eqref{eq:MinimalEntangledGround}, which can be seen from 
\begin{align}\label{eq:PrepUnitary}
    \hat U_{\text{p}}\ket{0_\textsc{a}0_{\textsc{a}_n}0_\textsc{b}} &= C_\textsc{not} \circ Y(\theta)\ket{0_\textsc{a}0_{\textsc{a}_n}0_\textsc{b}} \nonumber\\
    &= -\frac{1}{\sqrt{2}}\left(F_+\ket{0_\textsc{a}0_\textsc{b}}- F_-\ket{1_\textsc{a}1_\textsc{b}}\right)\otimes\ket{0_{\textsc{a}_n}}.
\end{align}
Thus, except for an irrelevant global phase, we have prepared qubits A and B in the max-rank entangled state ground state \mbox{$\ket{g} = \frac{1}{\sqrt{2}}\left(F_+\ket{0_\textsc{a}0_\textsc{b}}- F_-\ket{1_\textsc{a}1_\textsc{b}}\right)$}. 

In practice, the experiment is performed at room temperature (298K), so if one wants to properly prepare this state, we must ensure that $\hat U_\text{p}$ is indeed acting on the correct pseudopure state. Since the system is in thermal equilibrium at room temperature, we must first create a pseudopure state from the current thermal state. 

For an $n$-qubit NMR system, the Hamiltonian is given by 
\begin{equation}
    \hat H = \hat H_\textsc{z} + \hat H_\textsc{j},
\end{equation}
where $\hat H_\textsc{z}$ and $\hat H_\textsc{j}$ are the Zeeman and indirect spin-spin coupling Hamiltonians, respectively. Specifically, they are given by
\begin{align}
    \hat H_\textsc{z} &= \pi\sum_j\omega_j \hat \sigma^j_z,\\
    \hat H_\textsc{j} &= \frac{\pi}{2} \sum_{j,k}^{j<k}J_{jk}\,\hat{\bm{\sigma}}^j\!\cdot\!\hat{\bm{\sigma}}^k,
\end{align}
where  $\{\omega_j\}$ are the Larmor frequencies, $J_{jk}$ are the coupling values given in the table of Fig. \ref{fig:Molecule}, and the Pauli vectors are \mbox{$\hat{\bm{\sigma}}^j = (\hat \sigma_x^j,\hat\sigma_y^j,\hat\sigma_z^j)$} (see, e.g.,~\cite{LevittSpinDynamics}).

At room temperature, the density matrix for an $n$-qubit NMR system can be approximated to first order in $\xi = \hbar \omega_i /k_{\textsc{b}}T$ as~\cite{PsuedopureStates}
\begin{equation}
    \hat{\rho}_{T}=\frac{1}{2^n}\exp(-\beta \hat  H) \approx \frac{1}{2^n}\left(\openone - \xi \bar{\rho}_{T}\right),
\end{equation}
where $\beta=1/k_\textsc{b}T$ and $\bar{\rho}_T = \sum_{j}\hat\sigma^j_z/2$ is the deviation density matrix. Specifically, in the experimental setup used in~\cite{IQCQETExperiment}, with a magnetic field of \mbox{16.4 T}, for typical values of $\omega_i$ and in the room temperature conditions in the lab it resulted in \mbox{$\xi \approx 10^{-5}$ (see~\cite{IQCQETExperiment} for further details)}. 

\begin{figure*}[!hbtp]
    \centering
    \includegraphics[width = 1\linewidth]{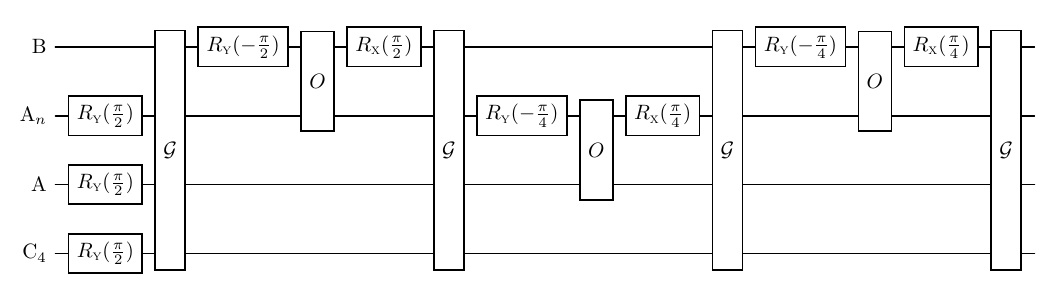}
    \caption{Gate sequence used to prepare the pseudopure state $\ket{0_\textsc{b}0_{\textsc{a}_n}0_\textsc{a}}\otimes\openone$ from the deviation density matrix. Here we recall that $\hat R_j(\theta)$ is a single qubit rotation about the $j^{\text{th}}$ axis by an angle of $\theta$, $\hat O = \exp(-\ii\frac{\pi}{4}\s_z^j\s_z^k)$, and $\hat{\mathcal{G}}$ is the pulse-field gradient used to eliminate all coherences except for the diagonal terms. This is reconstructed circuit diagram of Fig. 5 in~\cite{IQCQETExperiment}}
    \label{fig:PseudopurePreparation}
\end{figure*}

With this in mind, the pseudopure state is prepared using a spatial averaging method~\cite{PsuedopureStates} whose gate sequence is given in Fig.~\ref{fig:PseudopurePreparation}.
In this sequence, $R_j(\theta)$ corresponds to a rotation about the $j^\text{th}$ axis by an angle of $\theta$, and the two qubit operator $\hat O$ is given by
\begin{equation}
    \hat{O} = e^{-\ii\frac{\pi}{4} \hat{\sigma}^j_z\hat{\sigma}^k_z}.
\end{equation}
Additionally, $\mathcal{G}$ represents a pulse-field gradient across all four qubits to destroy all coherences leaving only the diagonal terms~\cite{LevittSpinDynamics}. This gate sequence  results in the pseudopure state $\ket{0_\textsc{a}0_{\textsc{a}_n}0_\textsc{b}}\otimes\openone_{C_4}$ which is our starting point. 

\begin{figure*}[!hbtp]
    \centering
    \includegraphics[width = 1\linewidth]{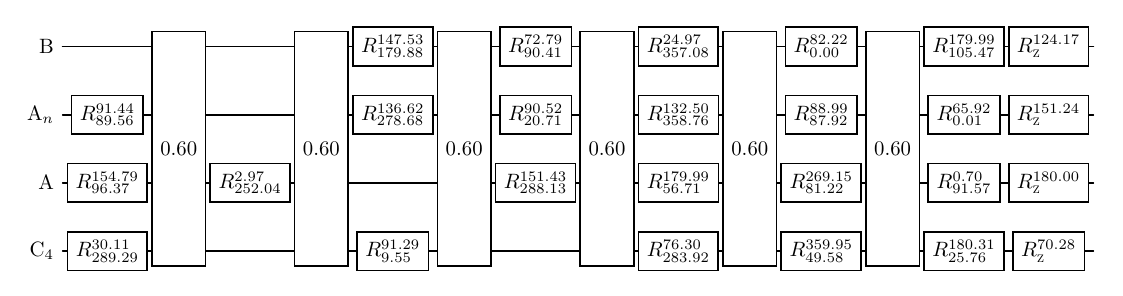}
    \caption{Decomposition of the $\text{A}_n$-mediated CNOT gate used in the preparation step of the ground state from the pseudopure state. In this figure, $\hat R_\phi^\theta = \exp(-\ii\theta(\cos(\phi)\s_x + \sin(\phi)\s_y)/2)$, $\hat R_\textsc{z}^\theta = \exp(-\ii\theta\s_z/2)$, and the multi-qubit operations are the amount of time the system evolves for under the natural Hamiltonian in milliseconds. This is a reconstructed circuit diagram of Fig. 7 in~\cite{IQCQETExperiment}}
    \label{fig:UPrepSequence}
\end{figure*}

In the state preparation step of the protocol, a CNOT gate between qubits A and B is required as detailed by Eq.~\eqref{eq:PrepUnitary}. While it is possible to implement this directly between qubits A and B, an $\text{A}_n$ mediated CNOT gate is used instead. This is due to the fact that a direct CNOT between A and B would have an approximate duration of 431ms, while the $\text{A}_n$-mediated CNOT had a duration of approximately 26ms. The faster implementation is helpful to show that the QET protocol works, as well as lower decoherence in the system and as a result has an improved fidelity in the preparation of the ground state. The actual ground state preparation specified in Eq.~\eqref{eq:PrepUnitary} is further decomposed into a mixture of single qubit unitaries and natural time evolution of the system. For example, the $\text{A}_n$-mediated CNOT (implemented as a SWAP$_{\textsc{aa}_n}$CNOT$_{\textsc{b}{\textsc{a}_n}}$)
was decomposed as seen in Fig.~\ref{fig:UPrepSequence}, where in the diagram  
\begin{equation}
   \hat R^\theta_\phi = \exp(-\ii\theta(\cos(\phi)\s_x + \sin(\phi)\s_y)/2),
\end{equation}
\begin{equation}
    \hat R_\textsc{z}^\theta = \exp(-\ii\theta\s_z/2),
\end{equation}
and the multi-qubit gates of duration 0.60 ms represents time evolution under the natural Hamiltonian.

Notice that the actual ground state preparation circuit in Fig.~\ref{fig:UPrepSequence} and the theoretical ground state preparation circuit in Fig.~\ref{fig:FullyUnitaryCircuit} differ. This is because the theoretical fully unitary protocol in Fig.~\ref{fig:FullyUnitaryCircuit} only requires qubit A, an ancilla qubit, and qubit B, while Carbon C$_4$ does not play a role in the experiment, it is still coupled to the other carbons via the natural Hamiltonian. Therefore, one needs to design the gates so that they also undo the effect of its interaction with the other three carbons in the time evolution at every step of the protocol.

\subsection{Measurement of A and communication to B}

The next step of the protocol is for Alice to measure her qubit and transmit this information to Bob. As mentioned in Sec.~\ref{sec:FullyUnitaryQET}, in the fully unitary version of the protocol this is accomplished through the coupling of qubit A to the ancillary system $\text{A}_n$, and was implemented with the gate decomposition in Eq.~\eqref{eq:MeasurementDecomp}. The measurement unitary was performed in approximately 10ms. Finally, Bob implements a unitary that uses the measurement information from Alice. In the case of this experiment, the unitary $\hat U_{\textsc{ba}_n}$ was decomposed into two operations so that
\begin{equation}\label{eq:BobsUnitary}
    \hat U_{\textsc{ba}_n} = \hat U_{\text{rot}}\hat U_\text{diag},
\end{equation}
where
\begin{equation}
    \hat U_\text{rot} = \frac{1}{\sqrt{2}}
    \begin{pmatrix}
        F_{2+} & F_{2-} & 0 & 0  \\
        0 & 0 & -F_{2+} & F_{2-} \\
        0 & 0 & F_{2-} & F_{2+}  \\
        -F_{2-} & F_{2+} & 0 & 0
    \end{pmatrix},
\end{equation}
\begin{equation}
    \hat U_\text{diag} = \frac{1}{\sqrt{2}}
    \begin{pmatrix}
        0 & F_+ & F_- & 0  \\
        F_- & 0 & 0 & -F_+ \\
        F_+ & 0 & 0 & F_-  \\
        0 & -F_- & F_+ & 0 \\
    \end{pmatrix},
\end{equation}
and $F_{2\pm} = \sqrt{1\pm h_\textsc{b}/\sqrt{h^2_\textsc{b}+4k^2}}$. It can be shown that $\hat U_{\textsc{ba}_n}$ achieves the upper bound in~\eqref{eq:UnitaryMaxEnergy} on the amount of energy that can be extracted from qubit B. The unitary in Eq.~\eqref{eq:BobsUnitary} was performed in 4ms. 

\begin{figure}[h!]
    \centering
    \textbf{\shortstack{Circuit Diagram for the Fully Unitary QET \\Protocol}}
    \includegraphics[width=1\linewidth]{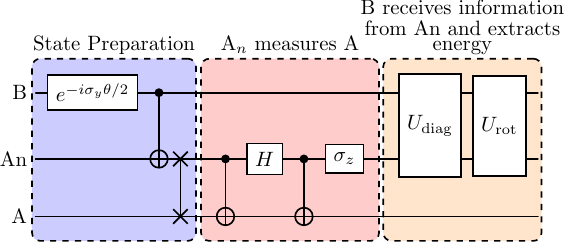}
    \caption{This figure displays the circuit diagram for the fully unitary QET protocol. We have broken the circuit up into three distinct components. In blue, we have the gates used for the preparation of the ground state when qubits B, An, and A start in the joint state $\ket{0_\textsc{b}0_{\textsc{a}_n}0_\textsc{a}}$. In red, we have the gates used to allow the ancilla system to gain information about qubit A. In orange, we have the gates that allow qubit B to gain information about qubit A through the ancilla, and extract energy from qubit B. This is reconstructed version of Fig. 3 in~\cite{IQCQETExperiment}}. 
    \label{fig:FullyUnitaryCircuit}
\end{figure}

The gates for the protocol, shown in Fig.~\ref{fig:FullyUnitaryCircuit}, were implemented using GRAPE pulses~\cite{GRAPEPulse} incorporating the technique described in~\cite{GRAPEPulses2}. As a result, smooth radio frequency pulses had to be designed. The theoretical fidelity of these pulses was checked to be over 0.998.  As a summary, Fig.~\ref{fig:FullyUnitaryCircuit} shows the full circuit that performs the fully unitary QET protocol.

\subsection{Analysis of the experimental timescales}
An important part of the QET protocol is ensuring that Alice's information is communicated in a timescale that is faster than the timescale in which energy from her measurement can propagate to Bob's subsystem through their natural Hamiltonian. In order to verify that the experiment was in fact a valid demonstration of QET, we must first upper-bound the full duration of the experiment. For a successful QET demonstration, Eq.~\eqref{eq:MaxTime} has to hold. 

Now, each of the unitaries in the QET protocol has a duration associated with the coupling strength between the qubit and the ancilla. A conservative theoretical bound for the time the protocol takes can be quickly computed for the molecule used: $t_{\textsc{a}_n\textsc{a}} = 1/J_{\textsc{a}_n\textsc{a}} \approx 13.8$ms and $t_{\textsc{a}_n\textsc{b}} = 1/J_{\textsc{a}_n\textsc{b}}\approx 14.3$ms. Finally, we must also include the duration of the pulses necessary to achieve the unitaries whose theoretical maximum duration is $t_p\approx 9.5$ms. This gives a theoretical total experiment time of 
\begin{equation}
    t_{\text{tot}} = t_{\textsc{a}_n\textsc{a}}+t_{\textsc{a}_n\textsc{b}}+t_p\approx 37.6\text{ms},
\end{equation} 
while the time it takes for energy to propagate from A to B is $t_{\textsc{a}\textsc{b}}\approx 862$ms. Thus we have that 
\begin{equation}
    t_\textsc{qet}<t_{\text{tot}}\approx 37.6\text{ms}\ll t_{\textsc{a}\textsc{b}}\approx 862\text{ms}.
\end{equation}
This gives us a strong indication that the selection of transcrotonic acid as a testbed for QET satisfies all of the necessary conditions. 

These theoretical bounds are quite conservative, and in practice, while each run of the experiment took a slightly different amount of time (of the same order) the actual average duration of the protocol was smaller than the bound. We recall that, in practice, the unitaries $U_{\textsc{a}_n\textsc{a}}$ and $U_{\textsc{b}\textsc{a}_n}$ were performed in roughly 10ms and 4ms, respectively, so that the entire protocol from measurement to energy extraction lasted $t_\textsc{qet} \approx 14$ms. This analysis shows  conclusively that if any energy is extracted through the protocol it has to come from QET.

\subsection{Energy extraction}

Now that we have demonstrated the fully unitary model and verified that the timescales for the experiment are short enough that no energy can naturally flow from A to B before energy is extracted from B, we move on to how the extracted energy was measured. From  Eq.~\eqref{eq:EnergyCost} we see that the extracted energy can be determined by the results of $-\langle \s_z^\textsc{b}\rangle$ and $\langle\s_x^\textsc{a}\s_x^\textsc{b}\rangle$. Each of these values was measured directly at the end of the circuit for various values of $k/h_\textsc{a}$ (taking $h_\textsc{b} = 0.4h_\textsc{a}$). 

For this particular setup, measurements are performed on first-order coherence terms of the density matrix using the NMR techniques described in~\cite{LevittSpinDynamics}. We make note of the fact that, for a two-qubit system, the directly measurable observables are $\langle\s_x^\textsc{a}\otimes\openone\rangle$, $\langle\s_x^\textsc{a}\otimes\s_z^\textsc{b}\rangle$, $\langle\s_y^\textsc{a}\otimes \openone\rangle$, $\langle\s_y^\textsc{a}\otimes \s_z^\textsc{b}\rangle$, 
$\langle\openone\otimes\s_x^\textsc{b}\rangle$, $\langle\s_z^\textsc{a}\otimes\s_x^\textsc{b}\rangle$,
$\langle\openone\otimes\s_y^\textsc{b}\rangle$, and
$\langle\s_z^\textsc{a}\otimes \s_y^\textsc{b}\rangle$. Thus, if one wanted to measure any other quantity a rotation must be applied to one of the qubits in order to make it fall within the subset of measurable observables.  

In order to make $\langle\s_z^\textsc{b}\rangle$ and $\langle\s_x^\textsc{a}\otimes\s_x^\textsc{b}\rangle$ quantities directly observable, an additional rotation $\hat R_y(\pi/2)$ is applied on qubit B.

\begin{figure}[h!]
    \centering
    \includegraphics[width=1\linewidth]{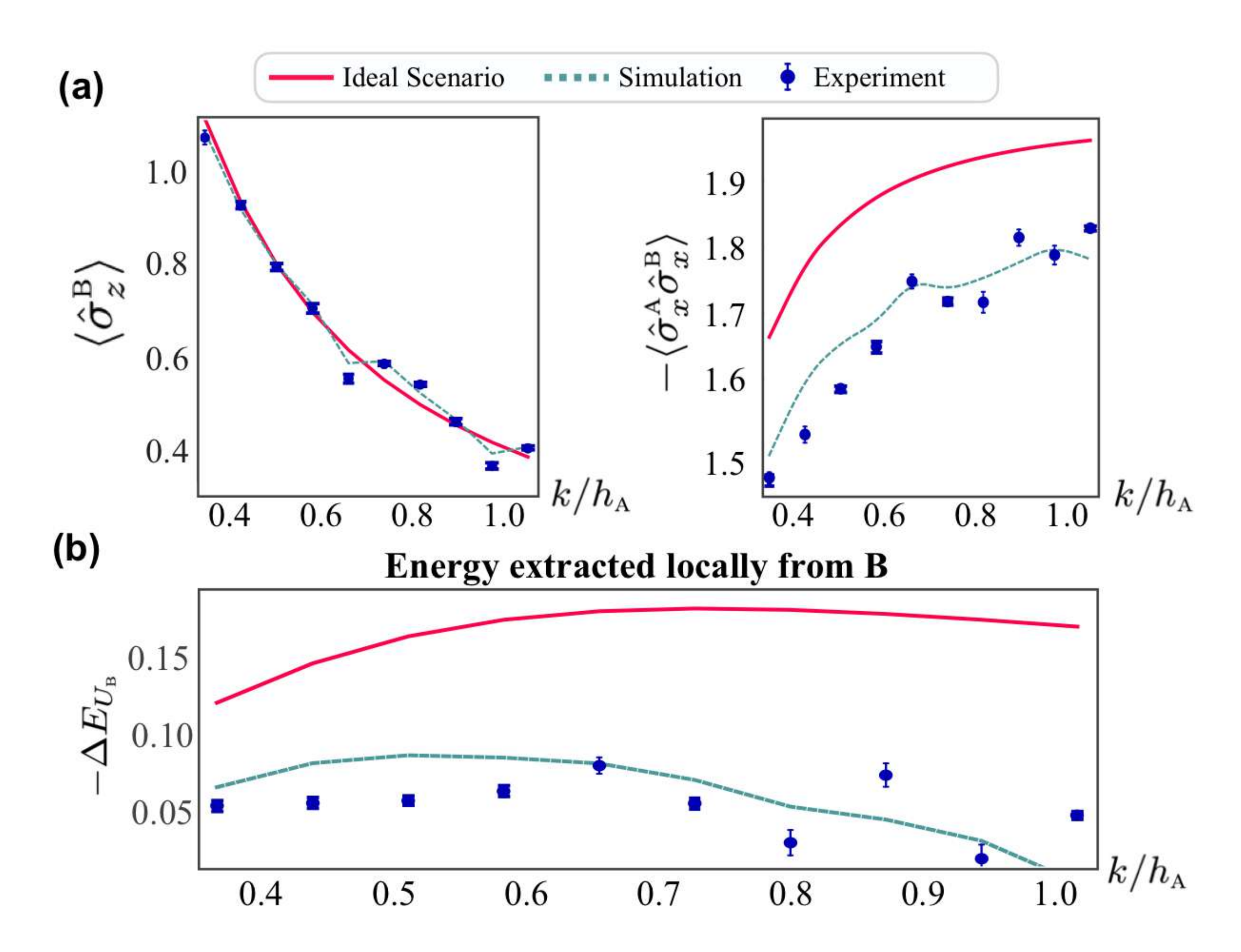}
    \caption{Experimental results of the protocol compared with ideal theoretical predictions and an open system simulation of the actual pulse sequence that accounts for decoherence. In both (a) and (b),  $h_\textsc{b} = 0.4h_\textsc{a}$. (a) gives the results of the measurements of both $\langle \s_z^\textsc{b}\rangle$ and $-\langle\s_x^\textsc{a}\s_x^\textsc{b}\rangle$ after the protocol as been completed for various values of $k/h_\textsc{a}$. (b) shows the experimentally obtained values of $-E_{U_\textsc{b}}$ showing the energy extraction for various values of $k/h_\textsc{a}$. Figure taken from  Fig. 4 of~\cite{IQCQETExperiment}.}
    \label{fig:ExperimentalData}
\end{figure}

In Fig.~\ref{fig:ExperimentalData}, we show the experimental results of these measurements, as well as an ideal simulation with no decoherence, and a simulation where optimized GRAPE pulses that account for decoherence are used. Each of the plots are given as functions of the coupling strengths between qubits A and B. In order to design the decoherence simulations, the following assumptions are used: the environment is Markovian, the qubit relaxations are independent of one another, and the dissipator commutes with the system Hamiltonian for the time discretization of GRAPE pulses. These assumptions are made in order to more easily implement the master equations for each time step as well as the evolution under propagator of the GRAPE pulses and dissipator. 

In all of the experimental trials, energy is extracted from qubit B. This marks the first implementation that activates a strong local passive state. We note that the discrepancies between the simulation and the experimental values are caused by two main differences between the two. The first is due to the decoherence assumptions used for the simulations, and the second is the way in which the GRAPE pulses are executed. In particular, the GRAPE pulses are executed slightly differently than what was assumed in the simulation. Due to this, the error bars shown in Fig.~\ref{fig:ExperimentalData} are the statistical error of the experiment.

\section{Superconducting quantum hardware simulations}\label{sec:SCImplementation}
In this section we discuss how the minimal QET protocol from Sec.~\ref{sec:minimalQET} was simulated using IBM's superconducting quantum computers~\cite{Ikeda2023}. In this study, complete quantum circuits are provided to simulate the QET protocol with a maximum circuit depth of six using two qubits. This verifies that the protocol can indeed be implemented in current publicly available quantum hardware. 

\subsection{Quantum circuits and results of the simulations}

\begin{figure}[!hbtp]
    \centering
    \includegraphics[width = 1\linewidth]{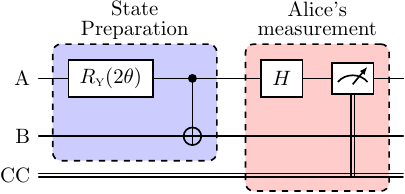}
    \caption{Quantum circuit used for the state preparation and Alice's measurement of qubit A. The wire denoted as CC is the classical channel will be used to transmit the information of the measurement result to Bob. This is a reconstructed version of Fig. 1 in~\cite{Ikeda2023}.}
    \label{fig:SCPrepCircuit}
\end{figure}

Same as in the NMR setup in Sec.~\ref{sec:IQCImplementation}, one must prepare the ground state, $\ket{g}$ of the Hamiltonian in Eq.~\eqref{eq:minimalHamiltonian}. This is accomplished through acting on the joint state $\ket{0_\textsc{a}0_\textsc{b}}$ of the two qubits in the following way
\begin{align}
    \ket{g} &= C_{\textsc{not}} (\hat R_{\textsc{y},\textsc{a}}(2\theta)\otimes\openone_\textsc{b})\ket{0_\textsc{a}0_\textsc{b}}\nonumber\\
    &= \cos(\theta)\ket{0_\textsc{a}0_\textsc{b}} + \sin(\theta)\ket{1_\textsc{a}1_\textsc{b}}\nonumber\\
    &= \frac{1}{\sqrt{2}}\left(\sqrt{1-g(h,k)}\ket{0_\textsc{a}0_\textsc{b}} - \sqrt{1+g(h,k)}\ket{1_\textsc{a}1_\textsc{b}}\right),
\end{align}
where $\hat R_{\textsc{y},\textsc{a}}(2\theta)$ is a rotation of qubit A by an angle of $2\theta$ around the $y$-axis and
\begin{align}
    \theta &= -\arccos\left(\frac{1}{\sqrt{2}}\sqrt{1-g(h,k)}\right),\\
    &g(h,k) = \frac{h}{\sqrt{h^2+k^2}}.
\end{align}
The quantum circuit that implements the $C_\textsc{not}$ and $\hat R_{\textsc{y},\textsc{a}}(2\theta)$ operations is given in (the first block of) Fig.~\ref{fig:SCPrepCircuit}. 

Now that we have the required ground state, the next stage of the protocol is for Alice to make a measurement of her qubit. Alice's measurement is implemented by the circuit in (The second block of) Fig.~\ref{fig:SCPrepCircuit}, where the classical channel storing the result of the measurement is denoted as CC. 

Moreover, (if we assume\footnote{Notice that there is no specified natural Hamiltonian or simulated Hamiltonian in this implementation. This will be discussed in more detail in the next section.} that the qubits are coupled by the Hamiltonian in Eq.~\eqref{eq:minimalHamiltonian})  Alice deposits a mean energy given by Eq.~\eqref{eq:MinAliceDeposit} into the system. This value was calculated on the quantum hardware and compared to the analytical value for different choices of $h$ and $k$, along with the values obtained by running this circuit on different quantum computers. The results of these calculations are shown in the first block of Table~\ref{table:SCResults}. The values can be compared against the analytical value of $E_{P_\textsc{a}}$. 

\begin{figure}
    \centering
    \includegraphics[width = 1\linewidth]{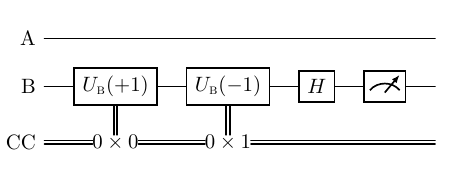}
    \caption{Quantum circuit used to measure $\hat V_\textsc{ab}$. Here we explicitly make note of the fact that Bob's operation is conditioned on the classical information he receives from Alice. This is a reconstructed Fig. 1 in~\cite{Ikeda2023}.}
    \label{fig:CircuitVab}
\end{figure}

\begin{figure}
    \centering
    \includegraphics[width=1\linewidth]{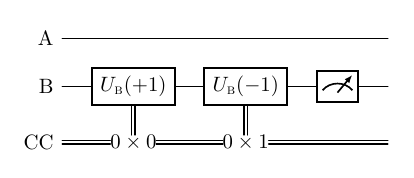}
    \caption{Quantum circuit used to measure $\hat H_\textsc{b}$. Here we make note of the fact that Bob's operation is conditioned on the result of Alice's measurement. This is a reconstructed version of Fig. 1 in~\cite{Ikeda2023}.}
    \label{fig:CircuitHb}
\end{figure}

Following Alice's measurement, she uses classical communication to tell Bob the result of her measurement. With this information, Bob performs the unitary $\hat U_\textsc{b}(\alpha)$, where 
\begin{equation}
    \hat U_\textsc{b}(\alpha) = \cos(\phi)\openone - \ii\alpha\sin(\phi)\s_y^\textsc{b} = \hat R_{\textsc{y},\textsc{b}}(2\alpha\phi),
\end{equation}
and $\phi$ is given by 
\begin{align}
    \cos(2\phi) &= \frac{h^2 + 2k^2}{\sqrt{(h^2+2k^2)^2+h^2k^2}},\\
    \sin(2\phi) &= \frac{hk}{\sqrt{(h^2+2k^2)^2+h^2k^2}}.
\end{align}
In order to determine the amount of energy extracted from Bob's qubit, one must measure both $\hat H_\textsc{b}$ and $\hat V_\textsc{ab}$\footnote{In~\cite{Ikeda2023}, it is claimed that ``the amount of energy available to Bob is greater if only $\hat V_\textsc{ab}$ is observed since $\langle\hat H_\textsc{b}\rangle$ is always positive." While it is true that $\langle\hat V_\textsc{ab}\rangle$ is responsible for the ability to extract energy, in practice, one cannot only calculate $\langle\hat V_\textsc{ab}\rangle$ to determine the amount of available energy to Bob. Rather, one must ensure that $|\langle\hat V_\textsc{ab}\rangle| > |\langle\hat H_\textsc{b}\rangle|$ in order to conclude that Bob extracted energy from his operation on qubit B.}. The quantum circuit that is used to calculate $\langle\hat V_\textsc{ab}\rangle$ is given in Fig.~\ref{fig:CircuitVab}, and the circuit used to calculate $\langle \hat H_\textsc{b}\rangle$ is given in Fig.~\ref{fig:CircuitHb}. 

As discussed in Sec.~\ref{sec:minimalQET}, if we make $\phi$ small enough, then the expected energy of Bob's operation becomes negative. This value was calculated using the \texttt{qasm\_simulator} that is capable of mimicking the gate operations of the quantum circuits for various values of $h$ and $k$ and was compared to the analytical values in Table~\ref{table:SCResults}. In looking at the results of the calculations, we see that all values computed using the implementations of the quantum circuits in Figs.~\ref{fig:CircuitVab} and~\ref{fig:CircuitHb} are quite close to those values predicted by the minimal QET protocol. 

\begin{figure}[h!]
    \centering
    \includegraphics[width = 1\linewidth]{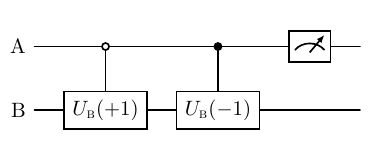}
    \caption{Quantum circuit that defers Alice's measurement until the conclusion of the simulation. This circuit is entirely equivalent to the one in Fig.~\ref{fig:CircuitVab}. Conveniently, with this circuit, we no longer have a need for carrying along the classical information in our circuit diagrams. This is a reconstructed version of Fig. 1 in~\cite{Ikeda2023}.}
    \label{fig:CircuitEquivalentVab}
\end{figure}

Importantly, for the QET protocol, Bob must be able to perform a conditional unitary based on the outcome of Alice's measurement. On IBM's publicly available quantum hardware however, these conditional unitaries are not natively supported operations, so they need to be implemented from other elementary operations. The author of~\cite{Ikeda2023} accomplishes this through a deferred measurement technique. In particular, Alice's measurement is delayed until the end of the circuit implementing the QET protocol. While executing the QET protocol the measurement of Alice does not commute with any of the other steps of the protocol, for the implementation in the IBM quantum computers one can substitute the conditional unitaries in Figs.~\ref{fig:CircuitVab} and \ref{fig:CircuitHb} by the equivalent circuit in Fig.~\ref{fig:CircuitEquivalentVab}. Fig.~\ref{fig:CircuitEquivalentVab} uses a sequence of a controlled and anticontrolled unitary gates given by 
\begin{equation}
    \hat \Lambda(U) = \ket{0_\textsc{a}}\!\bra{0_\textsc{a}}\otimes\openone_\textsc{b} + \ket{1_\textsc{a}}\!\bra{1_\textsc{a}}\otimes\hat U_\textsc{b},
\end{equation}
and 
\begin{equation}
    \hat{\tilde{\Lambda}}(U) = (\s_x^\textsc{a}\otimes\openone_\textsc{b})\hat \Lambda(U)(\s_x^\textsc{a}\otimes\openone_\textsc{b}),
\end{equation}
respectively. The quantum circuit for these operations is given in Fig.~\ref{fig:CircuitEquivalentVab}. 

\begin{figure}[h!]
    \centering
    \begin{tabular}{cc}
    \includegraphics[width=0.2\textwidth]{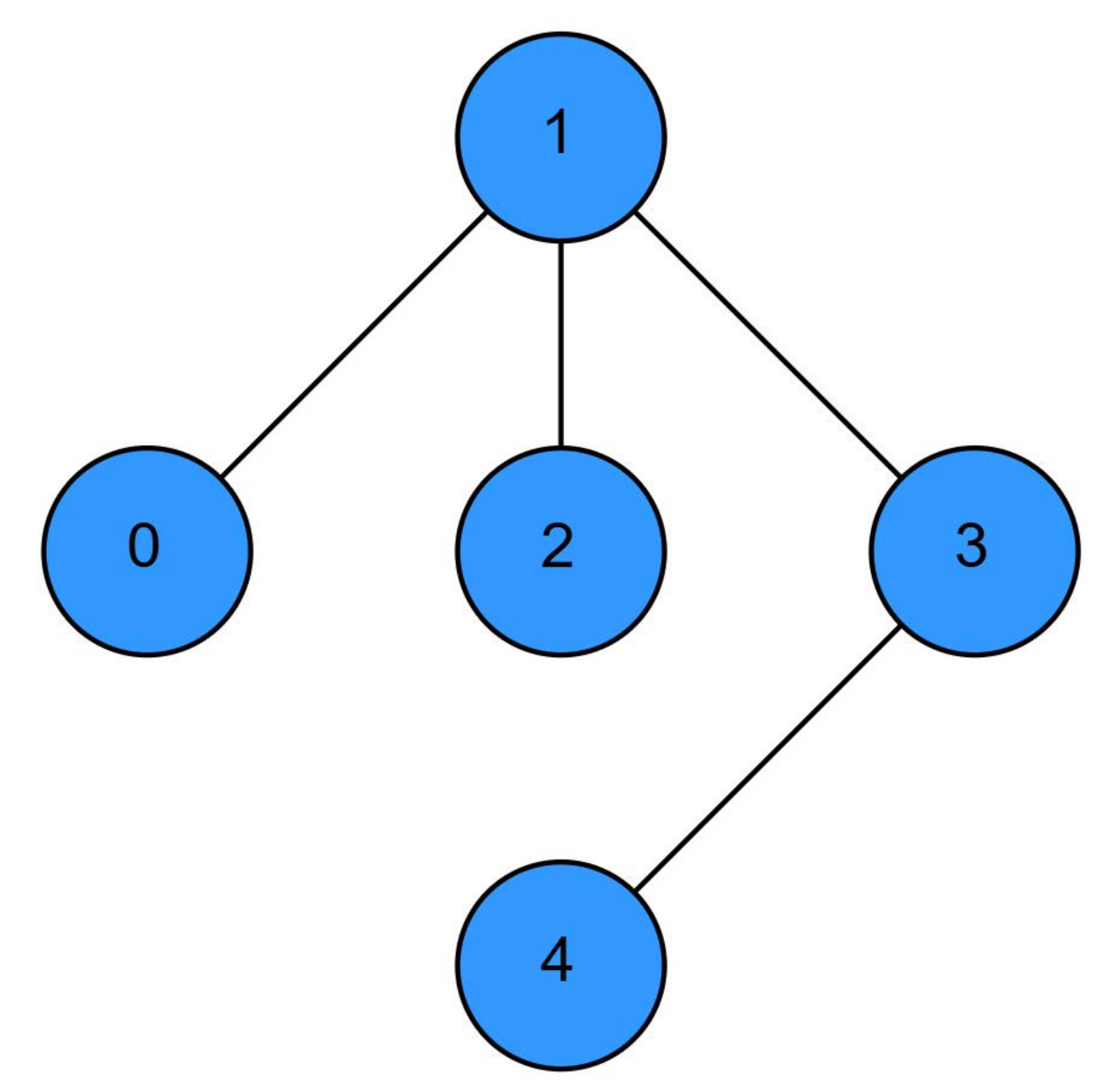} &
    \includegraphics[width=0.2\textwidth]{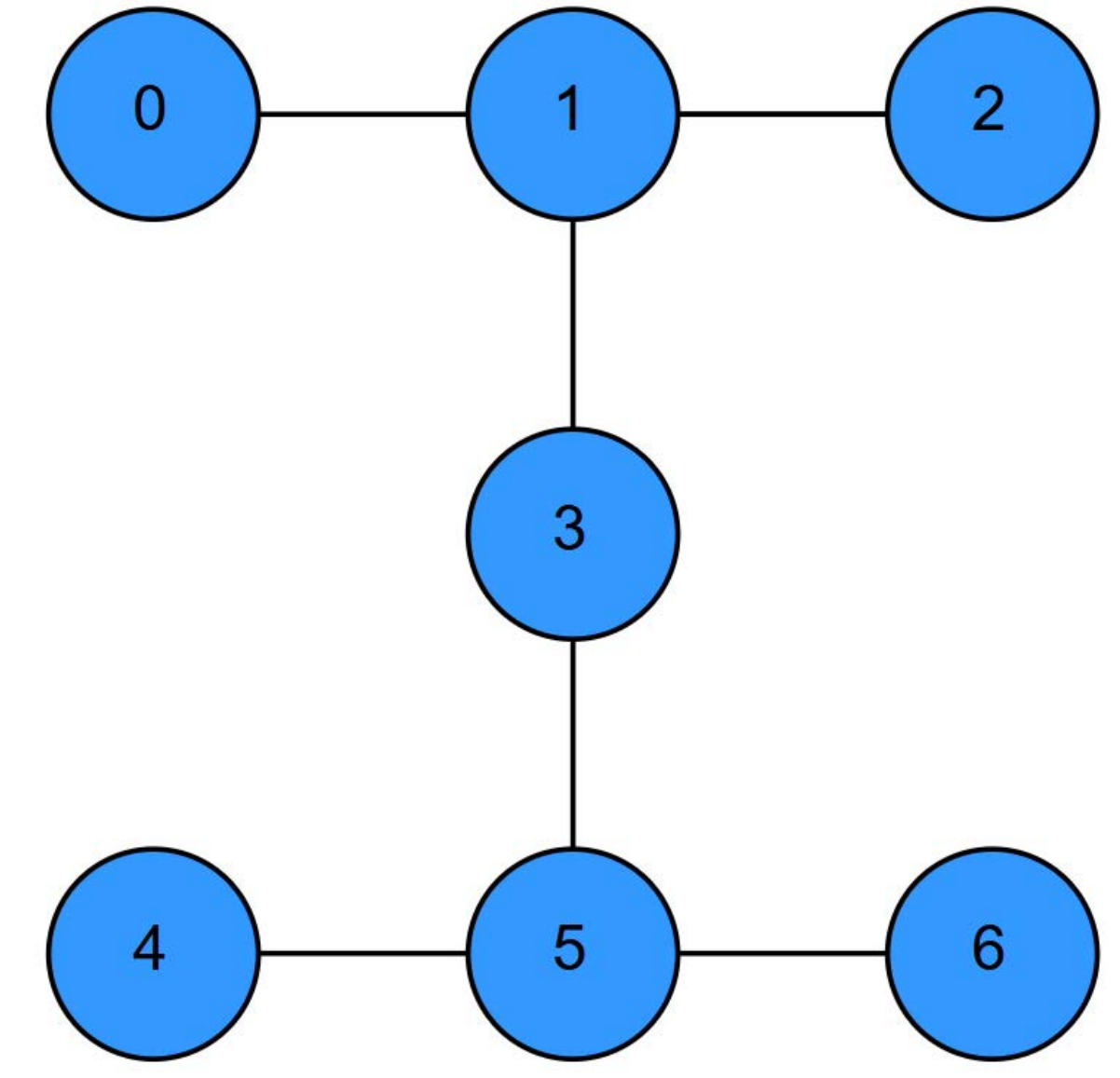}
  \end{tabular}
  \caption{(Left) Graph structure of \texttt{ibmq\_lima}. (Right) Graph structure of \texttt{ibmq\_jakarta}.}\label{fig:LimaJakarta}
\end{figure}

\begin{figure}[h!]
    \centering
    \includegraphics[width = 1\linewidth]{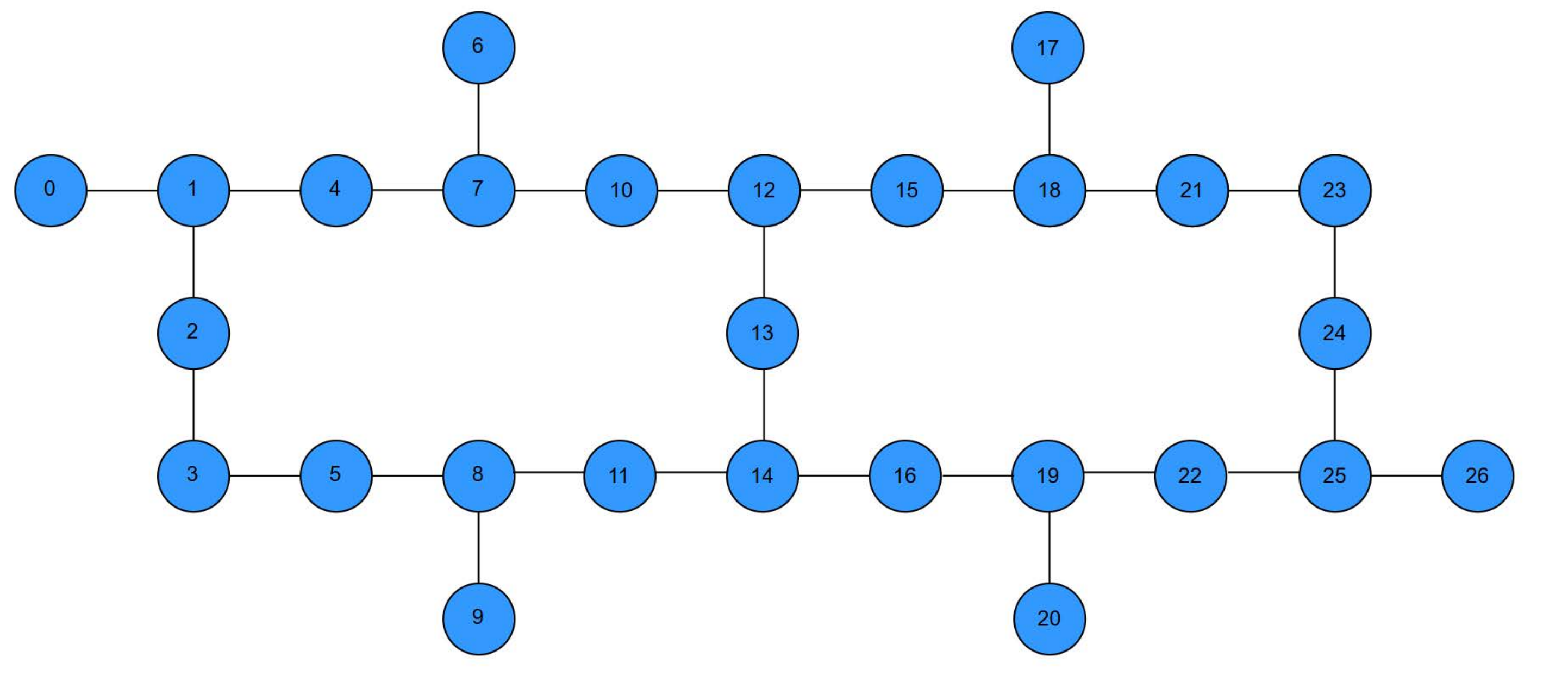}
    \caption{Graph structure of \texttt{ibm\_cairo}.}
    \label{fig:Cairo}
\end{figure}

The full circuit for the QET protocol (Fig.~\ref{fig:SCPrepCircuit} followed by \ref{fig:CircuitEquivalentVab}) was run on six of IBM's quantum computers: \texttt{ibmq\_lima}, \texttt{ibmq\_jakarta}, \texttt{ibm\_hanoi}, \texttt{ibm\_cairo}, \texttt{ibm\_auckland}, and \texttt{ibmq\_montreal}. It should be noted that all six of these devices are retired at the time this review was written, and therefore we are unable to reproduce the results of~\cite{Ikeda2023}.  Figs.~\ref{fig:LimaJakarta} and \ref{fig:Cairo} show the structures of each of \texttt{ibmq\_lima}, \texttt{ibmq\_jakarta}, and \texttt{ibm\_cairo}, respectively. In these figures, each numbered circle represents a single qubit, and the edges connecting qubits represent where a direct CNOT can be implemented between two qubits. We show only these three as \texttt{ibm\_hanoi}, \texttt{ibm\_auckland}, and \texttt{ibmq\_montreal} have the same structure as \texttt{ibm\_cairo}. In general, any two qubits within the graph can be selected to perform quantum computation; however, in this particular study, two qubits connected by an edge with relatively small CNOT and Hadamard gate errors were selected. Finally, in Table~\ref{table:SCResults} the calculated values of $E_{P_\textsc{a}}$, $\langle \hat H_\textsc{b}\rangle$, $\langle \hat V_\textsc{ab}\rangle$, and $ E_{U_\textsc{b}}$ are presented for various different hardwares and different coupling strengths $h$ and $k$. 

In comparing the values obtained through the quantum computers with the analytical values, as well as with the values calculated by \texttt{qasm\_simulator}, we can see that each of the quantum computers are in fact computing each of the expected values to a relatively high degree of accuracy. In analyzing each of these calculations, \texttt{ibmq\_jakarta} performed the best for coupling constants $h = 1.5$ and $k = 1$ with an accuracy of approximately 76\%. 

\begin{figure*}[htbp]
  \centering
  \begin{tabular}{cc}
    \includegraphics[width=0.5\textwidth]{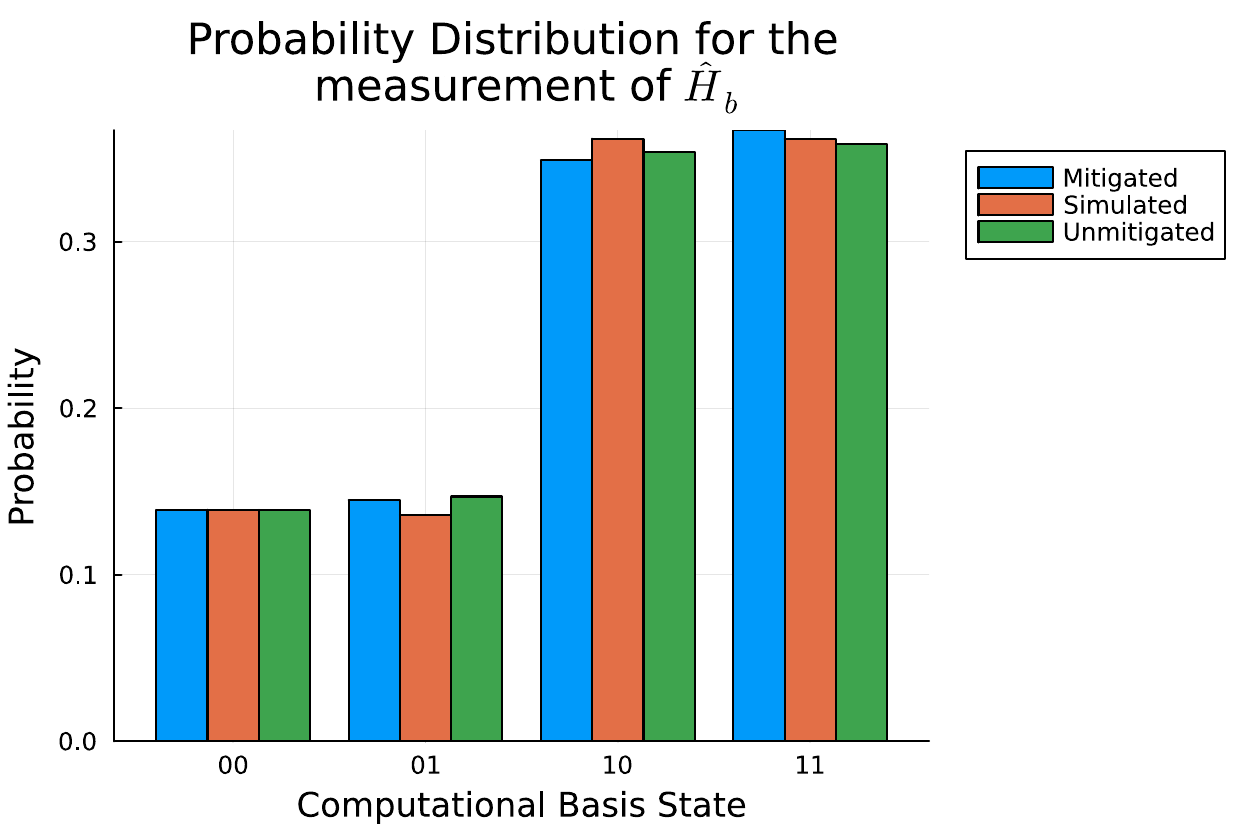} &
    \includegraphics[width=0.5\textwidth]{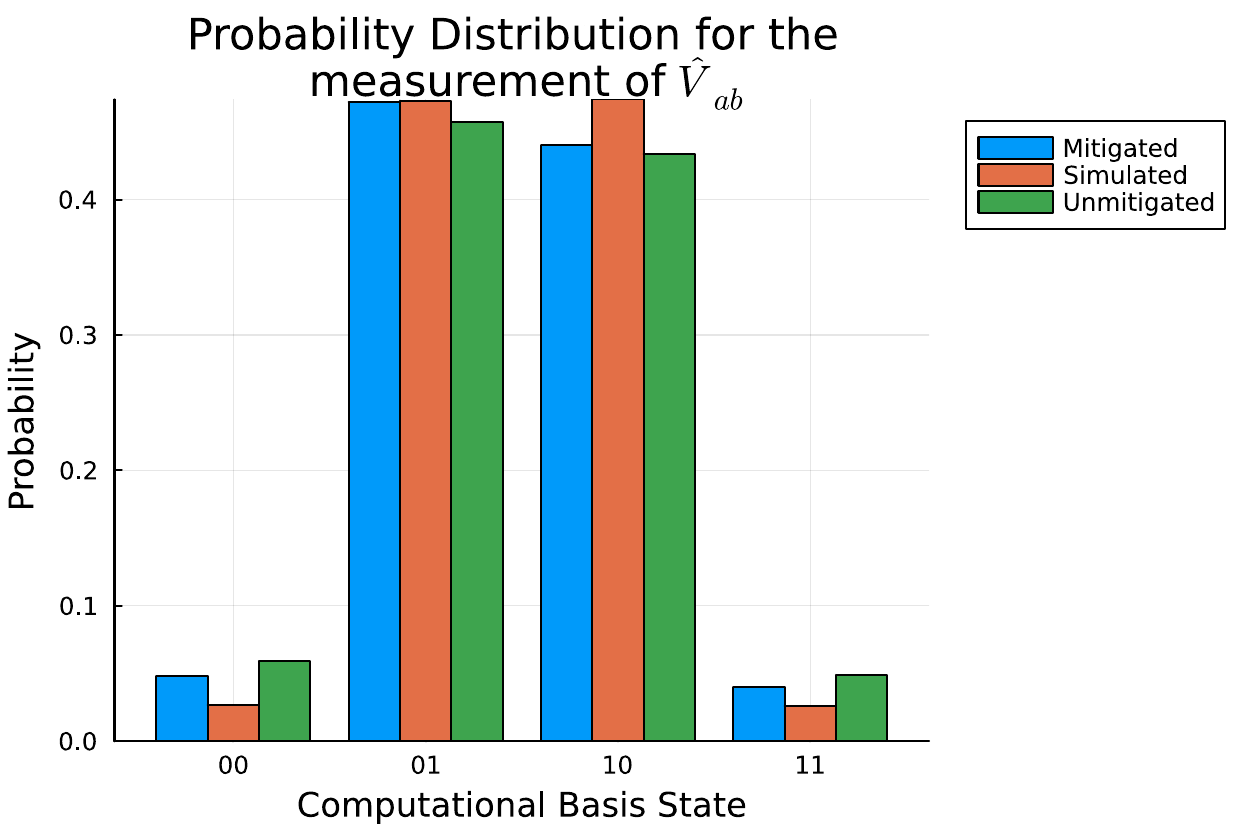}
  \end{tabular}
  \caption{Probability distribution for each of the computational basis states after applying the measurement calibration circuits. The left plot shows the results for the calculation of $\hat{V}_\textsc{ab}$ and the right plot shows the results for the calculation of $\hat{H}_\textsc{b}$. In each plot, we show the unmitigated, mitigated, and simulated distributions, and the distributions are only shown for calculations performed on \texttt{ibm\_cairo}. This is a reconstructed plot of Fig. 2 in~\cite{Ikeda2023} using the data from~\cite{Ikeda2023}.}
  \label{fig:MeasurementDistribution}
\end{figure*}

As in Sec.~\ref{sec:minimalQET},  the energy extracted by Bob was computed for an ensemble of repetitions of the protocol. Therefore, an important point of~\cite{Ikeda2023} is understanding the distribution of states obtained through identical repetitions the QET protocol, and how the measurement results vary when error-mitigation is used in each of the calculations. In order to apply the error-mitigation, the first step was to understand the effects of measurement errors within the quantum circuit. Four measurement calibration circuits were constructed for the full Hilbert space and were immediately measured in order to determine the probability distribution of each of the four computational basis states of the two qubits. In Fig.~\ref{fig:MeasurementDistribution} we show the results of these measurements obtained by the author from the \texttt{ibm\_cairo} quantum computer. Each of these values was compared to the \texttt{qasm\_simulator} to determine the extent of measurement error within the actual hardware. Given the results of the measurement errors, a calibration matrix was then constructed and applied to the results of all measurements. These corrected values can be seen in the rows labeled ``error mitigation" in Table~\ref{table:SCResults}.

\begin{table*}[htbp]
  \centering
  \begin{tabular}{|l|l|
                  S[table-format=-1.4(4)]|
                  S[table-format=-1.4(4)]|
                  S[table-format=-1.4(4)]|
                  S[table-format=-1.4(4)]|}
    \hline
    \multicolumn{1}{|c|}{\multirow{2}{*}{\textbf{Backend}}}
      & \multicolumn{1}{|c|}{\multirow{2}{*}{\textbf{Mode}}}
      & \multicolumn{4}{c|}{$(h,k)$} \\
    \cline{3-6}
      & 
      & {(1,0.2)} & {(1,0.5)} & {(1,1)} & {(1.5,1)} \\
    \hline
    Analytical value
      & \multicolumn{1}{|c|}{$E_{P_\textsc{a}}$}
      & \multicolumn{1}{|c|}{0.9806}
      & \multicolumn{1}{|c|}{0.8944}
      & \multicolumn{1}{|c|}{0.7071}
      & \multicolumn{1}{|c|}{1.2481} \\
    \hline
    \multirow{1}{*}{\texttt{qasm\_simulator}}
      & {}     
      & 0.9827\pm0.0031 & 0.8941\pm0.0001 & 0.7088\pm0.0001 & 1.2437\pm0.0047 \\
    \hline
    \multirow{2}{*}{\texttt{ibmq\_lima}}
      & error mitigated  
      & 0.9423\pm0.0032 & 0.8169\pm0.0032 & 0.6560\pm0.0031 & 1.2480\pm0.0047 \\
    \cline{2-6}
      & unmitigated      
      & 0.9049\pm0.0017 & 0.8550\pm0.0032 & 0.6874\pm0.0031 & 1.4066\pm0.0047 \\
    \hline
    \multirow{2}{*}{\texttt{ibmq\_jakarta}}
      & error mitigated  
      & 0.9299\pm0.0056 & 0.8888\pm0.0056 & 0.7039\pm0.0056 & 1.2318\pm0.0084 \\
    \cline{2-6}
      & unmitigated      
      & 0.9542\pm0.0056 & 0.9089\pm0.0056 & 0.7232\pm0.0056 & 1.2624\pm0.0083 \\
    \hline
    \multirow{2}{*}{\texttt{ibm\_cairo}}
      & error mitigated  
      & 0.9571\pm0.0032 & 0.8626\pm0.0031 & 0.7277\pm0.0031 & 1.2072\pm0.0047 \\
    \cline{2-6}
      & unmitigated      
      & 0.9578\pm0.0031 & 0.8735\pm0.0031 & 0.7362\pm0.0031 & 1.2236\pm0.0047 \\
    \hline
    \hline
    Analytical value
      & \multicolumn{1}{|c|}{$\langle \hat H_\textsc{b}\rangle$}
      & \multicolumn{1}{|c|}{0.0521}
      & \multicolumn{1}{|c|}{0.1873}
      & \multicolumn{1}{|c|}{0.2598}
      & \multicolumn{1}{|c|}{0.3480} \\
    \hline
    \multirow{1}{*}{\texttt{qasm\_simulator}}
      & {}  
      & 0.0547\pm0.0012 & 0.1857\pm0.0022 & 0.2550\pm0.0028 & 0.3487\pm0.0038 \\
    \hline
    \multirow{2}{*}{\texttt{ibmq\_lima}}
      & error mitigated  
      & 0.0733\pm0.0032 & 0.1934\pm0.0032 & 0.2526\pm0.0032 & 0.3590\pm0.0047 \\
    \cline{2-6}
      & unmitigated      
      & 0.1295\pm0.0053 & 0.2422\pm0.0024 & 0.2949\pm0.0028 & 0.4302\pm0.0039 \\
    \hline
    \multirow{2}{*}{\texttt{ibmq\_jakarta}}
      & error mitigated  
      & 0.0736\pm0.0055 & 0.2018\pm0.0056 & 0.2491\pm0.0056 & 0.3390\pm0.0084 \\
    \cline{2-6}
      & unmitigated      
      & 0.0852\pm0.0022 & 0.2975\pm0.0045 & 0.3365\pm0.0052 & 0.4871\pm0.0073 \\
    \hline
    \multirow{2}{*}{\texttt{ibm\_cairo}}
      & error mitigated  
      & 0.0674\pm0.0032 & 0.1653\pm0.0031 & 0.2579\pm0.0031 & 0.3559\pm0.0047 \\
    \cline{2-6}
      & unmitigated      
      & 0.0905\pm0.0014 & 0.1825\pm0.0022 & 0.2630\pm0.0027 & 0.3737\pm0.0037 \\
    \hline
        \hline
    Analytical value
      & \multicolumn{1}{|c|}{$\langle \hat V_\textsc{ab}\rangle$}
      & \multicolumn{1}{|c|}{-0.0701}
      & \multicolumn{1}{|c|}{-0.2598}
      & \multicolumn{1}{|c|}{-0.3746}
      & \multicolumn{1}{|c|}{-0.4905} \\
    \hline
    \multirow{1}{*}{\texttt{qasm\_simulator}}
      & {}  
      & -0.0708\pm0.0012 & -0.2608\pm0.0032 & -0.3729\pm0.0063 & -0.4921\pm0.0038 \\
    \hline
    \multirow{2}{*}{\texttt{ibmq\_lima}}
      & error mitigated  
      & -0.0655\pm0.0012 & -0.2041\pm0.0031 & -0.2744\pm0.0063 & -0.4091\pm0.0063 \\
    \cline{2-6}
      & unmitigated      
      & -0.0538\pm0.0011 & -0.1471\pm0.0025 & -0.1233\pm0.0041 & -0.2737\pm0.0046 \\
    \hline
    \multirow{2}{*}{\texttt{ibmq\_jakarta}}
      & error mitigated  
      & -0.0515\pm0.0022 & -0.2348\pm0.0056 & -0.3255\pm0.0112 & -0.4469\pm0.0112 \\
    \cline{2-6}
      & unmitigated      
      & -0.0338\pm0.0021 & -0.1371\pm0.0046 & -0.0750\pm0.0075 & -0.2229\pm0.0083 \\
    \hline
    \multirow{2}{*}{\texttt{ibm\_cairo}}
      & error mitigated  
      & -0.0497\pm0.0013 & -0.1968\pm0.0031 & -0.2569\pm0.0063 & -0.3804\pm0.0063 \\
    \cline{2-6}
      & unmitigated      
      & -0.0471\pm0.0012 & -0.1682\pm0.0026 & -0.1733\pm0.0038 & -0.3089\pm0.0045 \\
    \hline
    \hline
    Analytical value
      & \multicolumn{1}{|c|}{$E_{U_\textsc{b}}$}
      & \multicolumn{1}{|c|}{-0.0180}
      & \multicolumn{1}{|c|}{-0.0726}
      & \multicolumn{1}{|c|}{-0.1147}
      & \multicolumn{1}{|c|}{-0.1425} \\
    \hline
    \multirow{1}{*}{\texttt{qasm\_simulator}}
      & {}  
      & -0.0161\pm0.0017 & -0.0751\pm0.0040 & -0.1179\pm0.0069 & -0.1433\pm0.0054 \\
    \hline
    \multirow{2}{*}{\texttt{ibmq\_lima}}
      & error mitigated  
      & 0.0078\pm0.0034 & -0.0107\pm0.0045 & -0.0217\pm0.0071 & -0.0501\pm0.0079 \\
    \cline{2-6}
      & unmitigated      
      & 0.0757\pm0.0054 & 0.0950\pm0.0035 & 0.1715\pm0.0050 & 0.1565\pm0.0060 \\
    \hline
    \multirow{2}{*}{\texttt{ibmq\_jakarta}}
      & error mitigated  
      & 0.0221\pm0.0059 & -0.0330\pm0.0079 & -0.0764\pm0.0125 & -0.1079\pm0.0140 \\
    \cline{2-6}
      & unmitigated      
      & 0.0514\pm0.0030 & 0.1604\pm0.0064 & 0.2615\pm0.0091 & 0.2642\pm0.0011 \\
    \hline
    \multirow{2}{*}{\texttt{ibm\_cairo}}
      & error mitigated  
      & 0.0177\pm0.0035 & -0.0315\pm0.0044 & 0.0010\pm0.0070 & -0.0245\pm0.0079 \\
    \cline{2-6}
      & unmitigated      
      & 0.0433\pm0.0018 & 0.0143\pm0.0034 & 0.0897\pm0.0047 & 0.0648\pm0.0058 \\
    \hline
  \end{tabular}
  \caption{This table is a reconstructed version of Table 1 in~\cite{Ikeda2023} showing the results for the calculation of each step of the QET process obtained by the author. The data is presented for three of the quantum computers, and for each of these computers, the results are shown when the error mitigation scheme is applied, and when it is not. The results are shown for multiple values of $(h,k)$.}
  \label{table:SCResults}
\end{table*}

While Fig.~\ref{fig:MeasurementDistribution} only shows the probability distributions measured for \texttt{ibm\_cairo}, the results of the measurement calibrations on the other five quantum computers showed similar behaviors to that of \texttt{ibm\_cairo}. In Fig.~\ref{fig:MeasurementDistribution}, we can see that the results of the \texttt{qasm\_simulator} agree quite well with the results obtained in the measurement calibration for the actual hardware. Moreover, we see that the inclusion of the error-mitigation scheme discussed improves the overall accuracy of each of the desired measurements, as can be seen in Table~\ref{table:SCResults}. Importantly, for all of the quantum computers used, and for all combinations of $h$ and $k$, it was found that $\langle\hat V_\textsc{ab}\rangle$ was, in fact, negative, and that $|\langle\hat V_\textsc{ab}\rangle| > |\langle\hat H_\textsc{b}\rangle|$. 

\subsection{Comparison with NMR based QET}

As discussed in Sec.~\ref{sec:IQCImplementation}, the QET protocol relies on Alice communicating her measurements to Bob faster than the natural energy flow within the system. This inherently requires that any implementation of QET takes into account the Hamiltonian describing the two coupled qubits, and how it affects the natural energy flow between the parts of the system. 


One key difference between the two implementations is that the one in~\cite{Ikeda2023} has no specified natural Hamiltonian for the system. Rather, two qubits within a quantum computer (starting in a logical $\ket{0}$ state) are manipulated to prepare them in a state that matches the ground state of Eq.~\eqref{eq:MinimalEntangledGround}. The time evolution according to the Hamiltonian in~\eqref{eq:minimalHamiltonian} was never simulated. Unlike in the NMR implementation where the actual coupling between the different nuclei is known, in~\cite{Ikeda2023}, it is unclear what (if any) the natural Hamiltonian is within this system. While this natural Hamiltonian may very well exist and be connected to the two-qubit gate time on the IBM computers, we do not know (nor is specified in~\cite{Ikeda2023}) how such a connection would come about. This makes it more challenging to assess the speed of the QET protocol to make sure that no energy from A to B propagated naturally through the lattice before the protocol is finished.

Due to the fact that there is no specified natural time evolution in the IBM simulation, in the spectrum of possible implementations from quantum simulations to true verifications of the QET protocol, \cite{Ikeda2023} feels closer to the former than the NMR implementation~\cite{IQCQETExperiment}. Perhaps a way in which \cite{Ikeda2023} could be upgraded to be on equal footing with \cite{IQCQETExperiment} would be to include some form of time discretization and pulse generation that induces a flow of energy between the two qubits according to some simulated `natural' Hamiltonian. Having these time dynamics could allow one to claim that, if $\langle \hat V_\textsc{ab} + \hat H_\textsc{b}\rangle<0$, then energy was extracted from the ground state before Alice's measurement energy can flow to Bob under the natural time evolution of the system. 



\section{Applications of QET to algorithmic cooling}\label{sec:AlgorithmicCooling}

We now move on to reviewing a practical application of the QET protocol: its use in quantum thermodynamics as a tool for subsystem purification. In particular, we focus on the work of~\cite{QETAlgorithmicCooling}, where the authors propose a method for leveraging QET to purify a designated subsystem within a many-body interacting quantum system. We review the operational principles underlying this QET-based cooling scheme, discuss its implementation, and provide a resource-based analysis that highlights how and why the protocol can outperform conventional cooling strategies—especially those designed for systems with weak or negligible interactions.


\subsection{Algorithmic cooling and QET}

Algorithmic cooling is a quantum information processing technique used to reduce the entropy (or increase the purity) of a subset of qubits in a multipartite quantum system, typically to improve their initialization for quantum computation~\cite{BoykinMorAC,FernandezMorAC,BaughLaflamme2005,LaflammeParkBook}. It involves redistributing entropy from some qubits to others through a combination of unitary operations (reversible logic gates) and--typically--rethermalization steps (irreversible interactions with a heat bath or resettable qubit). The goal is to cool a few qubits (make them closer to the ground state, i.e., more pure), even at the expense of making others more mixed or noisy.  In this section, we review the use of QET as a protocol in the context of algorithmic cooling for systems where there are internal interactions. We will review the work~\cite{QETAlgorithmicCooling} where it was shown how to utilize the correlations between qubit chains in thermal contact with a heat bath to improve the efficiency of heat bath algorithmic cooling (HBAC). 

The challenge with cooling strongly interacting qubits is that if the interaction is such that the ground state of the whole system is entangled, the individual qubits--even in the global ground state---can be highly non-pure. QET-based algorithmic cooling excels at cooling individual qubits when interactions between the qubits are non-negligible precisely by using the multi-qubit correlations to remove energy and cool down individual qubits. In particular, this will be demonstrated by showing that subsystem purity can be improved further than what is currently achieved~\cite{BoykinMorAC,NayeliACPRL,RaeisiMoscaAC} by algorithmic cooling methods that are optimal for non-interacting systems. 

Before discussing how QET can improve over current algorithmic cooling methods, we will present a brief discussion of two HBAC algorithms that will be used as a testbed against QET-based cooling methods. The first is known as the partner pairing algorithm (PPA)~\cite{PPA1,LaflammeParkBook} in which we have a target qubit (the one to be cooled, or equivalently to be purified), a system of ``reset" qubits, and a thermal bath. Specifically, if we have $n$ qubits we will denote the PPA method of $n$ qubits as PPA-$n$. The PPA-$n$ method aims to purify the target qubit by iterating over two steps; the first is entropy compression, followed by pumping entropy out of the system and into the heat bath. The entropy compression step is a unitary operation that removes some of the entropy from the target qubit and moves it to the reset qubits. Then, by allowing the reset qubits to thermalize with the heat bath, the entropy is pumped out of the reset qubit and the process can be iterated to purify the target qubit up to a threshold~\cite{NayeliACPRL}. This allows one to cool the target qubits to a temperature below the temperature of the cool bath. 

The second algorithmic cooling method that we will compare our results against is known as \mbox{SR$\Gamma_n$-HBAC~\cite{SRHBAC}}.  This method differs from PPA in that the coupling to the environment is not limited to just rethermalization, but could also include correlations between the qubits of the system and the bath. The usage of those correlations allows for more efficient state resets. In more detail, inspired by the Nuclear Overhauser
Eﬀect~\cite{NuclearOverhauser}, the authors of~\cite{SRHBAC} take advantage of the fact that there are directions in the subspace of the qubits that thermalize faster in particular directions to beat the theoretical limits of PPA-HBAC protocols. With this, we now present the QET assisted algorithmic cooling method and demonstrate an improvement over both of the previously discussed methods. 

\subsection{QET algorithmic cooling}

The cooling method presented in~\cite{QETAlgorithmicCooling} begins with a generalization of the minimal QET model presented in Sec.~\ref{sec:minimalQET} in which we replace the projective measurement $\hat P_\textsc{a}(\alpha)$ in Eq.~\eqref{eq:Palpha} with a POVM given by the measurement operator
\begin{equation}
    \hat M_\textsc{a}(\alpha) = e^{\ii \delta_\alpha}\left(m_\alpha\openone + e^{\ii\gamma_\alpha}l_\alpha\s_x^\textsc{a}\right),
\end{equation}
which models the action of a generalized measurement of $\s_x^\textsc{a}$ yielding result $\alpha$. In this POVM, $\delta_\alpha$, $m_\alpha$, $\gamma_\alpha$, and $l_\alpha$ are real coefficients that satisfy 
\begin{align}
    \sum_{\alpha}(m_\alpha^2+l_\alpha^2) = 1,\\
    \sum_{\alpha}m_\alpha l_\alpha \cos(\gamma_\alpha) = 0.
\end{align}
Given this POVM\footnote{Notice that the particular choice $\delta_\alpha = \gamma_\alpha = 0$, $l_\alpha = \alpha/2$, and $m_\alpha = 1/2$ yields the PVM case presented in Sec.~\ref{sec:minimalQET}.} the unitary that optimizes Bob's energy extraction is the following 
\begin{equation}
    \hat U_\textsc{b}(\alpha) = \cos(\Omega_\alpha)\openone +\ii \sin(\Omega_\alpha)\s_y^\textsc{b},
\end{equation}
where
\begin{align}
    \cos(2\Omega_\alpha) = \frac{(h^2+2k^2)p_\textsc{a}(\alpha)}{\sqrt{(h^2+2k^2)^2p_\textsc{a}^2(\alpha) + h^2k^2q_\textsc{a}^2(\alpha)}},\\
    \sin(2\Omega_\alpha) = - \frac{hkq_\textsc{a}(\alpha)}{\sqrt{(h^2+2k^2)^2p_\textsc{a}^2(\alpha) + h^2k^2q_\textsc{a}^2(\alpha)}},
\end{align}
and $p_\textsc{a}(\alpha) = m_\alpha^2+l_\alpha^2$ and $q_\textsc{a}(\alpha) = 2l_\alpha m_\alpha\cos(\gamma_\alpha)$.

Applying this protocol boosts the purity of Bob's qubit. To see it, we first calculate the initial purity, $\mathcal{P}_i^\textsc{b}$, of $\hat{\rho}_\textsc{b}$. The initial purity is 
\begin{align}
    \mathcal{P}_0^\textsc{b} = \text{Tr}(\hat{\rho}_\textsc{b}^2),
\end{align}
where $\hat{\rho}_\textsc{b} = \text{Tr}_\textsc{a}(\hat{\rho}_{\textsc{a}\textsc{b}})$ and $\hat{\rho}_{\textsc{a}\textsc{b}}$ is the density operator of the ground state defined in Eq.~\eqref{eq:MinimalEntangledGround}. We find that 
\begin{equation}
    \hat{\rho}_\textsc{b} = \frac{1}{2}\left(C_-^2\ket{1_\textsc{b}}\!\bra{1_\textsc{b}}+C_+^2\ket{0_\textsc{a}}\!\bra{0_\textsc{b}}\right)
\end{equation}
which leads to the initial purity of subsystem B being given by
\begin{equation}
    \mathcal{P}_0^\textsc{b} = \frac{2h^2+k^2}{2(h^2+k^2)}.
\end{equation}
In order to calculate the final purity of subsystem B after applying the QET protocol, we will first have to determine the state of subsystem B after the protocol. Analogously to Sec.~\ref{sec:minimalQET}, the final state of the full system will be given by 
\begin{equation}
    \hat{\rho}_f = \sum_{\alpha = \pm 1} \hat U_\textsc{b}(\alpha)\hat M_\textsc{a}(\alpha)\ket{g}\!\bra{g}\hat M^\dagger_\textsc{a}(\alpha) \hat U_\textsc{b}^\dagger(\alpha)
\end{equation}
Following the same procedure for the initial purity, it is possible, though lengthy, to evaluate the final purity of subsystem B. Assuming for simplicity that $\gamma_\alpha = 0$, the final purity of subsystem B after the QET protocol is given by 
\begin{align}
    &\mathcal{P}_f^\textsc{b} = \frac{2}{h^2+k^2}\bigg(\frac{h^2}{2}+\frac{k^2}{4} - hkl_1m_1\sin(2(\Omega_0 - \Omega_1))\nonumber\\
    &+ \!\big(4k^2l_1^2m_1^2+h^2(l_1^2+m_1^2-1)(l_1^2+m_1^2)\big)\sin^2(\Omega_0-\Omega_1)\bigg).
\end{align}
One can check that in cases the unitary is chosen to have a negative energy cost through the QET protocol, Bob's qubit is also purified. 

To compare how efficiently QET performs as an algorithm cooling protocol, let us now compare how it purifies a Gibbs state of an interacting system with how PPA-HBAC and SR$\Gamma_n$-HBAC perform under the same conditions. Consider a two-qubit state with an interaction defined by Eq.~\eqref{eq:minimalHamiltonian} in a thermal state of inverse temperature, $\beta$, given by 
\begin{equation}\label{eq:thermalstate}
    \hat{\rho}_\beta = \frac{1}{\text{Tr}(e^{-\beta\hat{H}})}e^{-\beta\hat{H}}.
\end{equation}

\begin{figure}
    \centering
    \includegraphics[width=1\linewidth]{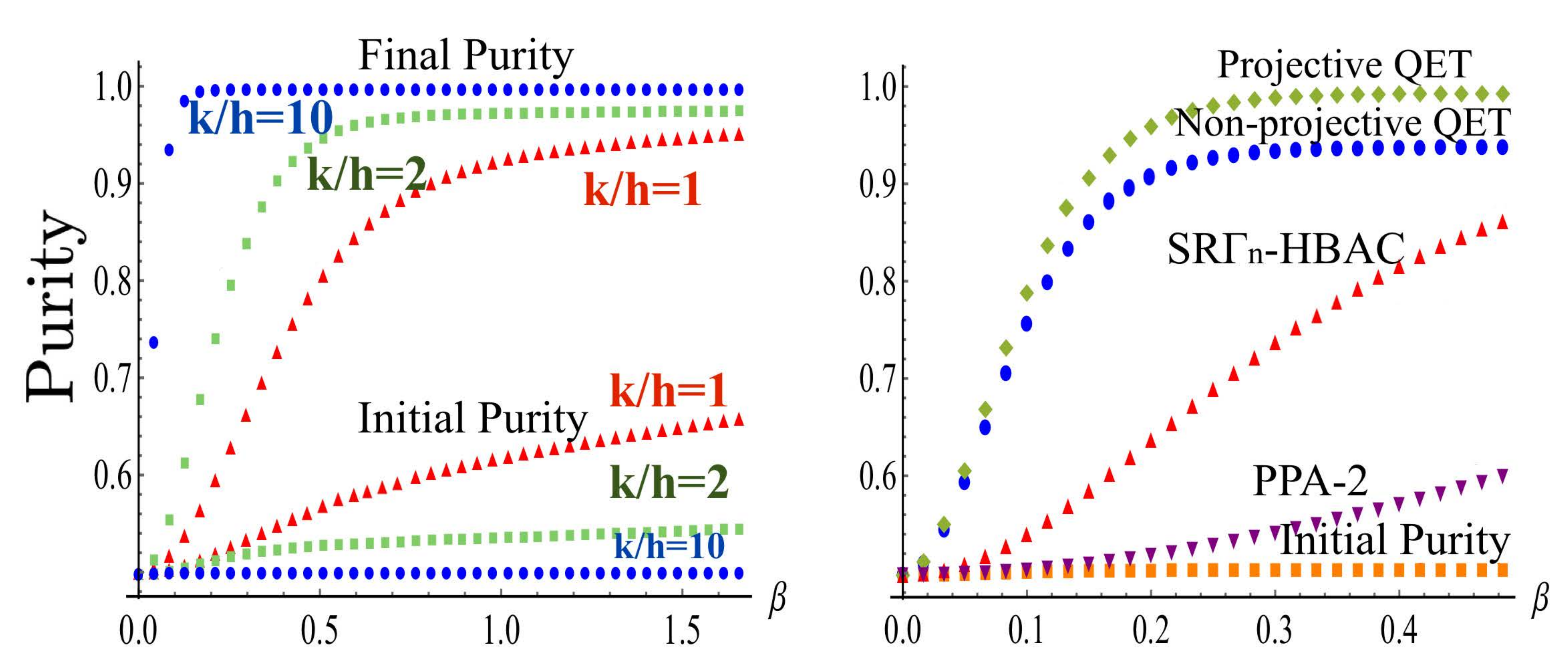}    
    \caption{(Left) Initial and final purities of qubit B before and after the minimal QET protocol has been implemented. The purity is given as a function of the inverse temperature $\beta$ for various values of $k/h$. (Right) Comparison of the final purity of qubit B as a function of $\beta$ for QET-based algorithmic cooling using projective POVMs (green), QET-based algorithmic cooling using non-projective measurements (blue), SR$\Gamma_n$-HBAC (red), and PPA-2 (purple) for $k/h = 5$. The initial purity of qubit B is also given as a reference for the purity enhancement of each algorithmic cooling method. This is a slightly modified version of Fig. 1 from~\cite{QETAlgorithmicCooling}.}
    \label{fig:PurityComparison}
\end{figure}

The measurement that maximizes the purification of Bob's qubit is precisely the case where the POVM corresponds to a PVM. We see in this case in Fig.~\ref{fig:PurityComparison}a that the purity achievable for Bob's qubit through the QET protocol increases quickly as the relative coupling strength  $k/h$ increases. 

However, we can see that one does not need to consider the optimal case (which is pretty idealized) in order for the QET protocol to yield high purification, outperforming other available algorithmic cooling methods. To show this, the authors of~\cite{QETAlgorithmicCooling} repeated the study for a partial optimization over POVMs such that the measurement operators were at least a distance of 1/2 from projective measurements in the Frobenius norm. This guarantees that the chosen measurements are far from the ideal PVMs case. The results of this are shown in Fig.~\ref{fig:PurityComparison}b. This figure shows the purity of subsystem B as a function of the inverse temperature for different algorithmic cooling methods for a fixed relative coupling strength of $k/h = 5$ in the case of the POVM partial optimization. We see that, even when a non-PVM is used, there is still a marked increased in the purity enhancement over the other baseline algorithmic cooling methods used for the same system of two qubits. A deeper comparison with other algorithmic cooling methods follows in the next subsection. 

\subsection{Comparison of QET and other methods from a resource perspective}

We can better understand how QET assisted algorithmic cooling measures to other methods by performing a comparison from a resource perspective. In the PPA-$n$ method, for the entropy compression step, non-local unitaries are applied across all available $n$ qubits. Then a thermalization step is applied to the reset qubits turning them back to a completely uncorrelated thermal state through the use of a bath. For the protocol, it is assumed that both of these steps can be repeated an arbitrary number of times until the fixed point of the target qubit's purity is reached. From the resource perspective, one needs access to $n$ qubits, non-local unitaries among them and the ability to reset some subset of them by destroying their correlations through (independent) thermalization with an external bath. 

In contrast, rather than destroying the correlations in the system, QET uses LOCC with single qubit operations and does not require a thermal bath as a resource. Instead, QET utilizes the correlations that are already present in the system due to the interaction. However, when performing the protocol, one would need the ability to perform the right POVMs for the protocol to yield purification.

In principle this is possible through LOCC, but as we showed in Sec.~\ref{sec:FullyUnitaryQET}, QET can also be implemented through quantum communication. This has advantages (easier to perform optimizations in practice) and disadvantages (requires access and control of an extra ancillary qubit). 

Indeed, in \cite{QETAlgorithmicCooling} it was shown how to use an ancilla to implement purification through a QET protocol. As discussed in Subsection~\ref{sec:FullyUnitaryQET}, we use an ancilla to gain information about subsystem A, and then make the ancilla interact with subsystem B to deliver the information. Concretely in the simple implementation discussed in~\cite{QETAlgorithmicCooling}, the ancilla gains information about subsystem A through the application of a joint unitary on A and $\text{A}_n$ given by
\begin{equation}\label{eq:ACAAnUnitary}
    \hat U_\textsc{a} = e^{\ii\s_y^\textsc{a}\s_y^{\textsc{a}_n}}.
\end{equation} 
Once this joint unitary has been applied, one then applies the following unitary on the ancilla and subsystem B 
\begin{equation}\label{eq:ACBAnUnitary}
    \hat U_\textsc{b} = e^{\ii\s_x^\textsc{b}\s_z^{\textsc{a}_n}}.
\end{equation}

If we assume that the ancilla has some internal Hamiltonian given by 
\begin{equation}
    \hat H_{\textsc{a}_n} = h_{\textsc{a}_n}\s_z^{\textsc{a}_n}
\end{equation}
one can derive an expression for the final purity of subsystem B. Specifically, it is found that
\begin{align}
    &\mathcal{P}_f^\textsc{b} = \frac{1}{2} + \frac{h_-S_+^2\left[(h_\textsc{a}^2+h_\textsc{b}^2)+k^2\sin^4(2)\tanh^2(\beta h_{\textsc{a}_n})\right]}{2h_-h_+(C_-+C_+)^2}\nonumber\\
    & + \frac{S_-^2\left[h_+\left[(h_\textsc{a}-h_\textsc{b})^2 + k^2\sin^4(2)\tanh^2(\beta h_{\textsc{a}_n})\right]+2h_\textsc{b}^2h_r\right]}{2h_-h_+(C_-+C_+)^2}\nonumber\\
    & - \frac{2h_rS_+S_-\left[h_\textsc{a}^2+k^2\sin^4(2)\tanh^2(\beta h_{\textsc{a}_n})\right]}{2h_-h_+(C_-+C_+)^2},
\end{align}
where
\begin{align}
    &h_\pm \coloneqq (h_\textsc{a}\pm h_\textsc{b})^2 + k^2, && h_r \coloneqq \sqrt{\frac{1}{2}(h_-^2+h_+^2)-8h_\textsc{a}^2h_\textsc{b}^2},\nonumber\\
    &S_\pm \coloneqq \sinh(\sqrt{h_\pm \beta}), && C_\pm \coloneqq \cosh(\sqrt{h_\pm \beta}).
\end{align}
\begin{figure}
    \centering
    \includegraphics[width=\linewidth]{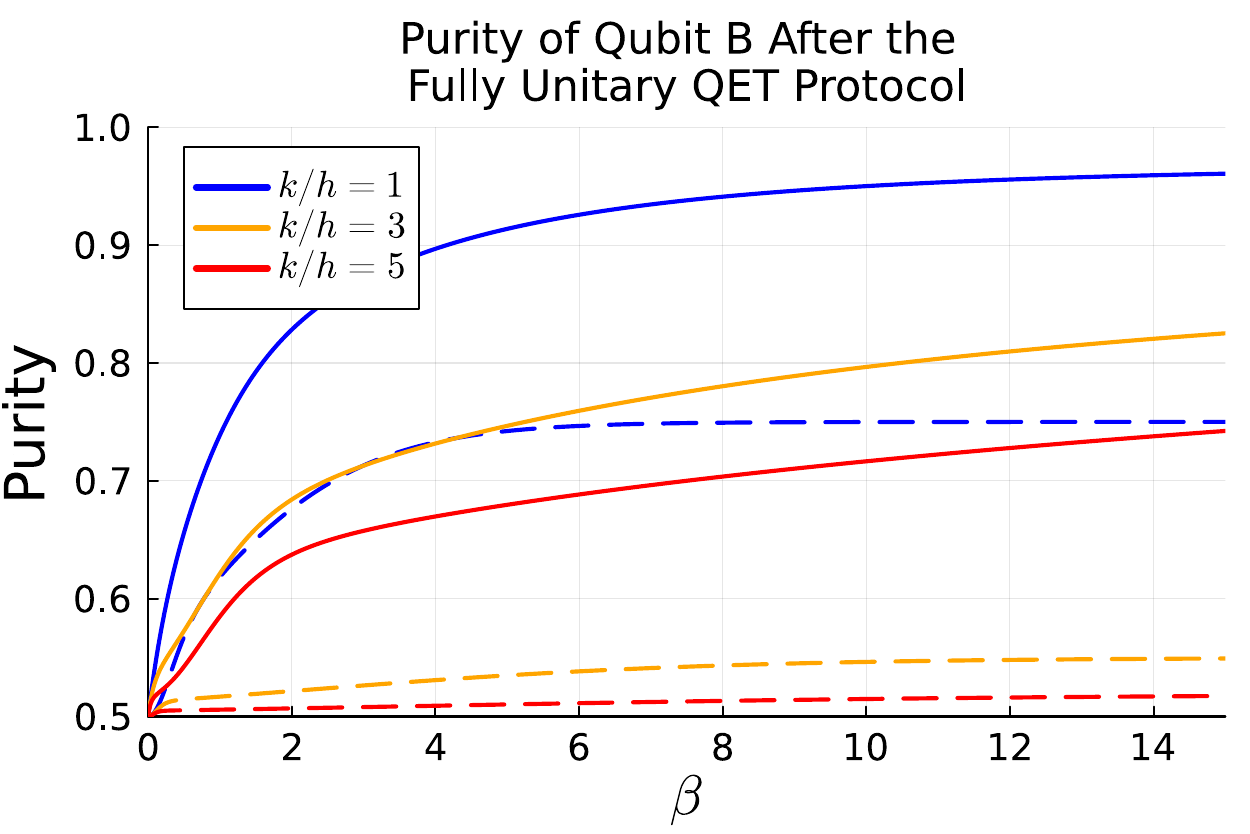}
    \caption{Final purity of qubit B as a function of the inverse temperature $\beta$ after the fully unitary QET protocol has been implemented with the analytic example presented. That is, $\hat U_\textsc{a} = e^{\ii\s_y^\textsc{a}\s_y^{\textsc{a}_n}}$, and $\hat U_\textsc{b} = e^{\ii\s_x^\textsc{b}\s_z^{\textsc{a}_n}}$. We show the plot for three different values of $k/h$. This figure is a reconstructed version of Fig. 2 in~\cite{QETAlgorithmicCooling}.}
    \label{fig:AnalyticPurity}
\end{figure}
 Fig.~\ref{fig:AnalyticPurity} shows the final purity of qubit B as a function of the inverse temperature $\beta$, for $k/h\in\{1,3,5\}$. Notice that this is simply a non-optimized example to illustrate that a fully unitary QET protocol can also purify subsystem B. In fact, one can check that the unitary in Eq.~\eqref{eq:ACAAnUnitary} does not maximize the mutual information between qubit A and the ancilla. Thus, (as disucssed in Subsection~\ref{sec:FullyUnitaryQET}) this example is not optimal neither for energy extraction nor for purification, and can be improved if we optimized the joint unitaries between each subsystem and the ancilla. 

The authors of~\cite{QETAlgorithmicCooling} do precisely this, and optimize the choice of Ancilla-subsystem unitaries to maximize purification of qubit B. The optimization can be performed numerically to find values of $\hat U_\textsc{a}$ and $\hat U_\textsc{b}$, where we define each as follows 
\begin{align}\label{eq:ACProbeHam}
    &\hat U _\textsc{a} = e^{\ii\hat H_\text{probe}^\textsc{a}}, && \hat U_\textsc{b} = e^{\ii \hat H_\text{probe}^\textsc{b}},\nonumber\\
    &H_\text{probe}^\textsc{a} = \sum_{i,j}\s_i^\textsc{a}\mathcal{J}^{ij}\s_j^{\textsc{a}_n}, && H_\text{probe}^\textsc{b} = \sum_{i,j}\s_i^\textsc{b}\mathcal{K}^{ij}\s_j^{\textsc{a}_n},
\end{align}
where $\mathcal{J}^{ij}$ and $\mathcal{K}^{ij}$ are both Hermitian matrices. 

With the numerically optimized local unitaries, the authors compare the results to a PPA-3 method applied to the exact same scenario. This is done to guarantee that they are comparing two methods that use the same number of qubits as an available resource.  We note, however, that PPA-3 fully utilizes all three qubits of the system for entropy compression, unlike the QET method which uses one of the three qubits as a measurement apparatus and it is not used to improve the performance of the QET protocol. 
\begin{figure}[h!]
    \centering
    \includegraphics[width=1\linewidth]{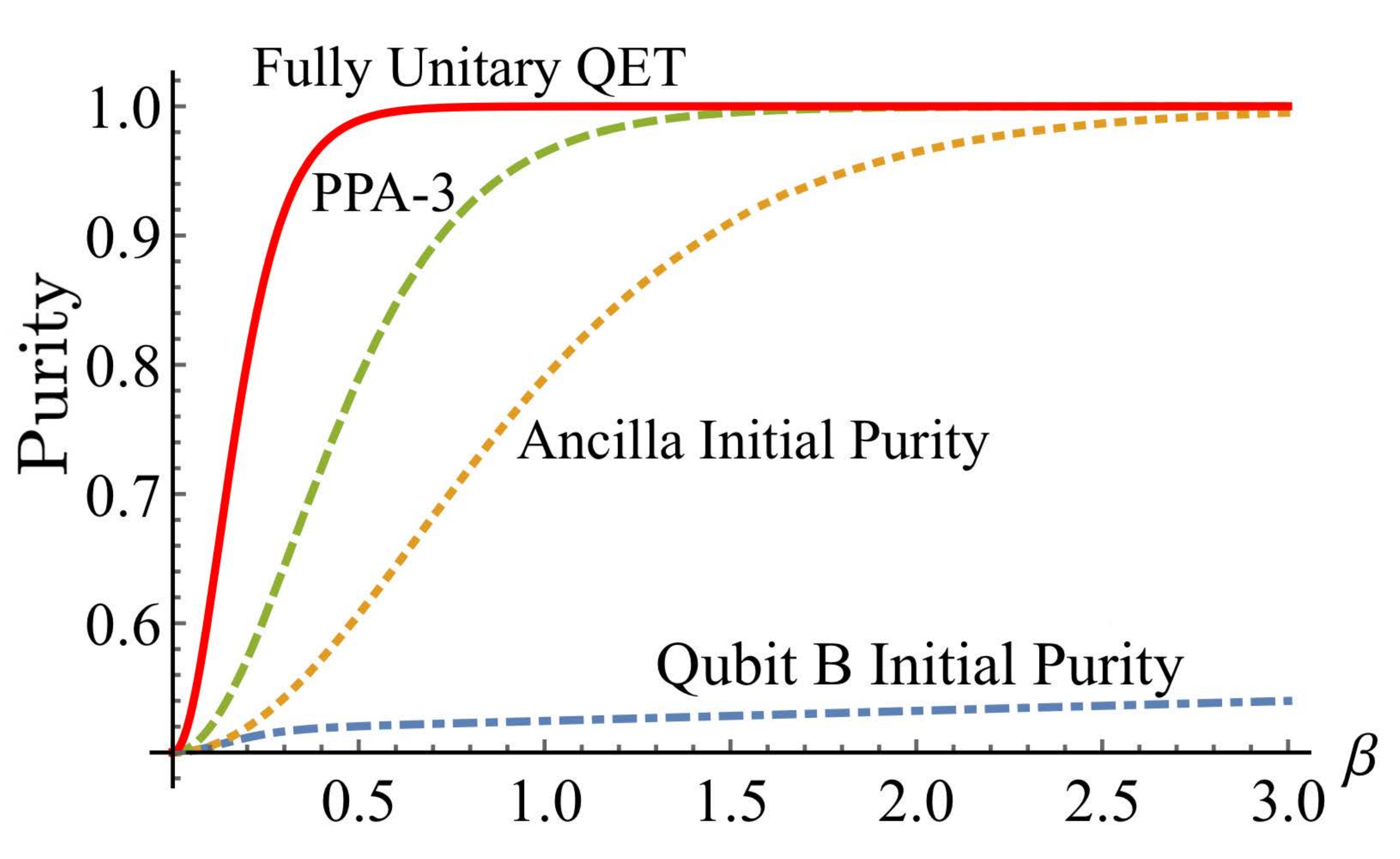}
    \caption{Comparison between the fully unitary QET-based algorithmic cooling method and PPA-3 as a function of $\beta$. In this plot, $k/h = 5$ and $h_\textsc{a}=h_\textsc{b}=h_\textsc{c} = h$. The initial purity of the ancilla qubit and qubit B are included to demonstrate the purity enhancement obtained by each method. This is a modified version of Fig. 3 in~\cite{QETAlgorithmicCooling}}
    \label{fig:ACFullyUnitary}
\end{figure}
In Fig.~\ref{fig:ACFullyUnitary}, the final purity of qubit B as a function of $\beta$ is shown for the fully unitary QET-based algorithmic cooling method and the PPA-3 algorithmic cooling method. Even though PPA-3 utilizes an additional qubit to cool the target qubit, we can see that fully unitary QET outperforms the PPA-3 method. When optimized it yields the same purity as in the LOCC case. 

The reason that QET with two qubits can outperform the PPA-3 method even though the latter fully utilizes the ability to fully manipulate the three qubits for entropy compression is due to the fact that the method does not use the system's correlations in the cooling process. Indeed, PPA-3  can outperform the QET method for weak couplings; however, for systems with strong couplings between the subsystems (and therefore higher correlations in the corresponding thermal states), QET is a more efficient cooling algorithm. This behaviour can be exploited for efficient cooling of strongly interacting systems, for example for ultra-strongly coupled superconducting qubits~\cite{Superconducting1,Superconducting2,Superconducting3}.

\section{Engineering negative energy densities}\label{sec:Engineering}

We have seen so far how QET has emerged as a compelling framework for understanding how information and local operations can be used to extract energy from correlated quantum systems without direct energy transfer. In the previous sections, we limited ourselves to QET protocols in discrete systems such as systems of $n$ qubits, revealing rich connections between entanglement, local operations, and energy flow in quantum many-body setups.

In this section, we are going to move to the continuum limit, focusing on a QET protocol in the setting of quantum field theory (QFT). Our goal is to explore how light-matter interaction models, where localized particle detectors couple to a quantum field—can be used not just to study energy transfer, but to engineer regions of negative stress-energy in a controlled, operational way.

This transition from discrete to field-theoretic systems opens the door to qualitatively new physics. In particular, while classical energy conditions like the weak energy condition (WEC) play a foundational role in general relativity—underpinning key results such as singularity theorems and prohibiting exotic geometries like traversable wormholes~\cite{HawkingGR,Alcubierre1994,Wormholes}—these conditions are known to be violated in quantum field theory. Effects like the Casimir force and squeezed states of light exhibit local negative energy densities~\cite{DynamicalCasimir1,SqueezedState}, albeit within strict quantum bounds known as quantum inequalities~\cite{Ford1978,PfenningsThesis}.

We will review how QET protocols, when implemented with local detector–field interactions, provide an operational route to generating negative energy densities that saturate these quantum bounds. This suggests that QET is not just a theoretical tool for understanding energy and information flow, but also a mechanism for constructing physically realizable stress-energy configurations with exotic properties.

We will focus in more detail on the case of relativistic scalar fields in both (1+1) dimensions, interacting with localized, inertial particle detectors—models closely related to atoms in quantum optics. We demonstrate how the spatial profile and magnitude of the resulting negative energy can be precisely shaped using detector design and interaction timing. We will also briefly comment on the case of (3+1) dimensions and show how QET can be used to achieve optimal violations of classical energy conditions, within the limits allowed by QFT. These results shed light on how quantum information protocols might be harnessed to shape the stress-energy landscape of quantum fields in regimes relevant to both quantum optics and semiclassical gravity and possibly create the conditions to observe gravitational effects of exotic matter. 

The content of this section is a review of the paper~\cite{nichoTeleport}, where all these studies are carried out in detail.

\subsection{The QET protocol for a scalar field in 1+1 dimensions}

We first begin by reviewing the QET protocol for a massless scalar field in a 1+1-dimensional Minkowski spacetime. We choose our metric signature such that $\eta_{\mu\nu} = \text{diag}(-1,1)$ and adopt inertial frame coordinates $\mf x = (t,x)$. We will assume that Alice and Bob each have access to a particle detector that is coupled to the field, each of which can be turned on and off in a controlled manner through a switching function $\chi(t)$. Additionally, each of the detectors, A and B, have some spatial localization governed by the smearing functions $\lambda(x)$ and $\mu(x)$, respectively. 

Alice and Bob's particle detectors are modeled as Unruh-DeWitt (UDW) detectors~\cite{Unruh1976,DeWitt}. The UDW model has been proven to be a simple yet effective way to capture the most relevant features of the light-matter interaction between atomic probes and the electromagnetic field~\cite{richard,Pozas2016,DelReyMonteroEdu}.

Let us consider that Alice's detector is  a two-level quantum system that couples to a scalar field $\hat\phi(\mathsf{x})$ through the following Unruh-DeWitt (UDW) interaction Hamiltonian
\begin{equation}\label{eq:EngAliceCouple}
    \hat H_\textsc{i} = \delta(t-t_0)\,\s_x^\textsc{a}\int \dd  x\,\lambda(x)\hat \pi(\mf x),
\end{equation}
where we have particularized the switching function, $\chi(t) = \delta(t-t_0)$ which couples the detector to the field for an instant at an arbitrary time $t_0$, and $\hat \pi(\mf x) = \partial_t\phihat(\mf x)$ is the field amplitude's conjugate momentum. Moreover, we define $\ket{\pm}$ as the eigenstates of $\s_x$, so that $\s_x\ket{\pm} = \pm\ket{\pm}$. After this interaction, Alice's detector is sent to Bob (using an LOQC scheme analogously to that presented in Subsection~\ref{sec:FullyUnitaryQET}). From this Bob's detector then interacts through a similar UDW interaction at some later time $T>t_0$ given by 
\begin{equation}\label{eq:EngBobCouple}
   \hat H_\textsc{i} = \delta(t-T-t_0)\s_z^\textsc{b}\int \dd x \,\mu(x)\hat \phi(\mf x).
\end{equation}
We will define $\ket{e}$ and $\ket{g}$ as the eigenstates of $\s_z$ such that $\s_z\ket{g} = -\ket{g}$ and $\s_z\ket{e} = \ket{e}$. In this protocol, it is important to choose $\lambda(x)$, $\mu(x)$, and $T$ in such a way that Bob only gains information about Alice's detector after he has entered her lightcone. We also choose the smearing functions to have compact support as this ensures the detectors have no spatial overlap. Lastly, the use of $\delta$-switching allows for non-perturbative calculations using the coherent state basis.  The detector observables that Alice and Bob each couple to the field in Eqs.~\eqref{eq:EngAliceCouple} and \eqref{eq:EngBobCouple} correspond to non-commuting generators of SU(2), $(\s_x,\s_z)$. This is necessary for the QET protocol to work as intended. Regarding the field observables each detector couples to, the reason why Alice couples to the field momentum and Bob to the amplitude is taken for illustration purposes and is meant to simplify some of the math in the setup~\cite{DerivativeQET}. Similar results can be achieved with the detectors coupling to the same field observable for both Alice and Bob.

Once again for mathematical convenience, for the (1+1) dimensional case we will assume that there is only one detector that Alice and Bob share. This can be thought of as Alice transmitting the state of her detector through quantum communication (for example a regular quantum teleportation protocol~\cite{DerivativeQET,GuillaumeThesis}) where the quantum state of Alice's detector is sent to Bob's. This simplifies the notation since there is only one detector that Alice passes on to Bob, and reduces the computational burden. We note that this scenario is equivalent in negative energy yield to the scenario where Alice measures her detector and sends classical information to Bob who uses that information to couple his detector (as discussed in Appendix A of \cite{nichoTeleport}). We will revisit this discussion when we analyze the (3+1) dimensional case where this becomes more relevant.

In determining the effects of the QET protocol, we focus on the $\hat T_{00}(\mf x)$ component of the normal ordered stress-energy tensor, $: \hat T_{\mu\nu}(\mf x) :$, representing the energy density of the field. Specifically, we have that the stress-energy tensor of the field is given by
\begin{equation}
    \hat T_{\mu\nu}(\mf x) = \partial_\mu\hat\phi(\mf x)\partial_\nu\hat\phi(\mf x) - \frac{1}{2}\eta_{\mu\nu}(\partial_\rho\hat \phi(\mf x)\partial^\rho\hat\phi(\mf x)).
\end{equation}

Without loss of generality, we will assume that $t_0 = 0$, and the initial state of the system is given by 
\begin{equation}
    \ket{\psi(0)} = \ket{0}\otimes\ket{A_0},
\end{equation}
where $\ket{0}$ is the field's Minkowski vaccuum and $\ket{A_0}$ is an arbitrary pure state of the detector shared between Alice and Bob. The (interaction picture) time evolution of the system after Alice couples the detector to the field yields the following
\begin{align}\label{eq:EngAliceTime}
    \ket{\psi(t)} &= e^{-\ii \s_x^\textsc{a}\int\dd x\,\lambda(x)\hat \pi(\mf x)}\ket{0}\otimes\ket{A_0}\nonumber\\
    &=\ket{\bm \alpha(t)}\ket{+}\braket{+}{A_0} + \ket{-\bm \alpha(t)}\ket{-}\braket{-}{A_0},
\end{align}
where we have defined 
\begin{align}
    \alpha_k(t) &= e^{-\ii |k|t}\sqrt{\frac{|k|}{4\pi}}\int \dd x\, \lambda(x)e^{\ii kx},\\
    \ket{\bm \alpha(t)} &= \bigotimes_{k\in\mathbb{R}}\ket{\alpha_k(t)},
\end{align}
where $\ket{\bm \alpha(t)}$ is a tensor product of  coherent states. We can evaluate the renormalized stress-energy density after Alice's detector coupling and find that the expected value of the energy density of the state in Eq.~\eqref{eq:EngAliceTime} is given by 
\begin{equation}\label{eq:EngAliceEnergy}
    \bra{\psi(t)}: \hat T_{00}(\mf x) :\ket{\psi(t)} = \frac{1}{4}(\lambda'(x-t))^2 + \frac{1}{4}(\lambda'(x+t))^2.
\end{equation}
where 
\begin{equation}\bra{\psi(t)}: \hat T_{00}(\mf x) :\ket{\psi(t)} \coloneqq \bra{\psi(t)} \hat T_{00}(\mf x)\ket{\psi(t)} -\bra{0} \hat T_{00}(\mf x) \ket{0}.
\end{equation}
Since the expectation of the renormalized energy density in the vacuum state is always zero, we can see that the result of Alice's interaction with the field is an injection of energy that propagates away from Alice at the speed of light. We can see that when $t=0$ the energy density is only non-zero on the support of the detector's smearing function due to the local interaction of the detector. This means that, after her interaction, the energy in the field in Bob's region is the same as in the vacuum state. 

Following the same procedure, after Bob's interaction at a time $t>T$, we find that the system evolves to the state 
\begin{align}\label{eq:EngBobsState}
    &\ket{\psi(t)} = \nonumber\\
    &\hat D(\bm \beta(t))\left(\frac{\braket{+}{A_0}}{\sqrt{2}}\ket{\bm \alpha(t)}\ket{e} + \frac{\braket{-}{A_0}}{\sqrt{2}}\ket{-\bm \alpha(t)}\ket{e}\right)\nonumber\\
    +&\hat D(-\bm \beta(t))\left(\frac{\braket{+}{A_0}}{\sqrt{2}}\ket{\bm \alpha(t)}\ket{g} - \frac{\braket{-}{A_0}}{\sqrt{2}}\ket{-\bm \alpha(t)}\ket{g}\right),
\end{align}
where $\hat D(\bm \beta) \coloneqq \text{exp}\big(\int\dd k\, \beta_k(t)\hat a_k^\dagger - \beta^*_k(t)\hat a_k\big)$ is a displacement operator, and
\begin{equation}
    \beta_k(t) = -\frac{\ii e^{-\ii |k|(t-T)}}{\sqrt{4\pi|k|}}\int\dd x\,\mu(x)e^{\ii kx}.
\end{equation}
We can now compute the expectation of the energy density in the state of Eq.~\eqref{eq:EngBobsState}. The energy density after the completion of the protocol then becomes
\begin{widetext}
    \begin{align}
        \begin{split}
            &\left\langle :\hat T_{00}(x,t): \right\rangle = \frac{\left(\lambda'(x-t)\right)^2}{4} + \frac{\left(\lambda'(x+t)\right)^2}{4} + \frac{\left(\mu(x-(t-T))\right)^2}{4} + \frac{\left(\mu(x+(t-T))\right)^2}{4} \\
            &+ \underbrace{\frac{e^{-2\norm{\alpha}}\bra{A_0}\s_y\ket{A_0}}{2\pi} \mu(x-(t-T)) \int \dd y\, \lambda'(y) \frac{\text{P.P}}{y-x+t}}_{\text{Right moving QET term}} + \underbrace{\frac{e^{-2\norm{\alpha}}\bra{A_0}\s_y\ket{A_0}}{2\pi} \mu(x+(t-T)) \int \dd y\, \lambda'(y) \frac{\text{P.P}}{y-x-t}}_{\text{Left moving QET term}} .
        \end{split}\label{eq:Eng11T00}
    \end{align}
\end{widetext}
Here we have that $\norm{\alpha} = \int\dd k\,|\alpha_k|^2$, and P.P denotes Cauchy's principle value prescription. In this equation, we can see the two strictly positive energy density contributions: one that comes from Alice's coupling of the detector  \mbox{$\frac{1}{4}(\lambda'(x\pm t))^2$} and one from Bob's coupling of the detector  \mbox{$\frac{1}{4}(\mu(x\pm(t-T)))^2$}. The positive energy packets for each detector split into a left-propagating and right-propagating terms. 

The terms we dubbed QET terms depend on the spatial smearing of both detectors. These terms can have a positive or negative sign depending on the state of the detector after the protocol  (concretely on the expectation of $\hat\sigma_y$) and are initially localized at the position of Bob. Since these terms are linear in $\lambda'(x)$, we can find smearing functions that allow these QET terms to be negative enough that they overcome the positive contributions. 

While we are primarily focused on generating local negative energy densities, it is worth mentioning that the choice of initial state of the detector allows for freedom in designing the effect of the QET terms. One could instead use this QET protocol to inject energy into the field. For example, we could  choose $\ket{A_0}$ so that the effect of the QET terms is to increase the energy density to a greater amount than the sum of Alice and Bob's individual contributions. 

With the objective in mind to create as much energy density in the field, we need to find smearing functions so that we minimize the positive energy contributions from Alice and Bob in the first terms of~\eqref{eq:Eng11T00} while maximizing the magnitude of the QET terms since the amount of negative energy density comes from a competition between both terms.

The analysis becomes more clear by doing some work on the QET terms. In particular, the principal value integral can be written as
\begin{equation}
    \int\dd y\, \lambda'(y)\frac{\text{P.P}}{y-x+t} = \lim_{\epsilon\rightarrow0}\left[\int\limits_{-\infty}^{x-t-\epsilon}+\int\limits_{x-t+\epsilon}^{\infty}\right]\dd y\frac{\lambda'(y)}{y-x+t},
\end{equation}
where we use the notation
\begin{equation}
    \left[\int\limits_{a}^b+\int\limits_{c}^d\right]\dd x\,f(x) \coloneqq \int\limits_{a}^{b}\dd x\, f(x) + \int\limits_{c}^{d}\dd x\, f(x).
\end{equation}
We can subdivide the domain of integration into a part that does not have any singular values and a part that does as follows. For some $a>0$, we have
\begin{align}\label{eq:EngSubdivision}
    \int\dd y\, \lambda'(y)\frac{\text{P.P}}{y-x+t} = \Bigg[&\int\limits_{-\infty}^{x-t-a}+\int\limits_{x-t+a}^{\infty}\Bigg]\dd y\frac{\lambda'(y)}{y-x+t}\nonumber\\
    +\lim_{\epsilon\rightarrow0}\Bigg[\,&\int\limits_{x-t-a}^{x-t-\epsilon}+\int\limits_{x-t+\epsilon}^{x-t+a}\,\Bigg]\dd y\frac{\lambda'(y)}{y-x+t}\nonumber\\
\end{align}
With some analysis (see~\cite{nichoTeleport}) we are able to numerically evaluate the principal value integral, that approximately becomes \begin{align}\label{eq:EngIntuition}
    \int\dd y\, \lambda'(y)\frac{\text{P.P}}{y-x+t} &= \Bigg[\int\limits_{-\infty}^{x-t-a}+\int\limits_{x-t+a}^{\infty}\Bigg]\dd y\frac{\lambda'(y)}{y-x+t}\nonumber\\
    & + 2a\lambda''(x-t) + \frac{a^3}{9}\mathcal{O}(\lambda^{(4)}(\xi)),
\end{align}
where $\mathcal{O}(\lambda^{(4)}(\xi))$ is notation for terms that go with the fourth and higher derivatives of the smearing function. Since $|\lambda^{(4)}(\xi)|$, where $\xi\in[x-t,y]$, is bounded on our domain, this justifies the approximation if  $a$ is small enough. How small is small enough for a numerical evaluation will depend on the shape of the smearing function that is chosen. The authors of~\cite{nichoTeleport} only used this approximation for numerical integration purposes and also made sure numerically that any corrections to the approximation provide negligible contributions with respect to the contribution proportional to  $a\lambda''$. 

With the objective in mind to create as much energy density in the field, in order to rig the competition between the positive terms and the QET terms in Eq.~\eqref{eq:Eng11T00} in favour of the latter, we need to find smearing functions whose first derivatives are small, and second derivatives are large. This is not feasible over the entire support of the smearing functions. However---since for physical detectors the smearing functions are strongly localized---it is enough that the second derivative dominates the first one around the functions' maxima. A consequence of not being able to do this over the whole domain is that any amount of negative energy density that we generate like this is necessarily going to be paired with regions of positive energy density in the surrounding area. Moreover, the more negative energy we have in a particular region the thinner its region will have to be. These limitations on the amount of negative energy allowed in a particular region are linked to the quantum energy conditions~\cite{Ford1978,PfenningsThesis} and the quantum interest conjecture~\cite{InterestConjecture}, and will be discussed in more detail in a later section. Finally, we comment on the fact that, while we have presented the QET protocol in $(1+1)$-dimensional Minkowski spacetime, the same prescription can be generalized to $(3+1)$-D. While this generalization is straightforward in some ways, there are some extra subtleties. For example, the need to consider more than one detector if one wants to create a physically meaningful scenario. We will comment on this in a later subsection. The calculations for the $(3+1)$-dimensional case are pretty analogous to the $(1+1)$-dimensional case, albeit significantly more mathematically involved, and can be found in full detail in~\cite{nichoTeleport}. 

\subsection{Generating negative energy densities in 1+1 dimensions}\label{sec:NegativeGeneration}

We recall that, in the QET protocol, there are three terms that contribute to the final energy density of the field:  Alice's local energy density contribution proportional to $\lambda'^2$, Bob's local contribution proportional to $\mu^2$, and the QET terms, which as per equation~\eqref{eq:EngIntuition}, are dominated by the product $\mu\lambda''$. The local terms are positive, thus the  only term that can create the negative energy density is the QET term. In order to create regions with negative energy density we should choose smearing functions $\lambda$ and $\mu$ such that there are regions in spacetime where the support of $\mu$ coincides with the region where $\lambda'$ is small as compared to $\lambda''$.

For an illustrative example, in~\cite{nichoTeleport} they choose three different sets smearing functions. They include a smooth compactly supported function, a Gaussian and a Lorentzian case. In particular they define
\begin{gather}
    f(z,\sigma,\delta) = 
        \begin{cases}
            S(\frac{\sigma/2 + \pi\delta + z}{\delta}) & -\pi\delta<z+\sigma/2<0, \\
            1 & -\sigma/2\leq z\leq \sigma/2,\\
            S(\frac{\sigma/2 + \pi\delta - z}{\delta}) & 0<z-\sigma/2<\pi\delta,\\
            0 &\text{otherwise},
        \end{cases}\\
    g(z,\delta) = \frac{1}{\sqrt{2\pi}}e^{-\frac{z^2}{2\delta^2}},\\
    h(z,\delta) = \frac{1}{\pi} \frac{1}{1+(\frac{z}{\delta})^2},
\end{gather}
where $S(x) = \frac{1}{2}(1-\tanh(\cot(x)))$ and $z = x-x_0$. Then, the spatial smearing of the detectors in the three cases will be given respectively by 
\begin{align}
   \lambda(x) = \lambda_0f(x,\sigma_\textsc{a},\delta_\textsc{a}), & \;\mu(x) = \mu_0f(x-x_\textsc{b},\sigma_\textsc{b},\delta_\textsc{b}),\label{eq:EngLambda1}\\
    \lambda(x) = \lambda_0g(x,\delta_\textsc{a}), &\; \mu(x) = \mu_0g(x-x_\textsc{b},\delta_\textsc{b}),\label{eq:EngLambda2}\\
    \lambda(x) = \lambda_0h(x,\delta_\textsc{a}), & \;\mu(x) = \mu_0 h(x-x_\textsc{b},\delta_\textsc{b})\label{eq:EngLambda3},
\end{align}
where $\lambda_0$ has units of $\text{[Length]}^{-(n-1)/2}$ and $\mu_0$ has units of $\text{[Length]}^{-n/2}$, where $n$ is the number of spacetime dimensions.
In each of the cases~\eqref{eq:EngLambda1}, \eqref{eq:EngLambda2}, and \eqref{eq:EngLambda3}, the parameters $x_\textsc{b}$, $\sigma_{\textsc{a},\textsc{b}}$, and $\delta_{\textsc{a},\textsc{b}}$ were optimized for the maximum negative energy generation.


Once the optimal parameters have been found, it is possible to time evolve the energy density through Eq.~\eqref{eq:Eng11T00}.
\begin{figure}[h!]
    \centering
    \includegraphics[width=1\linewidth]{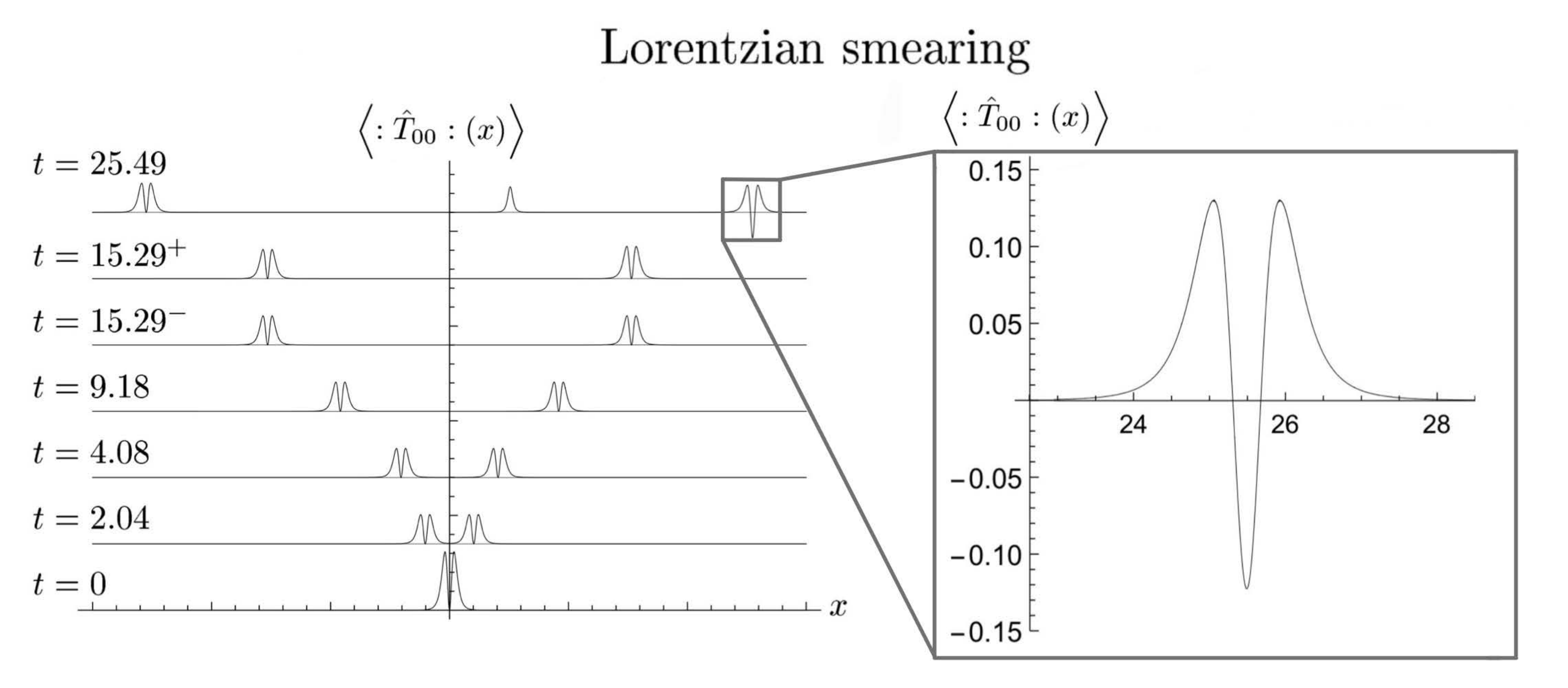}
    \caption{(Left) This figure shows the energy density at different time slices when the smearing function for the detector is given by the Lorentzian in Eq.~\eqref{eq:EngLambda1}. Here we include the notation $t=15.29^\pm$ time units denoting the energy density before and after Bob performs his unitary on the qubit. We see that as his positive contribution moves away, the result is a local negative energy density. (Right) This figure is a zoomed in plot of the negative energy density that is generated once Bob's positive energy contribution has had time to move away, leaving the desired negative energy density. This is a modified version of Figs. 1 and 2 in\cite{nichoTeleport}.}
    \label{fig:LorentzianQET}
\end{figure}
For illustration, we plot here the case of the Lorentzian smearing. In Fig.~\ref{fig:LorentzianQET} we show the time evolution of the negative energy density at different time slices. Initially, we see how the energy deposited by Alice's switching propagates. Two plots are given at both sides of $t=T = 15.29$ time units to represent the energy right before and right after Bob's interaction with the field. Bob's interaction creates a negative energy density region within the support of his smearing function that is revealed as the left moving positive energy density wavepacket from Bob's interaction propagates away from the negative energy region. 

From this figure, we see what we learned from Eq.~\eqref{eq:EngIntuition}. Namely, that the negative energy region will be deepest where $\lambda''$ is the largest. This is better demonstrated by the right plot in Fig.~\ref{fig:LorentzianQET} where we see the negative energy that is left behind once the left-moving half of Bob's local contribution to the energy density has had time to propagate away from the right-propagating region containing the negative energy density. In~\cite{nichoTeleport}, the authors include plots for all three smearing functions presented and it is shown that the Lorentzian smearing function creates the deepest negative energy well, which is inline with this smearing function having the largest second derivative with respect to its length scale, $\delta$. In these scenarios, the negative energy density region will always be within two positive energy peaks when the parameters of the protocol have been optimized. While it is possible to select parameters to have the negative energy region located behind (or, in the case of a massive field, in front) of the positive peaks, this will come with the trade off that a much smaller negative energy density region will be obtained. 

\subsection{A summary of QET in 3+1 dimensions}

In this section, we will provide a brief overview of how the QET protocol can be extended to the case of a (3+1)-dimensional Minkowski spacetime. In general, much of the protocol remains the same; however, there is a key difference in the (3+1)-case regarding how information is transmitted from Alice to Bob. 

In~\cite{nichoTeleport} the (3+1)-dimensional setup, an LOQC QET protocol is used with a ``non-local" detector for Bob. That is, a single (hollow) spherical detector that is delocalized in a large region surrounding Alice's detector. This is computationally convenient but arguably physically unsound since it does not make sense to ``delocalize" Bob to such extent as an agent in the QET protocol.

Realistically, if one wanted to utilize the QET in (3+1) in a more physically meaningful way, one would have many spatially distributed Bob-detectors that mimic Bob's smearing function, $\mu(\bm x)$ and Alice will communicate classically with each of them revealing the outcome of her measurement. The fact that there may be an equivalence between these two scenarios should  not be very difficult to accept since we showed in Sec.~\ref{sec:FullyUnitaryQET} that a full LOQC QET protocol is equivalent to a QET protocol that relies on LOCC. In Appendix A of~\cite{nichoTeleport} it is proven that the two detector scenario with quantum communication and a scenario where Bob is substituted by a cloud of many small detectors distributed along the smearing function $\mu(\bm x)$ and classical communication are equivalent, provided the right choice of initial detector state. 

For the 3+1 dimensional case all computations are performed in the same fashion by first having Alice couple the detector to the field, communicating the results of this to Bob, and Bob then interacting with the field at a later time. One then finds that the stress-energy density of the field is given by
\begin{widetext}
\begin{align}
\begin{split}
&\bra{\psi(T+\Delta T)}:\hat{T}_{\mu\nu}:(\bm{x})
\ket{\psi(T+\Delta T)}=\frac{1}{4^4\pi^6}\Bigg[
\underbrace{\left(I_{\mu}^{1}I_{\nu}^{1}-\eta_{\mu\nu}\frac{I_{\lambda}^{1}I^{1,\lambda}}{2}\right)}_{\text{Bob's positive contribution}}
-\underbrace{\left(I_{\mu}^{2}I_{\nu}^{2}-\eta_{\mu\nu}\frac{I_{\lambda}^{2}I^{2,\lambda}}{2}\right)}_{\text{Alice's positive contribution}}\\
-&e^{-2\|\alpha\|}\bra{A_{0}}\hat{\sigma}_{y}\ket{A_{0}}\underbrace{\left(\left(I_{\mu}^{1}I_{\nu}^{3}-\eta_{\mu\nu}\frac{I_{\lambda}^{1}I^{3,\lambda}}{2}\right)+\left(I_{\mu}^{3}I_{\nu}^{1}-\eta_{\mu\nu}\frac{I_{\lambda}^{1}I^{3,\lambda}}{2}\right)\right)}_{\text{QET negative contribution}}\Bigg],
\end{split}\label{eq:31T00}
\end{align}
where
\begin{align}
I^{1}_{\mu}=&\int \dd ^{3}\bm{r} \dd^{3}\bm{k}\,\tilde{k}_{\mu}
\bigg(
e^{\left|\bm{k}\right|\left(-2\varepsilon+\ii\Delta T\right) +\ii\bm{k}\cdot\left(\bm{r}-\bm{x}\right)}
+e^{\left|\bm{k}\right|\left(-2\varepsilon-\ii\Delta T\right) -\ii\bm{k}\cdot\left(\bm{r}-\bm{x}\right)}\bigg)\mu\left(\bm{r}\right),\\
I^{2}_{\mu}=&\int \dd ^{3}\bm{r} \dd^{3}\bm{k}\,\tilde{k}_{\mu}
\bigg(
e^{\left|\bm{k}\right|\left(-2\varepsilon-\ii(\Delta T+T)\right) -\ii\bm{k}\cdot\left(\bm{r}-\bm{x}\right)}-e^{\left|\bm{k}\right|\left(-2\varepsilon+\ii(\Delta T+T)\right) +\ii\bm{k}\cdot\left(\bm{r}-\bm{x}\right)}\bigg)\left|\bm{k}\right|\lambda\left(\bm{r}\right),\\
I^{3}_{\mu}=&\int \dd ^{3}\bm{r} \dd^{3}\bm{k}\,\tilde{k}_{\mu}
\bigg(
e^{\left|\bm{k}\right|\left(-2\varepsilon-\ii(\Delta T+T)\right) -\ii\bm{k}\cdot\left(\bm{r}-\bm{x}\right)}+e^{\left|\bm{k}\right|\left(-2\varepsilon+\ii(\Delta T+T)\right) +\ii\bm{k}\cdot\left(\bm{r}-\bm{x}\right)}\bigg)\left|\bm{k}\right|\lambda\left(\bm{r}\right),
\end{align}
\end{widetext}
$\norm{\alpha} = \int\dd^3\bm k\, |\alpha_{\bm k}|^2$, $\Delta T>T$ is some time after Bob's coupling, $\epsilon$ is a UV soft cutoff regulator that is eventually set to zero, and $\tilde{k}_\mu \equiv k_\mu/|\bm k|$. Here we are again assuming Alice couples at $t_0=0$ and Bob couples at a later time $t=T$.

The authors of~\cite{nichoTeleport} again use three smearing functions (spherically symmetric versions of the ones used in the 1+1 dimensional case) and show that in all cases, with optimal parameters, it is possible to generate local negative energy density through the QET protocol. 

\begin{figure}[h!]
    \centering
    \includegraphics[width=1\linewidth]{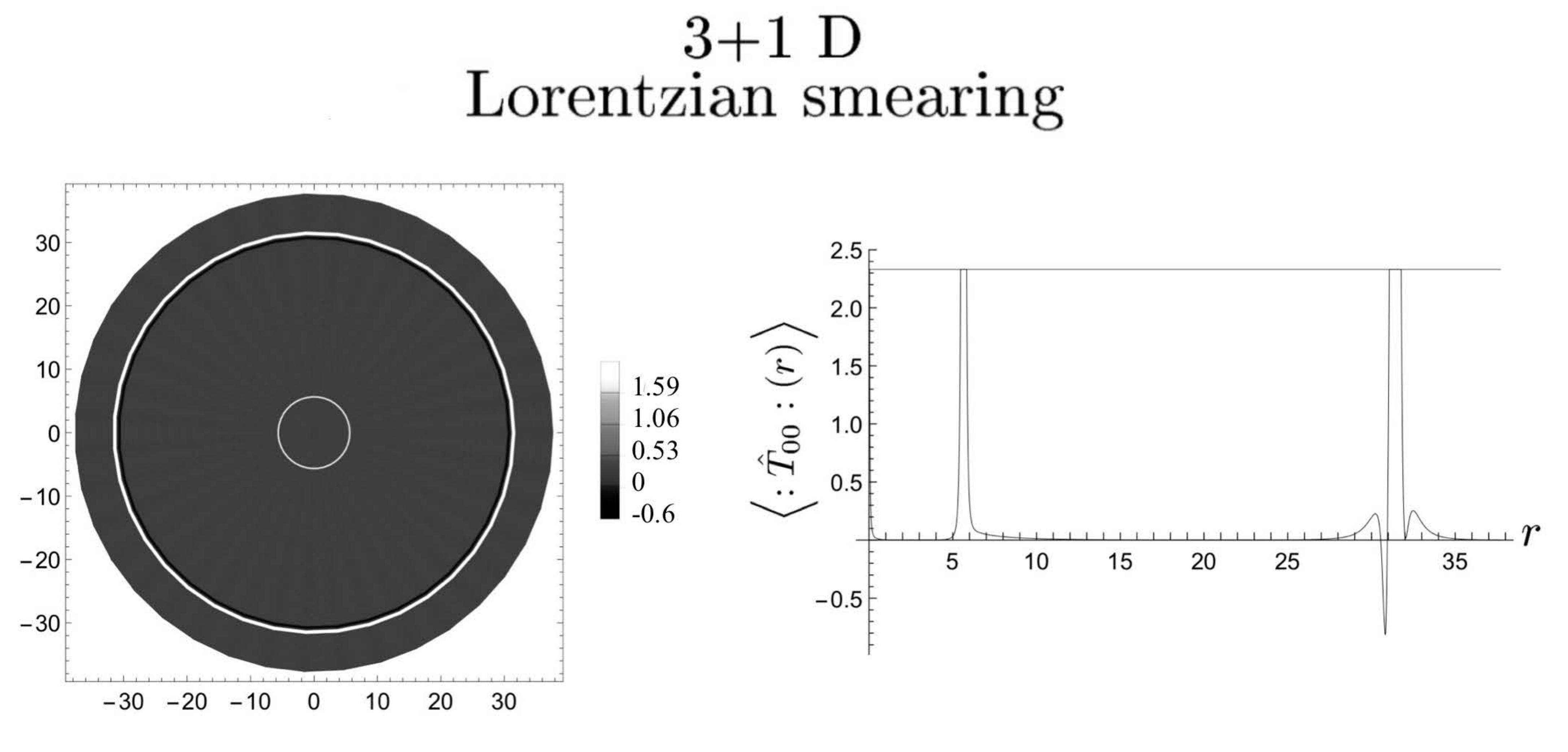}
    \caption{(Left) Contour plots showing the energy density generated through the QET protocol using the (3+1)-dimensional Lorentzian smearing function. (Right) Radial component of the energy density. This is a modified version of Fig. 5 in~\cite{nichoTeleport}.}
    \label{fig:3DLorentzianQET}
\end{figure}

In Fig.~\ref{fig:3DLorentzianQET} we show the contour plots and a radial cross-section of the 3D Lorentzian smearing function case. In comparison to the (1+1) QET protocol, the (3+1) QET protocol generates more total negative energy for negative energy wells of comparable widths. Additionally, in (3+1), the wavepackets (all positive and negative energy generated through the protocol) slowly decay over time due to dispersion governed by the inverse square law that is not present in the (1+1)-dimensional Minkowski spacetime. We also see that, since Bob's smearing is spherically symmetric, that the negative energy density is generated as a shell around Alice's position.

\subsection{Negative energy density scaling}


One key question left to address is how much negative energy we can generate with this protocol if we optimize it, and how it scales when compared to fundamental limits for violations of energy conditions such as the average weak energy condition quantum inequalities~\cite{Ford1978,PfenningsThesis} and the quantum interest conjecture~\cite{InterestConjecture}. These fundamental limitations hint towards the fact that one can increase the amount of negative energy by making the regions where it is concentrated smaller and smaller. 

In order to study how the amount of negative energy scales as we scale the protocol to concentrate the negative energy in smaller regions, let us consider that both Alice and Bob have arbitrary, compact smearing functions and coupling strengths. We will present the derivation in an $n$-dimensional spacetime, which is simply a generalization of the protocol that has already been presented in the 3+1 dimensional case and can be found in Appendix A of~\cite{nichoTeleport}.

If we analyze Eq.~\eqref{eq:31T00}, we see that, in order to have a positive scaling for the amount of negative energy, we should ensure that $\norm{\alpha}$ is kept constant. This is due to the fact that $\norm{\alpha}$ would otherwise cause an exponential decay in the QET terms that are solely responsible for the generation of negative energy.  With this in mind, we study how $\norm{\alpha}$ changes as we scale the width of Alice's smearing function by some factor $\Upsilon^{-1}$. The result of such a scaling is given by
\begin{align}\label{eq:EngScaleSubst}
    \norm{\alpha}_{\Upsilon} = \frac{1}{2(2\pi)^{n-1}}&\int\dd^{n-1}\bm x\int\dd^{n-1}\bm y\int\dd^{n-1}\bm k\nonumber\\
    &\!\!\!\!\!\!\!\!\!\!\!\!\times \lambda(\Upsilon\bm x)\lambda(\Upsilon\bm y)|\bm k|e^{-2\epsilon|\bm k| - \ii \bm k\cdot(\bm x-\bm y)}.
\end{align}
If one performs a change of variable given by $\tilde{\bm x} = \Upsilon\bm x$, $\tilde{\bm y} = \Upsilon\bm y$, and $\Upsilon\tilde{\bm k} = \bm k$, Eq.~\eqref{eq:EngScaleSubst} becomes

\begin{align}
    \norm{\alpha}_\Upsilon = \frac{1}{2(2\pi)^{n-1}}&\int\frac{\dd^{n-1}\tilde{\bm x}}{\Upsilon^{n-1}}\int\frac{\dd^{n-1}\tilde{\bm y}}{\Upsilon^{n-1}}\int\dd^{n-1}\tilde{\bm k}\Upsilon^{n-1}\nonumber\\
    &\!\!\!\!\!\!\!\!\!\!\!\!\times \lambda(\tilde{\bm x})\lambda(\tilde{\bm y})|\Upsilon\tilde{\bm k}|e^{-2\epsilon|\Upsilon\bm k| - \ii \tilde{\bm k}\cdot(\tilde{\bm x}-\tilde{\bm y})}\nonumber\\
    &=\frac{1}{\Upsilon^{n-2}}\norm{\alpha}_{\Upsilon = 1},
\end{align}
where $\epsilon$ is a UV cutoff that is eventually set to zero. This means, that if we scale Alice's smearing function by a factor of $\Upsilon^{-1}$ we must scale the coupling strength by a factor of $\Upsilon^{\frac{n-2}{2}}$ in order for $\norm{\alpha}$ to remain constant. Specifically, $\lambda(\bm x)\rightarrow \Upsilon^{\frac{n-2}{2}}\lambda(\Upsilon\tilde{\bm x})$ leaves $\norm{\alpha}$ invariant as desired. 

We now consider rescaling Bob's smearing function, again by a factor of $\Upsilon^{-1}$, and its coupling strength by a factor of $\Upsilon^\xi$. Specifically, $\mu(\bm x) \rightarrow \Upsilon^\xi\mu(\Upsilon\bm x)$. Overall, this rescaling only affects the $I^1_\mu$ terms in the energy density and results in the following 
\begin{equation}
    \left(I^1_\mu(\Upsilon\bm x,\Upsilon\Delta T\right)_\Upsilon = \Upsilon^{2\xi}\left(I^1_\mu(\bm x,\Delta T)\right)_{\Upsilon=1}.
\end{equation}
Here we are also rescaling the time interval by a factor of $\Upsilon$ as well (i.e $t\rightarrow\Upsilon t$) in order to ensure that the width of Bob's smearing function scales as $\Upsilon^{-1}$. The fact that we are rescaling
the time interval between Alice and Bob's interaction in the same way as their spatial support imposes that Bob's
smearing function width also scales by $\Upsilon^{-1}$ due to the optimization constraints.

In turn, we find that the rescaled stress-energy density becomes 
\begin{widetext}
\begin{align}
\begin{split}
\bra{\psi(\Upsilon t)}:\hat{T}_{\mu\nu}:(\Upsilon\bm{x})
\ket{\psi(\Upsilon t)}_{\Upsilon}&=\Bigg[\Upsilon^{2 \xi}
\underbrace{\left(I_{\mu}^{1}I_{\nu}^{1}-\eta_{\mu\nu}\frac{I_{\lambda}^{1}I^{1,\lambda}}{2}\right)}_{\text{Bob's positive contribution}}
-\Upsilon^{n}\underbrace{\left(I_{\mu}^{2}I_{\nu}^{2}-\eta_{\mu\nu}\frac{I_{\lambda}^{2}I^{2,\lambda}}{2}\right)}_{\text{Alice's positive contribution}}\\
-&\Upsilon^{\frac{n}{2}+\xi}\braket{A_{0}|\hat{\sigma}_{y}|A_{0}}e^{-2\|\alpha\|}\underbrace{\left(\left(I_{\mu}^{1}I_{\nu}^{3}-\eta_{\mu\nu}\frac{I_{\lambda}^{1}I^{3,\lambda}}{2}\right)+\left(I_{\mu}^{3}I_{\nu}^{1}-\eta_{\mu\nu}\frac{I_{\lambda}^{1}I^{3,\lambda}}{2}\right)\right)}_{\text{QET negative contribution}}\Bigg].
\end{split}\label{eq:EngScaledT00}
\end{align}
\end{widetext}

In analyzing Eq.~\eqref{eq:EngScaledT00}, we see that Alice and Bob's positive energy density contribution scale as $\Upsilon^n$ and $\Upsilon^{2\xi}$, respectively, while the QET terms responsible for the negative energy density scale as $\Upsilon^{\frac{n}{2}+\xi}$. Thus, we can see that choosing $\xi\geq n/2$, we can ensure that the QET terms are not being overpowered by Alice's positive energy contributions. On the other hand, if $\xi\leq n/2$ we would have that the QET terms are able to compete with Bob's positive energy density contributions. Hence, we would require that $\xi = n/2$ which results in all three contributions, Alice and Bob's positive along with the QET terms, to all scale with $\Upsilon^n$. This condition means that the best one can do is have all the terms scale with the same law. This means that it is possible to obtain an arbitrarily deep negative energy density well at the cost of narrowing the support of the negative energy density region. 

Specifically, if we adopt the following scaling relations
\begin{align}
    \lambda(\bm x)&\rightarrow \Upsilon^{\frac{n-2}{2}}\lambda(\Upsilon\bm x)\label{eq:EngOptimalLambda},\\
    \mu(\bm x)&\rightarrow\Upsilon^{\frac{n}{2}}\mu(\Upsilon\bm x)\label{eq:EngOptimalMu},
\end{align}
the resulting stress-energy density follows the scaling relation given by 
\begin{align}
    \bra{\psi(\Upsilon t)}:&\hat T_{\mu\nu}(\Upsilon \bm x):\ket{\psi(\Upsilon t)} = \nonumber\\
    &\Upsilon^n\bra{\psi(t)}:\hat T_{\mu\nu}(\bm x):\ket{\psi(t)}_{\Upsilon = 1}.
\end{align}
That is, the maximum positive and negative energy densities that one achieves in the field after the protocol is performed increase linearly with the scaling constant $\Upsilon$. 
Therefore, we are free to rescale our smearing functions to arbitrarily increase the total amount of negative energy created, albeit in a more and more concentrated region of spacetime.

To make it a bit more concrete let us see how the optimal scaling plays out in the 1+1 dimensional case that we analyzed above. The scaling relations \eqref{eq:EngOptimalLambda} and \eqref{eq:EngOptimalMu} imply that given a negative energy well of `width' $w$ and `depth' $d$ we can rescale the setup to a narrower well of `width' $w/\Upsilon$ and `depth' $\Upsilon^{2} d$, where $\Upsilon>1$. In particular, as the well approaches zero-width, the total amount of negative energy present in the well diverges as $\Upsilon$. A particularly useful way of expressing this scaling behaviour is to express it in terms of  the separation between the center of the well and the point where the closest region of positive energy density starts ($\Delta x$) and the total amount of negative energy in the well ($\Delta E$). $\Delta x$ scales as $w$ and $\Delta E$ scales as $w\cdot d$. Therefore under rescaling $\Delta x\sim \frac{1}{\Delta E}$. Similar to the 1+1-dimensional case, in 3+1-dimensional Minkowski spacetime, the rescaling of a negative energy density well satisfies the scaling relationships $w\to\frac{w}{\Upsilon}$ and $d\to\Upsilon^4 d$. Thus we have the relationship that $\Delta E \propto \frac{1}{\Delta r^3}$, where $\Delta r = \frac{w}{\Upsilon}$.

One can wonder how efficient the QET protocol is at creating negative energy densities. The authors of ~\cite{nichoTeleport} discuss how the scaling laws obtained above saturate the scaling limits imposed by the quantum interest conjecture~\cite{InterestConjecture}. Consequently, the QET protocol is (at least theoretically) as efficient as it can be at creating negative energy density distributions and how much positive energy one has to create in the surrounding areas to make up for it. 

\section{Discussion and outlook}\label{sec:Conclusions}

In this manuscript, we provided a review of some recent applications of the protocol of Quantum Energy Teleportation. Our aim has been to underscore the versatility of QET as a tool not only for theoretical exploration but also for practical applications. In particular, QET offers promising perspectives for addressing fundamental challenges in many-body physics, where entanglement and energy transport are deeply intertwined. It also contributes to our understanding of quantum thermodynamic processes, such as algorithmic cooling, by offering protocols that go beyond conventional cooling schemes to exploit informational and energetic correlations.

Perhaps most intriguingly, QET provides a concrete operational pathway to engineer quantum states of exotic matter, including those that exhibit negative average stress-energy densities—objects typically constrained by classical energy conditions. This positions QET as a unique theoretical and potentially experimental framework for probing the quantum limits of semiclassical gravity.

We hope this review encourages further research into the wide-ranging implications of QET. Future directions may include exploring its role in the entanglement dynamics of interacting systems, resource theory,  non-equilibrium thermodynamics, and even the structure of spacetime itself. As our understanding of QET deepens, it may serve as a foundational protocol through which novel connections between energy, information, and geometry can be forged.

\begin{acknowledgements}
    The authors thank Nicholas Funai for helpful discussions. EMM acknowledges the support of the NSERC Discovery program. BR acknowledges the Ontario Provincial Government for financial support from the Ontario Graduate Scholarship. Research at Perimeter Institute and the University of Waterloo is supported in part by the Government of Canada through the Department of Innovation, Science and Industry Canada and by the Province of Ontario through the Ministry of Colleges and Universities. 
\end{acknowledgements}

\twocolumngrid
\bibliography{references.bib}

\begin{thebibliography}{47}%
\makeatletter
\providecommand \@ifxundefined [1]{%
 \@ifx{#1\undefined}
}%
\providecommand \@ifnum [1]{%
 \ifnum #1\expandafter \@firstoftwo
 \else \expandafter \@secondoftwo
 \fi
}%
\providecommand \@ifx [1]{%
 \ifx #1\expandafter \@firstoftwo
 \else \expandafter \@secondoftwo
 \fi
}%
\providecommand \natexlab [1]{#1}%
\providecommand \enquote  [1]{``#1''}%
\providecommand \bibnamefont  [1]{#1}%
\providecommand \bibfnamefont [1]{#1}%
\providecommand \citenamefont [1]{#1}%
\providecommand \href@noop [0]{\@secondoftwo}%
\providecommand \href [0]{\begingroup \@sanitize@url \@href}%
\providecommand \@href[1]{\@@startlink{#1}\@@href}%
\providecommand \@@href[1]{\endgroup#1\@@endlink}%
\providecommand \@sanitize@url [0]{\catcode `\\12\catcode `\$12\catcode `\&12\catcode `\#12\catcode `\^12\catcode `\_12\catcode `\%12\relax}%
\providecommand \@@startlink[1]{}%
\providecommand \@@endlink[0]{}%
\providecommand \url  [0]{\begingroup\@sanitize@url \@url }%
\providecommand \@url [1]{\endgroup\@href {#1}{\urlprefix }}%
\providecommand \urlprefix  [0]{URL }%
\providecommand \Eprint [0]{\href }%
\providecommand \doibase [0]{https://doi.org/}%
\providecommand \selectlanguage [0]{\@gobble}%
\providecommand \bibinfo  [0]{\@secondoftwo}%
\providecommand \bibfield  [0]{\@secondoftwo}%
\providecommand \translation [1]{[#1]}%
\providecommand \BibitemOpen [0]{}%
\providecommand \bibitemStop [0]{}%
\providecommand \bibitemNoStop [0]{.\EOS\space}%
\providecommand \EOS [0]{\spacefactor3000\relax}%
\providecommand \BibitemShut  [1]{\csname bibitem#1\endcsname}%
\let\auto@bib@innerbib\@empty
\bibitem [{\citenamefont {Hotta}(2008)}]{Hotta2008}%
  \BibitemOpen
  \bibfield  {author} {\bibinfo {author} {\bibfnamefont {M.}~\bibnamefont {Hotta}},\ }\bibfield  {title} {\bibinfo {title} {Quantum measurement information as a key to energy extraction from local vacuums},\ }\href {https://doi.org/10.1103/PhysRevD.78.045006} {\bibfield  {journal} {\bibinfo  {journal} {Phys. Rev. D}\ }\textbf {\bibinfo {volume} {78}},\ \bibinfo {pages} {045006} (\bibinfo {year} {2008})}\BibitemShut {NoStop}%
\bibitem [{\citenamefont {Hotta}(2009)}]{Hotta2009}%
  \BibitemOpen
  \bibfield  {author} {\bibinfo {author} {\bibfnamefont {M.}~\bibnamefont {Hotta}},\ }\bibfield  {title} {\bibinfo {title} {Quantum energy teleportation in spin chain systems},\ }\href {https://doi.org/10.1143/JPSJ.78.034001} {\bibfield  {journal} {\bibinfo  {journal} {J. Phys. Soc. Jpn.}\ }\textbf {\bibinfo {volume} {78}},\ \bibinfo {pages} {034001} (\bibinfo {year} {2009})}\BibitemShut {NoStop}%
\bibitem [{\citenamefont {Hotta}(2011)}]{HottaMinimal}%
  \BibitemOpen
  \bibfield  {author} {\bibinfo {author} {\bibfnamefont {M.}~\bibnamefont {Hotta}},\ }\href {https://arxiv.org/abs/1101.3954} {\bibinfo {title} {Quantum energy teleportation: An introductory review}} (\bibinfo {year} {2011}),\ \Eprint {https://arxiv.org/abs/1101.3954} {arXiv:1101.3954 [quant-ph]} \BibitemShut {NoStop}%
\bibitem [{\citenamefont {Frey}\ \emph {et~al.}(2014)\citenamefont {Frey}, \citenamefont {Funo},\ and\ \citenamefont {Hotta}}]{FreyFunoHottaSLP}%
  \BibitemOpen
  \bibfield  {author} {\bibinfo {author} {\bibfnamefont {M.}~\bibnamefont {Frey}}, \bibinfo {author} {\bibfnamefont {K.}~\bibnamefont {Funo}},\ and\ \bibinfo {author} {\bibfnamefont {M.}~\bibnamefont {Hotta}},\ }\bibfield  {title} {\bibinfo {title} {Strong local passivity in finite quantum systems},\ }\href {https://doi.org/10.1103/PhysRevE.90.012127} {\bibfield  {journal} {\bibinfo  {journal} {Phys. Rev. E}\ }\textbf {\bibinfo {volume} {90}},\ \bibinfo {pages} {012127} (\bibinfo {year} {2014})}\BibitemShut {NoStop}%
\bibitem [{\citenamefont {Alhambra}\ \emph {et~al.}(2019)\citenamefont {Alhambra}, \citenamefont {Styliaris}, \citenamefont {Rodr\'{i}guez-Briones}, \citenamefont {Sikora},\ and\ \citenamefont {Mart\'{i}n-Mart\'{i}nez}}]{iffSLP}%
  \BibitemOpen
  \bibfield  {author} {\bibinfo {author} {\bibfnamefont {A.~M.}\ \bibnamefont {Alhambra}}, \bibinfo {author} {\bibfnamefont {G.}~\bibnamefont {Styliaris}}, \bibinfo {author} {\bibfnamefont {N.~A.}\ \bibnamefont {Rodr\'{i}guez-Briones}}, \bibinfo {author} {\bibfnamefont {J.}~\bibnamefont {Sikora}},\ and\ \bibinfo {author} {\bibfnamefont {E.}~\bibnamefont {Mart\'{i}n-Mart\'{i}nez}},\ }\bibfield  {title} {\bibinfo {title} {Fundamental limitations to local energy extraction in quantum systems},\ }\href {http://dx.doi.org/10.1103/PhysRevLett.123.190601} {\bibfield  {journal} {\bibinfo  {journal} {Phys. Rev. Lett.}\ }\textbf {\bibinfo {volume} {123}} (\bibinfo {year} {2019})}\BibitemShut {NoStop}%
\bibitem [{\citenamefont {Rodr\'{i}guez-Briones}\ \emph {et~al.}(2023)\citenamefont {Rodr\'{i}guez-Briones}, \citenamefont {Katiyar}, \citenamefont {Mart\'{i}n-Mart\'{i}nez},\ and\ \citenamefont {Laflamme}}]{IQCQETExperiment}%
  \BibitemOpen
  \bibfield  {author} {\bibinfo {author} {\bibfnamefont {N.~A.}\ \bibnamefont {Rodr\'{i}guez-Briones}}, \bibinfo {author} {\bibfnamefont {H.}~\bibnamefont {Katiyar}}, \bibinfo {author} {\bibfnamefont {E.}~\bibnamefont {Mart\'{i}n-Mart\'{i}nez}},\ and\ \bibinfo {author} {\bibfnamefont {R.}~\bibnamefont {Laflamme}},\ }\bibfield  {title} {\bibinfo {title} {Experimental activation of strong local passive states with quantum information},\ }\href {http://dx.doi.org/10.1103/PhysRevLett.130.110801} {\bibfield  {journal} {\bibinfo  {journal} {Phys. Rev. Lett.}\ }\textbf {\bibinfo {volume} {130}} (\bibinfo {year} {2023})}\BibitemShut {NoStop}%
\bibitem [{\citenamefont {Ikeda}(2023)}]{Ikeda2023}%
  \BibitemOpen
  \bibfield  {author} {\bibinfo {author} {\bibfnamefont {K.}~\bibnamefont {Ikeda}},\ }\bibfield  {title} {\bibinfo {title} {Demonstration of quantum energy teleportation on superconducting quantum hardware},\ }\href {http://dx.doi.org/10.1103/PhysRevApplied.20.024051} {\bibfield  {journal} {\bibinfo  {journal} {Phys. Rev. Appl.}\ }\textbf {\bibinfo {volume} {20}} (\bibinfo {year} {2023})}\BibitemShut {NoStop}%
\bibitem [{\citenamefont {Rodr\'{i}guez-Briones}\ \emph {et~al.}(2017)\citenamefont {Rodr\'{i}guez-Briones}, \citenamefont {Mart\'{i}n-Mart\'{i}nez}, \citenamefont {Kempf},\ and\ \citenamefont {Laflamme}}]{QETAlgorithmicCooling}%
  \BibitemOpen
  \bibfield  {author} {\bibinfo {author} {\bibfnamefont {N.~A.}\ \bibnamefont {Rodr\'{i}guez-Briones}}, \bibinfo {author} {\bibfnamefont {E.}~\bibnamefont {Mart\'{i}n-Mart\'{i}nez}}, \bibinfo {author} {\bibfnamefont {A.}~\bibnamefont {Kempf}},\ and\ \bibinfo {author} {\bibfnamefont {R.}~\bibnamefont {Laflamme}},\ }\bibfield  {title} {\bibinfo {title} {Correlation-enhanced algorithmic cooling},\ }\href {http://dx.doi.org/10.1103/PhysRevLett.119.050502} {\bibfield  {journal} {\bibinfo  {journal} {Phys. Rev. Lett.}\ }\textbf {\bibinfo {volume} {119}} (\bibinfo {year} {2017})}\BibitemShut {NoStop}%
\bibitem [{\citenamefont {Boykin}\ \emph {et~al.}(2002)\citenamefont {Boykin}, \citenamefont {Mor}, \citenamefont {Roychowdhury}, \citenamefont {Vatan},\ and\ \citenamefont {Vrijen}}]{BoykinMorAC}%
  \BibitemOpen
  \bibfield  {author} {\bibinfo {author} {\bibfnamefont {P.~O.}\ \bibnamefont {Boykin}}, \bibinfo {author} {\bibfnamefont {T.}~\bibnamefont {Mor}}, \bibinfo {author} {\bibfnamefont {V.}~\bibnamefont {Roychowdhury}}, \bibinfo {author} {\bibfnamefont {F.}~\bibnamefont {Vatan}},\ and\ \bibinfo {author} {\bibfnamefont {R.}~\bibnamefont {Vrijen}},\ }\bibfield  {title} {\bibinfo {title} {Algorithmic cooling and scalable nmr quantum computers},\ }\href {https://doi.org/10.1073/pnas.241641898} {\bibfield  {journal} {\bibinfo  {journal} {Proc. Natl. Acad. Sci.}\ }\textbf {\bibinfo {volume} {99}},\ \bibinfo {pages} {3388} (\bibinfo {year} {2002})}\BibitemShut {NoStop}%
\bibitem [{\citenamefont {Rodr\'{\i}guez-Briones}\ and\ \citenamefont {Laflamme}(2016)}]{NayeliACPRL}%
  \BibitemOpen
  \bibfield  {author} {\bibinfo {author} {\bibfnamefont {N.~A.}\ \bibnamefont {Rodr\'{\i}guez-Briones}}\ and\ \bibinfo {author} {\bibfnamefont {R.}~\bibnamefont {Laflamme}},\ }\bibfield  {title} {\bibinfo {title} {Achievable polarization for heat-bath algorithmic cooling},\ }\href {https://doi.org/10.1103/PhysRevLett.116.170501} {\bibfield  {journal} {\bibinfo  {journal} {Phys. Rev. Lett.}\ }\textbf {\bibinfo {volume} {116}},\ \bibinfo {pages} {170501} (\bibinfo {year} {2016})}\BibitemShut {NoStop}%
\bibitem [{\citenamefont {Raeisi}\ and\ \citenamefont {Mosca}(2015)}]{RaeisiMoscaAC}%
  \BibitemOpen
  \bibfield  {author} {\bibinfo {author} {\bibfnamefont {S.}~\bibnamefont {Raeisi}}\ and\ \bibinfo {author} {\bibfnamefont {M.}~\bibnamefont {Mosca}},\ }\bibfield  {title} {\bibinfo {title} {Asymptotic bound for heat-bath algorithmic cooling},\ }\href {https://doi.org/10.1103/PhysRevLett.114.100404} {\bibfield  {journal} {\bibinfo  {journal} {Phys. Rev. Lett.}\ }\textbf {\bibinfo {volume} {114}},\ \bibinfo {pages} {100404} (\bibinfo {year} {2015})}\BibitemShut {NoStop}%
\bibitem [{\citenamefont {Baugh}\ \emph {et~al.}(2005)\citenamefont {Baugh}, \citenamefont {Moussa}, \citenamefont {Ryan}, \citenamefont {Nayak},\ and\ \citenamefont {Laflamme}}]{BaughLaflamme2005}%
  \BibitemOpen
  \bibfield  {author} {\bibinfo {author} {\bibfnamefont {J.}~\bibnamefont {Baugh}}, \bibinfo {author} {\bibfnamefont {O.}~\bibnamefont {Moussa}}, \bibinfo {author} {\bibfnamefont {C.~A.}\ \bibnamefont {Ryan}}, \bibinfo {author} {\bibfnamefont {A.}~\bibnamefont {Nayak}},\ and\ \bibinfo {author} {\bibfnamefont {R.}~\bibnamefont {Laflamme}},\ }\bibfield  {title} {\bibinfo {title} {Experimental implementation of heat-bath algorithmic cooling using solid-state nuclear magnetic resonance},\ }\href {https://doi.org/10.1038/nature04272} {\bibfield  {journal} {\bibinfo  {journal} {Nature}\ }\textbf {\bibinfo {volume} {438}},\ \bibinfo {pages} {470–473} (\bibinfo {year} {2005})}\BibitemShut {NoStop}%
\bibitem [{\citenamefont {Fernandez}\ \emph {et~al.}(2004)\citenamefont {Fernandez}, \citenamefont {Lloyd}, \citenamefont {Mor},\ and\ \citenamefont {Roychowdhury}}]{FernandezMorAC}%
  \BibitemOpen
  \bibfield  {author} {\bibinfo {author} {\bibfnamefont {J.~M.}\ \bibnamefont {Fernandez}}, \bibinfo {author} {\bibfnamefont {S.}~\bibnamefont {Lloyd}}, \bibinfo {author} {\bibfnamefont {T.}~\bibnamefont {Mor}},\ and\ \bibinfo {author} {\bibfnamefont {V.}~\bibnamefont {Roychowdhury}},\ }\bibfield  {title} {\bibinfo {title} {Algorithmic cooling of spins: A practicable method for increasing polarization},\ }\href {https://doi.org/10.1142/S0219749904000419} {\bibfield  {journal} {\bibinfo  {journal} {Int. J. Quantum Inf.}\ }\textbf {\bibinfo {volume} {02}},\ \bibinfo {pages} {461} (\bibinfo {year} {2004})}\BibitemShut {NoStop}%
\bibitem [{\citenamefont {Schulman}\ \emph {et~al.}(2005)\citenamefont {Schulman}, \citenamefont {Mor},\ and\ \citenamefont {Weinstein}}]{PPA1}%
  \BibitemOpen
  \bibfield  {author} {\bibinfo {author} {\bibfnamefont {L.~J.}\ \bibnamefont {Schulman}}, \bibinfo {author} {\bibfnamefont {T.}~\bibnamefont {Mor}},\ and\ \bibinfo {author} {\bibfnamefont {Y.}~\bibnamefont {Weinstein}},\ }\bibfield  {title} {\bibinfo {title} {Physical limits of heat-bath algorithmic cooling},\ }\href {https://doi.org/10.1103/PhysRevLett.94.120501} {\bibfield  {journal} {\bibinfo  {journal} {Phys. Rev. Lett.}\ }\textbf {\bibinfo {volume} {94}},\ \bibinfo {pages} {120501} (\bibinfo {year} {2005})}\BibitemShut {NoStop}%
\bibitem [{\citenamefont {Schulman}\ \emph {et~al.}(2007)\citenamefont {Schulman}, \citenamefont {Mor},\ and\ \citenamefont {Weinstein}}]{Schulman2007}%
  \BibitemOpen
  \bibfield  {author} {\bibinfo {author} {\bibfnamefont {L.~J.}\ \bibnamefont {Schulman}}, \bibinfo {author} {\bibfnamefont {T.}~\bibnamefont {Mor}},\ and\ \bibinfo {author} {\bibfnamefont {Y.}~\bibnamefont {Weinstein}},\ }\bibfield  {title} {\bibinfo {title} {Physical limits of heat‐bath algorithmic cooling},\ }\href {https://doi.org/10.1137/050666023} {\bibfield  {journal} {\bibinfo  {journal} {SIAM J. Comput.}\ }\textbf {\bibinfo {volume} {36}},\ \bibinfo {pages} {1729} (\bibinfo {year} {2007})}\BibitemShut {NoStop}%
\bibitem [{\citenamefont {Elias}\ \emph {et~al.}(2006)\citenamefont {Elias}, \citenamefont {Fernandez}, \citenamefont {Mor},\ and\ \citenamefont {Weinstein}}]{Elias2006}%
  \BibitemOpen
  \bibfield  {author} {\bibinfo {author} {\bibfnamefont {Y.}~\bibnamefont {Elias}}, \bibinfo {author} {\bibfnamefont {J.~M.}\ \bibnamefont {Fernandez}}, \bibinfo {author} {\bibfnamefont {T.}~\bibnamefont {Mor}},\ and\ \bibinfo {author} {\bibfnamefont {Y.}~\bibnamefont {Weinstein}},\ }\bibfield  {title} {\bibinfo {title} {Optimal algorithmic cooling of spins},\ }\href {https://doi.org/https://doi.org/10.1560/IJC\_46\_4\_371} {\bibfield  {journal} {\bibinfo  {journal} {Isr. J. Chem.}\ }\textbf {\bibinfo {volume} {46}},\ \bibinfo {pages} {371} (\bibinfo {year} {2006})}\BibitemShut {NoStop}%
\bibitem [{\citenamefont {Funai}\ and\ \citenamefont {Mart\'{\i}n-Mart\'{\i}nez}(2017)}]{nichoTeleport}%
  \BibitemOpen
  \bibfield  {author} {\bibinfo {author} {\bibfnamefont {N.}~\bibnamefont {Funai}}\ and\ \bibinfo {author} {\bibfnamefont {E.}~\bibnamefont {Mart\'{\i}n-Mart\'{\i}nez}},\ }\bibfield  {title} {\bibinfo {title} {Engineering negative stress-energy densities with quantum energy teleportation},\ }\href {https://doi.org/10.1103/PhysRevD.96.025014} {\bibfield  {journal} {\bibinfo  {journal} {Phys. Rev. D}\ }\textbf {\bibinfo {volume} {96}},\ \bibinfo {pages} {025014} (\bibinfo {year} {2017})}\BibitemShut {NoStop}%
\bibitem [{\citenamefont {Hotta}(2010)}]{Hotta2010}%
  \BibitemOpen
  \bibfield  {author} {\bibinfo {author} {\bibfnamefont {M.}~\bibnamefont {Hotta}},\ }\bibfield  {title} {\bibinfo {title} {Energy entanglement relation for quantum energy teleportation},\ }\href {https://doi.org/10.1016/j.physleta.2010.06.058} {\bibfield  {journal} {\bibinfo  {journal} {Phys. Lett. A}\ }\textbf {\bibinfo {volume} {374}},\ \bibinfo {pages} {3416–3421} (\bibinfo {year} {2010})}\BibitemShut {NoStop}%
\bibitem [{\citenamefont {Verdon-Akzam}\ \emph {et~al.}(2016)\citenamefont {Verdon-Akzam}, \citenamefont {Mart\'{\i}n-Mart\'{\i}nez},\ and\ \citenamefont {Kempf}}]{DerivativeQET}%
  \BibitemOpen
  \bibfield  {author} {\bibinfo {author} {\bibfnamefont {G.}~\bibnamefont {Verdon-Akzam}}, \bibinfo {author} {\bibfnamefont {E.}~\bibnamefont {Mart\'{\i}n-Mart\'{\i}nez}},\ and\ \bibinfo {author} {\bibfnamefont {A.}~\bibnamefont {Kempf}},\ }\bibfield  {title} {\bibinfo {title} {Asymptotically limitless quantum energy teleportation via qudit probes},\ }\href {https://doi.org/10.1103/PhysRevA.93.022308} {\bibfield  {journal} {\bibinfo  {journal} {Phys. Rev. A}\ }\textbf {\bibinfo {volume} {93}},\ \bibinfo {pages} {022308} (\bibinfo {year} {2016})}\BibitemShut {NoStop}%
\bibitem [{\citenamefont {Verdon-Akzam}(2017)}]{GuillaumeThesis}%
  \BibitemOpen
  \bibfield  {author} {\bibinfo {author} {\bibfnamefont {G.}~\bibnamefont {Verdon-Akzam}},\ }\emph {\bibinfo {title} {Probing Quantum Fields: Measurements and Quantum Energy Teleportation}},\ \href@noop {} {Ph.D. thesis},\ \bibinfo  {school} {University of Waterloo} (\bibinfo {year} {2017})\BibitemShut {NoStop}%
\bibitem [{\citenamefont {Boyd}\ and\ \citenamefont {Vandenberghe}(2004)}]{Semidefinite1}%
  \BibitemOpen
  \bibfield  {author} {\bibinfo {author} {\bibfnamefont {S.}~\bibnamefont {Boyd}}\ and\ \bibinfo {author} {\bibfnamefont {L.}~\bibnamefont {Vandenberghe}},\ }\href@noop {} {\emph {\bibinfo {title} {Convex Optimization}}}\ (\bibinfo  {publisher} {Cambridge University Press},\ \bibinfo {year} {2004})\BibitemShut {NoStop}%
\bibitem [{\citenamefont {Watrous}(2018)}]{Semidefinite2}%
  \BibitemOpen
  \bibfield  {author} {\bibinfo {author} {\bibfnamefont {J.}~\bibnamefont {Watrous}},\ }\href@noop {} {\emph {\bibinfo {title} {The Theory of Quantum Information}}}\ (\bibinfo  {publisher} {Cambridge University Press},\ \bibinfo {year} {2018})\BibitemShut {NoStop}%
\bibitem [{\citenamefont {Rodr\'{i}guez-Briones}(2020)}]{NayeliThesis}%
  \BibitemOpen
  \bibfield  {author} {\bibinfo {author} {\bibfnamefont {N.~A.}\ \bibnamefont {Rodr\'{i}guez-Briones}},\ }\emph {\bibinfo {title} {Novel Heat-Bath Algorithmic Cooling methods}},\ \href@noop {} {Ph.D. thesis},\ \bibinfo  {school} {University of Waterloo} (\bibinfo {year} {2020})\BibitemShut {NoStop}%
\bibitem [{Bru()}]{BrukerNMR}%
  \BibitemOpen
  \href@noop {} {}\bibinfo {note} {\href{https://www.bruker.com/en/products-and-solutions/mr/nmr/avance-nmr-spectrometer.html}{https://www.bruker.com}}\BibitemShut {NoStop}%
\bibitem [{\citenamefont {Cory}\ \emph {et~al.}(1998)\citenamefont {Cory}, \citenamefont {Price},\ and\ \citenamefont {Havel}}]{PsuedopureStates}%
  \BibitemOpen
  \bibfield  {author} {\bibinfo {author} {\bibfnamefont {D.~G.}\ \bibnamefont {Cory}}, \bibinfo {author} {\bibfnamefont {M.~D.}\ \bibnamefont {Price}},\ and\ \bibinfo {author} {\bibfnamefont {T.~F.}\ \bibnamefont {Havel}},\ }\bibfield  {title} {\bibinfo {title} {Nuclear magnetic resonance spectroscopy: An experimentally accessible paradigm for quantum computing},\ }\href {https://doi.org/https://doi.org/10.1016/S0167-2789(98)00046-3} {\bibfield  {journal} {\bibinfo  {journal} {Physica D: Nonlinear Phenom.}\ }\textbf {\bibinfo {volume} {120}},\ \bibinfo {pages} {82} (\bibinfo {year} {1998})},\ \bibinfo {note} {proceedings of the Fourth Workshop on Physics and Consumption}\BibitemShut {NoStop}%
\bibitem [{\citenamefont {Levitt}(2013)}]{LevittSpinDynamics}%
  \BibitemOpen
  \bibfield  {author} {\bibinfo {author} {\bibfnamefont {M.~H.}\ \bibnamefont {Levitt}},\ }\href@noop {} {\emph {\bibinfo {title} {Spin Dynamics: Basics of Nuclear Magnetic Resonance}}}\ (\bibinfo  {publisher} {John Wiley \& Sons},\ \bibinfo {year} {2013})\BibitemShut {NoStop}%
\bibitem [{\citenamefont {Khaneja}\ \emph {et~al.}(2005)\citenamefont {Khaneja}, \citenamefont {Reiss}, \citenamefont {Kehlet}, \citenamefont {Schulte-Herbrüggen},\ and\ \citenamefont {Glaser}}]{GRAPEPulse}%
  \BibitemOpen
  \bibfield  {author} {\bibinfo {author} {\bibfnamefont {N.}~\bibnamefont {Khaneja}}, \bibinfo {author} {\bibfnamefont {T.}~\bibnamefont {Reiss}}, \bibinfo {author} {\bibfnamefont {C.}~\bibnamefont {Kehlet}}, \bibinfo {author} {\bibfnamefont {T.}~\bibnamefont {Schulte-Herbrüggen}},\ and\ \bibinfo {author} {\bibfnamefont {S.~J.}\ \bibnamefont {Glaser}},\ }\bibfield  {title} {\bibinfo {title} {Optimal control of coupled spin dynamics: design of nmr pulse sequences by gradient ascent algorithms},\ }\href {https://doi.org/https://doi.org/10.1016/j.jmr.2004.11.004} {\bibfield  {journal} {\bibinfo  {journal} {J. Magn. Reson.}\ }\textbf {\bibinfo {volume} {172}},\ \bibinfo {pages} {296} (\bibinfo {year} {2005})}\BibitemShut {NoStop}%
\bibitem [{\citenamefont {Peterson}\ \emph {et~al.}(2020)\citenamefont {Peterson}, \citenamefont {Sarthour},\ and\ \citenamefont {Laflamme}}]{GRAPEPulses2}%
  \BibitemOpen
  \bibfield  {author} {\bibinfo {author} {\bibfnamefont {J.~P.}\ \bibnamefont {Peterson}}, \bibinfo {author} {\bibfnamefont {R.~S.}\ \bibnamefont {Sarthour}},\ and\ \bibinfo {author} {\bibfnamefont {R.}~\bibnamefont {Laflamme}},\ }\bibfield  {title} {\bibinfo {title} {Enhancing quantum control by improving shaped-pulse generation},\ }\href {https://doi.org/10.1103/PhysRevApplied.13.054060} {\bibfield  {journal} {\bibinfo  {journal} {Phys. Rev. Appl.}\ }\textbf {\bibinfo {volume} {13}},\ \bibinfo {pages} {054060} (\bibinfo {year} {2020})}\BibitemShut {NoStop}%
\bibitem [{\citenamefont {Park}\ \emph {et~al.}(2016)\citenamefont {Park}, \citenamefont {Rodriguez-Briones}, \citenamefont {Feng}, \citenamefont {Rahimi}, \citenamefont {Baugh},\ and\ \citenamefont {Laflamme}}]{LaflammeParkBook}%
  \BibitemOpen
  \bibfield  {author} {\bibinfo {author} {\bibfnamefont {D.~K.}\ \bibnamefont {Park}}, \bibinfo {author} {\bibfnamefont {N.~A.}\ \bibnamefont {Rodriguez-Briones}}, \bibinfo {author} {\bibfnamefont {G.}~\bibnamefont {Feng}}, \bibinfo {author} {\bibfnamefont {R.}~\bibnamefont {Rahimi}}, \bibinfo {author} {\bibfnamefont {J.}~\bibnamefont {Baugh}},\ and\ \bibinfo {author} {\bibfnamefont {R.}~\bibnamefont {Laflamme}},\ }\bibinfo {title} {Heat bath algorithmic cooling with spins: Review and prospects},\ in\ \href {https://doi.org/10.1007/978-1-4939-3658-8_8} {\emph {\bibinfo {booktitle} {Electron Spin Resonance (ESR) Based Quantum Computing}}},\ \bibinfo {editor} {edited by\ \bibinfo {editor} {\bibfnamefont {T.}~\bibnamefont {Takui}}, \bibinfo {editor} {\bibfnamefont {L.}~\bibnamefont {Berliner}},\ and\ \bibinfo {editor} {\bibfnamefont {G.}~\bibnamefont {Hanson}}}\ (\bibinfo  {publisher} {Springer New York},\ \bibinfo {address} {New York, NY},\ \bibinfo {year} {2016})\ pp.\ \bibinfo {pages} {227--255}\BibitemShut
  {NoStop}%
\bibitem [{\citenamefont {Rodríguez-Briones}\ \emph {et~al.}(2017)\citenamefont {Rodríguez-Briones}, \citenamefont {Li}, \citenamefont {Peng}, \citenamefont {Mor}, \citenamefont {Weinstein},\ and\ \citenamefont {Laflamme}}]{SRHBAC}%
  \BibitemOpen
  \bibfield  {author} {\bibinfo {author} {\bibfnamefont {N.~A.}\ \bibnamefont {Rodríguez-Briones}}, \bibinfo {author} {\bibfnamefont {J.}~\bibnamefont {Li}}, \bibinfo {author} {\bibfnamefont {X.}~\bibnamefont {Peng}}, \bibinfo {author} {\bibfnamefont {T.}~\bibnamefont {Mor}}, \bibinfo {author} {\bibfnamefont {Y.}~\bibnamefont {Weinstein}},\ and\ \bibinfo {author} {\bibfnamefont {R.}~\bibnamefont {Laflamme}},\ }\bibfield  {title} {\bibinfo {title} {Heat-bath algorithmic cooling with correlated qubit-environment interactions},\ }\href {https://doi.org/10.1088/1367-2630/aa8fe0} {\bibfield  {journal} {\bibinfo  {journal} {New J. Phys.}\ }\textbf {\bibinfo {volume} {19}},\ \bibinfo {pages} {113047} (\bibinfo {year} {2017})}\BibitemShut {NoStop}%
\bibitem [{\citenamefont {Overhauser}(1953)}]{NuclearOverhauser}%
  \BibitemOpen
  \bibfield  {author} {\bibinfo {author} {\bibfnamefont {A.~W.}\ \bibnamefont {Overhauser}},\ }\bibfield  {title} {\bibinfo {title} {Paramagnetic relaxation in metals},\ }\href {https://doi.org/10.1103/PhysRev.89.689} {\bibfield  {journal} {\bibinfo  {journal} {Phys. Rev.}\ }\textbf {\bibinfo {volume} {89}},\ \bibinfo {pages} {689} (\bibinfo {year} {1953})}\BibitemShut {NoStop}%
\bibitem [{\citenamefont {Niemczyk}\ \emph {et~al.}(2010)\citenamefont {Niemczyk}, \citenamefont {Deppe}, \citenamefont {Menzel}, \citenamefont {Hocke}, \citenamefont {Schwarz}, \citenamefont {Garcia-Ripoll}, \citenamefont {Zueco}, \citenamefont {H\"{u}mmer}, \citenamefont {Solano}, \citenamefont {Marx},\ and\ \citenamefont {Gross}}]{Superconducting1}%
  \BibitemOpen
  \bibfield  {author} {\bibinfo {author} {\bibfnamefont {T.}~\bibnamefont {Niemczyk}}, \bibinfo {author} {\bibfnamefont {F.}~\bibnamefont {Deppe}}, \bibinfo {author} {\bibfnamefont {E.~P.}\ \bibnamefont {Menzel}}, \bibinfo {author} {\bibfnamefont {F.}~\bibnamefont {Hocke}}, \bibinfo {author} {\bibfnamefont {M.~J.}\ \bibnamefont {Schwarz}}, \bibinfo {author} {\bibfnamefont {J.~J.}\ \bibnamefont {Garcia-Ripoll}}, \bibinfo {author} {\bibfnamefont {D.}~\bibnamefont {Zueco}}, \bibinfo {author} {\bibfnamefont {T.}~\bibnamefont {H\"{u}mmer}}, \bibinfo {author} {\bibfnamefont {E.}~\bibnamefont {Solano}}, \bibinfo {author} {\bibfnamefont {A.}~\bibnamefont {Marx}},\ and\ \bibinfo {author} {\bibfnamefont {A.}~\bibnamefont {Gross}},\ }\bibfield  {title} {\bibinfo {title} {Circuit quantum electrodynamics in the ultrastrong-coupling regime},\ }\href {https://doi.org/10.1038/nphys1730} {\bibfield  {journal} {\bibinfo  {journal} {Nat. Phys.}\ }\textbf {\bibinfo {volume} {6}} (\bibinfo {year} {2010})}\BibitemShut {NoStop}%
\bibitem [{\citenamefont {Peropadre}\ \emph {et~al.}(2010)\citenamefont {Peropadre}, \citenamefont {Forn-D\'{\i}az}, \citenamefont {Solano},\ and\ \citenamefont {Garc\'{\i}a-Ripoll}}]{Superconducting2}%
  \BibitemOpen
  \bibfield  {author} {\bibinfo {author} {\bibfnamefont {B.}~\bibnamefont {Peropadre}}, \bibinfo {author} {\bibfnamefont {P.}~\bibnamefont {Forn-D\'{\i}az}}, \bibinfo {author} {\bibfnamefont {E.}~\bibnamefont {Solano}},\ and\ \bibinfo {author} {\bibfnamefont {J.~J.}\ \bibnamefont {Garc\'{\i}a-Ripoll}},\ }\bibfield  {title} {\bibinfo {title} {Switchable ultrastrong coupling in circuit qed},\ }\href {https://doi.org/10.1103/PhysRevLett.105.023601} {\bibfield  {journal} {\bibinfo  {journal} {Phys. Rev. Lett.}\ }\textbf {\bibinfo {volume} {105}},\ \bibinfo {pages} {023601} (\bibinfo {year} {2010})}\BibitemShut {NoStop}%
\bibitem [{\citenamefont {Forn-D\'{i}az}\ \emph {et~al.}(2017)\citenamefont {Forn-D\'{i}az}, \citenamefont {Garc\'{i}a-Ripoll}, \citenamefont {Peropadre}, \citenamefont {Orgiazzi}, \citenamefont {Yurtalan}, \citenamefont {Belyansky}, \citenamefont {Wilson},\ and\ \citenamefont {Lupascu}}]{Superconducting3}%
  \BibitemOpen
  \bibfield  {author} {\bibinfo {author} {\bibfnamefont {P.}~\bibnamefont {Forn-D\'{i}az}}, \bibinfo {author} {\bibfnamefont {J.~J.}\ \bibnamefont {Garc\'{i}a-Ripoll}}, \bibinfo {author} {\bibfnamefont {B.}~\bibnamefont {Peropadre}}, \bibinfo {author} {\bibfnamefont {J.-L.}\ \bibnamefont {Orgiazzi}}, \bibinfo {author} {\bibfnamefont {M.~A.}\ \bibnamefont {Yurtalan}}, \bibinfo {author} {\bibfnamefont {R.}~\bibnamefont {Belyansky}}, \bibinfo {author} {\bibfnamefont {C.~M.}\ \bibnamefont {Wilson}},\ and\ \bibinfo {author} {\bibfnamefont {A.}~\bibnamefont {Lupascu}},\ }\bibfield  {title} {\bibinfo {title} {Ultrastrong coupling of a single artificial atom to an electromagnetic continuum in the nonperturbative regime},\ }\href {https://doi.org/10.1038/nphys3905} {\bibfield  {journal} {\bibinfo  {journal} {Nat. Phys.}\ }\textbf {\bibinfo {volume} {13}} (\bibinfo {year} {2017})}\BibitemShut {NoStop}%
\bibitem [{\citenamefont {Hawking}(1972)}]{HawkingGR}%
  \BibitemOpen
  \bibfield  {author} {\bibinfo {author} {\bibfnamefont {S.~W.}\ \bibnamefont {Hawking}},\ }\bibfield  {title} {\bibinfo {title} {Black holes in general relativity},\ }\href {https://doi.org/10.1007/BF01877517} {\bibfield  {journal} {\bibinfo  {journal} {Commun. Math. Phys.}\ }\textbf {\bibinfo {volume} {25}} (\bibinfo {year} {1972})}\BibitemShut {NoStop}%
\bibitem [{\citenamefont {Alcubierre}(1994)}]{Alcubierre1994}%
  \BibitemOpen
  \bibfield  {author} {\bibinfo {author} {\bibfnamefont {M.}~\bibnamefont {Alcubierre}},\ }\bibfield  {title} {\bibinfo {title} {The warp drive: hyper-fast travel within general relativity},\ }\href {https://doi.org/10.1088/0264-9381/11/5/001} {\bibfield  {journal} {\bibinfo  {journal} {Class. Quantum Gravity}\ }\textbf {\bibinfo {volume} {11}},\ \bibinfo {pages} {L73} (\bibinfo {year} {1994})}\BibitemShut {NoStop}%
\bibitem [{\citenamefont {Morris}\ \emph {et~al.}(1988)\citenamefont {Morris}, \citenamefont {Thorne},\ and\ \citenamefont {Yurtsever}}]{Wormholes}%
  \BibitemOpen
  \bibfield  {author} {\bibinfo {author} {\bibfnamefont {M.~S.}\ \bibnamefont {Morris}}, \bibinfo {author} {\bibfnamefont {K.~S.}\ \bibnamefont {Thorne}},\ and\ \bibinfo {author} {\bibfnamefont {U.}~\bibnamefont {Yurtsever}},\ }\bibfield  {title} {\bibinfo {title} {Wormholes, time machines, and the weak energy condition},\ }\href {https://doi.org/10.1103/PhysRevLett.61.1446} {\bibfield  {journal} {\bibinfo  {journal} {Phys. Rev. Lett.}\ }\textbf {\bibinfo {volume} {61}},\ \bibinfo {pages} {1446} (\bibinfo {year} {1988})}\BibitemShut {NoStop}%
\bibitem [{\citenamefont {Moore}(1970)}]{DynamicalCasimir1}%
  \BibitemOpen
  \bibfield  {author} {\bibinfo {author} {\bibfnamefont {G.~T.}\ \bibnamefont {Moore}},\ }\bibfield  {title} {\bibinfo {title} {Quantum theory of the electromagnetic field in a variable‐length one‐dimensional cavity},\ }\href {https://doi.org/10.1063/1.1665432} {\bibfield  {journal} {\bibinfo  {journal} {J. Math. Phys.}\ }\textbf {\bibinfo {volume} {11}},\ \bibinfo {pages} {2679} (\bibinfo {year} {1970})}\BibitemShut {NoStop}%
\bibitem [{\citenamefont {Kuo}\ and\ \citenamefont {Ford}(1993)}]{SqueezedState}%
  \BibitemOpen
  \bibfield  {author} {\bibinfo {author} {\bibfnamefont {C.-I.}\ \bibnamefont {Kuo}}\ and\ \bibinfo {author} {\bibfnamefont {L.~H.}\ \bibnamefont {Ford}},\ }\bibfield  {title} {\bibinfo {title} {Semiclassical gravity theory and quantum fluctuations},\ }\href {https://doi.org/10.1103/PhysRevD.47.4510} {\bibfield  {journal} {\bibinfo  {journal} {Phys. Rev. D}\ }\textbf {\bibinfo {volume} {47}},\ \bibinfo {pages} {4510} (\bibinfo {year} {1993})}\BibitemShut {NoStop}%
\bibitem [{\citenamefont {Ford}(1978)}]{Ford1978}%
  \BibitemOpen
  \bibfield  {author} {\bibinfo {author} {\bibfnamefont {L.~H.}\ \bibnamefont {Ford}},\ }\bibfield  {title} {\bibinfo {title} {Quantum coherence effects and the second law of thermodynamics},\ }\href {http://www.jstor.org/stable/79760} {\bibfield  {journal} {\bibinfo  {journal} {Proc. R. Soc. Lond. Series A, Math. Phys. Sci.}\ }\textbf {\bibinfo {volume} {364}},\ \bibinfo {pages} {227} (\bibinfo {year} {1978})}\BibitemShut {NoStop}%
\bibitem [{\citenamefont {Pfenning}(1998)}]{PfenningsThesis}%
  \BibitemOpen
  \bibfield  {author} {\bibinfo {author} {\bibfnamefont {M.~J.}\ \bibnamefont {Pfenning}},\ }\emph {\bibinfo {title} {Quantum inequality restrictions on negative energy densities in curved space-times}},\ \href@noop {} {Ph.D. thesis},\ \bibinfo  {school} {Tufts University} (\bibinfo {year} {1998})\BibitemShut {NoStop}%
\bibitem [{\citenamefont {Unruh}(1976)}]{Unruh1976}%
  \BibitemOpen
  \bibfield  {author} {\bibinfo {author} {\bibfnamefont {W.~G.}\ \bibnamefont {Unruh}},\ }\bibfield  {title} {\bibinfo {title} {Notes on black-hole evaporation},\ }\href {https://doi.org/10.1103/PhysRevD.14.870} {\bibfield  {journal} {\bibinfo  {journal} {Phys. Rev. D}\ }\textbf {\bibinfo {volume} {14}},\ \bibinfo {pages} {870} (\bibinfo {year} {1976})}\BibitemShut {NoStop}%
\bibitem [{\citenamefont {DeWitt}(1980)}]{DeWitt}%
  \BibitemOpen
  \bibfield  {author} {\bibinfo {author} {\bibfnamefont {B.}~\bibnamefont {DeWitt}},\ }\href@noop {} {\emph {\bibinfo {title} {General Relativity; an Einstein Centenary Survey}}}\ (\bibinfo  {publisher} {Cambridge University Press},\ \bibinfo {address} {Cambridge, UK},\ \bibinfo {year} {1980})\BibitemShut {NoStop}%
\bibitem [{\citenamefont {Lopp}\ and\ \citenamefont {Mart\'{i}n-Mart\'{i}nez}(2021)}]{richard}%
  \BibitemOpen
  \bibfield  {author} {\bibinfo {author} {\bibfnamefont {R.}~\bibnamefont {Lopp}}\ and\ \bibinfo {author} {\bibfnamefont {E.}~\bibnamefont {Mart\'{i}n-Mart\'{i}nez}},\ }\bibfield  {title} {\bibinfo {title} {Quantum delocalization, gauge, and quantum optics: Light-matter interaction in relativistic quantum information},\ }\href {https://doi.org/10.1103/PhysRevA.103.013703} {\bibfield  {journal} {\bibinfo  {journal} {Phys. Rev. A}\ }\textbf {\bibinfo {volume} {103}},\ \bibinfo {pages} {013703} (\bibinfo {year} {2021})}\BibitemShut {NoStop}%
\bibitem [{\citenamefont {Pozas-Kerstjens}\ and\ \citenamefont {Mart\'{i}n-Mart\'{i}nez}(2016)}]{Pozas2016}%
  \BibitemOpen
  \bibfield  {author} {\bibinfo {author} {\bibfnamefont {A.}~\bibnamefont {Pozas-Kerstjens}}\ and\ \bibinfo {author} {\bibfnamefont {E.}~\bibnamefont {Mart\'{i}n-Mart\'{i}nez}},\ }\bibfield  {title} {\bibinfo {title} {Entanglement harvesting from the electromagnetic vacuum with hydrogenlike atoms},\ }\href {https://doi.org/10.1103/PhysRevD.94.064074} {\bibfield  {journal} {\bibinfo  {journal} {Phys. Rev. D}\ }\textbf {\bibinfo {volume} {94}},\ \bibinfo {pages} {064074} (\bibinfo {year} {2016})}\BibitemShut {NoStop}%
\bibitem [{\citenamefont {Montero}\ \emph {et~al.}(2012)\citenamefont {Montero}, \citenamefont {del Rey},\ and\ \citenamefont {Mart\'{\i}n-Mart\'{\i}nez}}]{DelReyMonteroEdu}%
  \BibitemOpen
  \bibfield  {author} {\bibinfo {author} {\bibfnamefont {M.}~\bibnamefont {Montero}}, \bibinfo {author} {\bibfnamefont {M.}~\bibnamefont {del Rey}},\ and\ \bibinfo {author} {\bibfnamefont {E.}~\bibnamefont {Mart\'{\i}n-Mart\'{\i}nez}},\ }\bibfield  {title} {\bibinfo {title} {Nonmonotonic entanglement of physical electromagnetic field states in noninertial frames},\ }\href {https://doi.org/10.1103/PhysRevA.86.012304} {\bibfield  {journal} {\bibinfo  {journal} {Phys. Rev. A}\ }\textbf {\bibinfo {volume} {86}},\ \bibinfo {pages} {012304} (\bibinfo {year} {2012})}\BibitemShut {NoStop}%
\bibitem [{\citenamefont {Ford}\ and\ \citenamefont {Roman}(1999)}]{InterestConjecture}%
  \BibitemOpen
  \bibfield  {author} {\bibinfo {author} {\bibfnamefont {L.~H.}\ \bibnamefont {Ford}}\ and\ \bibinfo {author} {\bibfnamefont {T.~A.}\ \bibnamefont {Roman}},\ }\bibfield  {title} {\bibinfo {title} {The quantum interest conjecture},\ }\href {https://doi.org/10.1103/PhysRevD.60.104018} {\bibfield  {journal} {\bibinfo  {journal} {Phys. Rev. D}\ }\textbf {\bibinfo {volume} {60}},\ \bibinfo {pages} {104018} (\bibinfo {year} {1999})}\BibitemShut {NoStop}%
\end{thebibliography}%

\end{document}